\documentclass[aps,prd,showpacs,twocolumn,floatfix,nofootinbib]{revtex4-1} 
\usepackage{graphicx}
\usepackage{amsmath,amssymb,siunitx}
\usepackage{color}
\usepackage{hyperref}

\newcommand{\addsf}[1]{{\color{black} #1}} 
\newcommand{\addms}[1]{{\color{black} #1}} 
\newcommand{\addkh}[1]{{\color{black} #1}} 

\begin{document}
\title{General-relativistic neutrino-radiation magnetohydrodynamic simulation of seconds-long black hole-neutron star mergers}
\author{
  Kota Hayashi$^{1}$,
  Sho Fujibayashi$^{2}$,
  Kenta Kiuchi$^{2,1}$,
  Koutarou Kyutoku$^{3,1,4}$,
  Yuichiro Sekiguchi$^{5,1}$,  Masaru Shibata$^{2,1}$
}
\affiliation{
  $^1$Center for Gravitational Physics, Yukawa Institute for Theoretical Physics,
  Kyoto University, Kyoto 606-8502, Japan\\
  $^2$Max Planck Institute for Gravitational Physics (Albert Einstein Institute),
  Am M{\"u}hlenberg 1, Postdam-Golm 14476, Germany\\
  $^3$Department of Physics, Kyoto University, Kyoto 606-8502, Japan\\
  $^4$Interdisciplinary Theoretical and Mathematical Sciences Program (iTHEMS), RIKEN, Wako, Saitama 351-0198, Japan \\
  $^5$Department of Physics, Toho University, Funabashi, Chiba 274-8510, Japan
}
\date{\today}

\begin{abstract}
Seconds-long numerical-relativity simulations for black hole-neutron star mergers are 
performed for the first time to obtain a self-consistent picture of the merger 
and post-merger evolution processes. To investigate the case that tidal disruption 
takes place, we choose the initial mass of the black hole to be $5.4M_\odot$ or 
$8.1M_\odot$ with 
the dimensionless spin of 0.75. The neutron-star mass is fixed to be $1.35M_\odot$. 
We find that after the tidal disruption, dynamical mass ejection takes place 
spending $\alt 10$\,ms together with the formation of a massive accretion disk. 
Subsequently, the magnetic field in the disk is amplified by the magnetic winding 
and magnetorotational instability, establishing a turbulent state and 
inducing the angular momentum transport. The post-merger mass ejection 
by the magnetically-induced viscous effect sets in 
at $\sim 300$--500\,ms after the tidal disruption, at which 
the neutrino luminosity drops below $\sim 10^{51.5}\,{\rm erg/s}$, and 
continues for several hundreds ms. A magnetosphere near the rotational axis of the 
black hole is developed after the matter and magnetic flux fall into the black hole 
from the accretion disk, and high-intensity Poynting flux generation sets in 
at a few hundreds ms after the tidal disruption. The intensity of the Poynting 
flux becomes low after the significant post-merger mass ejection, because the 
opening angle of the magnetosphere increases. The lifetime for the stage with 
the strong Poynting flux is $1$--2\,s, which agrees with the typical 
duration of short-hard gamma-ray bursts. 
\end{abstract}

\maketitle

\section{Introduction} \label{sec:intro}
The opening for the era of the gravitational-wave astronomy was heralded  
by the first observation of a binary black hole merger, referred to as 
GW150914~\cite{abbott2016feb}. 
To date, $\sim 80$ binary black hole merger events have been already observed 
by advanced LIGO and advanced Virgo~\cite{abbott2021jun,ligoo3b}. 
In addition to binary black holes, a couple of 
neutron-star binaries have been also observed. In particular, 
associated with the first binary neutron star merger event, GW170817~\cite{abbott2017oct1}, a wide variety of the signals of the electromagnetic counterpart were successfully detected~\cite{abbott2017oct2, abbott2017oct3}, and the multi-messenger astronomy including gravitational-wave observation was opened from this event. 

In addition, the gravitational-wave signals from the black hole-neutron star mergers 
referred to as GW200105 and GW200115 were detected in 2020~\cite{abbott2021jun2}.
These events surely \addms{indicate} that 
black hole-neutron star binaries exist in nature. 
Although any electromagnetic counterpart was not detected for these two events,
a number of numerical-relativity simulations for the black hole-neutron star mergers 
have predicted that tidal disruption of the neutron star and subsequent mass ejection 
should take place if the parameters of the source (black-hole mass, black-hole spin, 
and neutron-star compactness) are in an appropriate range~\cite{shibata2011aug,KST2021}. 
If the remnant black hole is rapidly spinning and is surrounded by a magnetized massive torus, an ultra-relativistic jet is 
likely to drive a short-hard gamma-ray burst~\cite{eichler1989,nakar2007apr,berger2014jun}.
In the presence of mass ejection, the $r$-process nucleosynthesis inevitably proceeds~\cite{lattimer1974,eichler1989},  
and subsequently, the ejecta should shine with a high luminosity associated with the 
thermal energy generated by the radioactive decay of neutron-rich heavy elements~\cite{li1998nov,metzger2010jun}. Since the sensitivity of gravitational-wave 
detectors and electromagnetic telescopes is improved year by year, it is quite natural to expect that electromagnetic counterparts of black hole-neutron star mergers are 
observed in the near future, if the distance to the source is within several hundreds Mpc, 
and thus, black hole-neutron star mergers are among 
the promising sources for the multi-messenger astronomy. 
In this situation, the theoretical studies to elucidate the entire process from 
the tidal disruption to the post-merger evolution
are required to predict the observable signals. 

In the last 15 years, a variety of numerical-relativity simulations have been performed for 
black hole-neutron star mergers~\cite{shibata2006dec,shibata2007may, shibata2008apr,etienne2008apr, duez2008nov, shibata2009feb, etienne2009feb, chawla2010sep, duez2010may, kyutoku2010aug, kyutoku2011sep, foucart2011jan, foucart2012feb, etienne2012mar, etienne2012oct, kyutoku2013aug, kyutoku2015aug, foucart2013apr, lovelace2013jun, deaton2013sep, foucart2014jul, paschalidis2015jun, kawaguchi2015jul, kiuchi2015sep, foucart2017jan, kyutoku2018jan, brege2018sep, ruiz2018dec, foucart2019feb, foucart2019may, hinderer2019sep, hayashi2021feb, foucart2021mar, chaurasia2021oct, most2021jul} improving the input physics and grid resolutions, 
and the processes of tidal disruption and subsequent accretion disk formation,  
merger remnants, dynamical mass ejection, gravitational waves, and neutrino emissivity have been extensively studied. 
However, all these work have focused primarily on the inspiral to early merger stages, and hence, 
the long-term (seconds-long) post-merger process has not been explored in these simulations. To compensate for this drawback, 
many numerical simulations (viscous hydrodynamics or magnetohydrodynamics simulations) have been also performed for exploring the long-term 
evolution of the accretion disks (or tori) around a black hole~\cite{fernamdez2013aug,metzger2014may,just2015feb,fernandez2015mar,fernandez2017jul,daniel2018may,fernandez2018oct,agnieszka2019sep,christie2019sep,miller2019jul,fujibayashi2020apr,fujibayashi2020dec,Li2021jun,fernandez2020jul,just2022jan,shibata2021sep}, and have clarified the post-merger mass ejection 
mechanisms and the properties of the post-merger ejecta. These work have reported 
that the post-merger mass ejection is driven primarily by a viscous hydrodynamics 
effect induced by the magnetohydrodynamics turbulence in the accretion disks 
and by a Lorentz force associated with the amplified magnetic fields. 
Although these work are important for understanding the post-merger mass ejection 
mechanisms, the initial conditions for the simulations were ad hoc or some of important physical inputs were absent, and hence, 
the conclusive quantitative details such as quantitative properties of the post-merger ejecta have not been fully understood yet. 

In order to acquire the full understanding of the black hole-neutron star mergers 
and associated mass ejection processes, 
we need to perform a self-consistent simulation starting from an inspiral stage 
throughout the post-merger stage. Specifically, the post-merger evolution 
has to be followed at least for a few seconds, because the post-merger 
mass ejection takes place spending the timescale of $\agt 1$\,s. 
Furthermore, to explore the generation mechanism of short-hard gamma-ray bursts, 
a simulation with the duration of $\agt 1$~s is needed because their typical duration is $\sim 1$\,s with the longest duration of $\sim 2$\,s~\cite{nakar2007apr,berger2014jun}. Keeping in mind these timescales, 
in this paper, we \addms{tackle this problem by performing} general-relativistic neutrino-radiation magnetohydrodynamics simulations of black hole-neutron star mergers for $\approx 1$--2~s.  Here, we 
emphasize that both the neutrino radiation transfer and magnetohydrodynamics effects are inevitable elements for determining the evolution of the merger remnant. In this long-term simulation with the relevant physics, the 
magnetohydrodynamics turbulence and associated angular-momentum transport 
in the accretion disk are naturally taken into account, 
and furthermore, a black-hole magnetosphere in the vicinity of the rotation 
axis of the remnant spinning black hole, which could be suitable for generating 
a short-hard gamma-ray burst, also naturally emerges. 

This paper is organized as follows.
In Sec.~\ref{sec:methods}, we briefly summarize the method and initial setup for the numerical simulation.
In Sec.~\ref{sec:results}, we present the numerical results focusing on 
the entire evolution process, mass ejection mechanisms, and collimated 
electromagnetic outflow developed near the rotation axis of the black hole. 
Finally, we  conclude this work in Sec.~\ref{sec:conclusion}. 
Throughout this paper, we adopt the geometrical units in which $G = c = 1$,
where $G$ and $c$ are the gravitational constant and the speed of light, respectively.

\section{Methods} \label{sec:methods}

Our numerical implementation for the present simulations is the same as 
that in Ref.~\cite{kyutoku2018jan} except for the ideal magnetohydrodynamics part 
for which we implement the scheme used in Ref.~\cite{kiuchi2015sep}. 
Specifically, we solve Einstein's equation by a puncture-Baumgarte-Shapiro-Shibata-Nakamura (BSSN) formalism
\cite{shibata1995nov,baumgarte1998dec,campanelli2006mar,baker2006mar,marronetti2008mar},
incorporating a Z4c-type constraint-propagation prescription~\cite{hilditch2013oct,kawaguchi2015jul}.
In this work, the original version of the BSSN formalism~\cite{shibata1995nov} is 
employed. 
The fourth-order finite-differencing scheme is applied to discretize the gravitational-field equation.
Magnetohydrodynamics equations are solved in a high-resolution shock capturing scheme \cite{shibata2005aug,shibata2007aug,kiuchi2012sep} together with the second-order 
constrained-transport scheme~\cite{evans1988sep} and Balsara's flux-preserving mesh refinement scheme~\cite{balsara2009}. 
Neutrino transfer is handled using a leakage-based scheme~\cite{sekiguchi2012oct}
together with a truncated moment formalism using a closure relation for the free-streaming component~\cite{thorne1981feb,shibata2011jun}.
Neutrino heating and absorption on free nucleons are incorporated using the updated numerical procedure~\cite{fujibayashi2020sep}. \addms{We do not take into account the neutrino pair annihilation effect in this paper}.

The simulation is performed using a fixed-mesh refinement (FMR) algorithm with the equatorial symmetry imposed at $z=0$. 
The $i$-th refinement level covers a half cubic region of 
$[-L_i:L_i] \times [-L_i:L_i] \times [0:L_i]$ where $L_i=N\Delta x_i$
and $\Delta x_i$ is the grid spacing for the $i$-th level.
The grid spacing for each level is determined by $\Delta x_i=2\Delta x_{i+1}$ 
($i=1,2, \cdots , i_{\mathrm{max}}-1$)
with $\Delta x_{i_{\mathrm{max}}}=400\,\mathrm{m}$ for low-resolution runs and $\Delta x_{i_{\mathrm{max}}}=270\,\mathrm{m}$ for high-resolution runs. 
$i_{\rm max}$ is chosen to be 9 or 10. The values of $N$ 
are $170$ or 192 for low-resolution runs and 234 or 282 for high-resolution runs, respectively (cf.~Table~\ref{tab:init_cond}). 

During the merger stage, the black hole is kicked mainly by the back reaction 
of the dynamical mass ejection and the resulting velocity is $v_{\rm kick}=200$--400\,km/s 
(\addms{which is estimated by $m_{\rm ej}v_{\rm ej}/M_{\rm BH}$ with $m_{\rm ej}$ dynamical ejecta mass, $v_{\rm ej}(\sim 0.2c)$ its absolute average velocity, and $M_{\rm BH}$ the remnant black hole mass}) in our present setting. 
Thus, the black hole moves toward a refinement boundary of the finest 
FMR level with time and eventually escapes from the highest-resolution level in the absence of any prescription. 
To keep the black hole in the highest-resolution level, we control the shift vector 
by modifying the evolution equation in the following prescription: 
\begin{eqnarray}
  \partial_{t} \beta^{i} &=& \frac{3}{4} \tilde{\gamma}^{i j}\left(F_{j}+\partial_{t} F_{j} \Delta t\right) - {v_{\rm BH}^i \over T_{\rm relax}} \nonumber  \\
  &&{\rm for}~T_{\rm sta} < t < T_{\rm sta}+T_{\rm relax}, \\
  \partial_{t} \beta^{i} &=& \frac{3}{4} \tilde{\gamma}^{i j}\left(F_{j}+\partial_{t} F_{j} \Delta t\right) \nonumber \\
  &&{\rm for~other~cases},
\end{eqnarray}
where $\beta^i$ is the shift vector,
$\tilde{\gamma}_{ij}$ is the conformal three-metric,
$F_{i}=\delta^{j k} \partial_{j} \tilde{\gamma}_{k i}$ is the auxiliary variable 
in the original version of the BSSN formalism,
$\Delta t$ is the time-step interval,
and $v_{\rm BH}^i$, $T_{\rm relax}$, and $T_{\rm sta}$ are constants which we 
determine appropriately based on the numerical result. $v_{\rm BH}^i$ is the coordinate velocity of the black-hole center (the location of the puncture) just 
before modifying the shift vector, which is of order $10^{-3}c$ as we already mentioned.
$T_{\rm relax}$ is the relaxation time, which we choose $T_{\rm relax}=40~{\rm ms}$.
$T_{\rm sta}$ is the starting time of this prescription, and it is set to be 
$T_{\rm sta} \approx 100$--$200~{\rm ms}$ \addms{to satisfy 
$v_{\rm kick}(T_{\rm sta}+T_{\rm relax}) \alt L_{i_{\rm max}}/2$}.

We stop the time evolution of the gravitational field at a certain moment
after the ratio of the rest mass of the remnant disk to the black-hole mass drops below $10^{-2}$. This prescription is reasonable because the self-gravity 
of the matter located outside the black hole can be safely neglected 
and the gravitational field is approximately stationary in such a low-mass disk stage.

For modeling the neutron-star matter, we employ a nuclear-theory-based finite-temperature equation of state (EOS) referred to as DD2~\cite{banik2014sep} for a high-density range and Helmholtz EOS~\cite{timmes2000} for a low-density range \addsf{(see Appendix~\ref{app:eos} for our method of the construction of the EOS and Appendix~\ref{appendB} for the heating effect due to the nuclear reactions)}.
Initial data are given by calculating a quasi-equilibrium state of black hole-neutron star binaries in a quasi-circular orbit assuming the neutrinoless beta-equilibrium cold state~\cite{kyutoku2018jan}.
The initial gravitational mass of the neutron star 
is set to be $M_{\rm NS}=1.35M_{\odot}$ following Ref.~\cite{kyutoku2018jan}.
The circumferential radius of the isolated spherical neutron star of mass 
$1.3$--$1.4M_{\odot}$ is $\approx 13.2\,\mathrm{km}$ with this EOS, \addms{and it satisfies constraints imposed by the observation of gravitational waves for GW170817~\cite{abbott2017oct1} and by the X-ray observation by NICER}~\addkh{\cite{miller2019dec}}.

For the initial black-hole mass, we choose $M_{\rm BH,0}=5.4M_{\odot}$ or $8.1M_{\odot}$; 
the mass ratio of the black hole to the neutron star is 
$Q:=M_{\rm BH,0}/M_{\rm NS}=4$ or 6. 
The initial dimensionless spin parameter of the black hole is set to be $0.75$. With 
such a spin, tidal disruption of the neutron star with $M_{\rm NS}=1.35M_\odot$ takes place 
for a wide range of $Q$. 
The initial orbital angular velocity $\Omega_0$ is set to be $m_0 \Omega_0 = 0.056$ 
for $Q=4$ and $0.064$ for $Q=6$,
where $m_0$ is the sum of the initial black-hole mass and neutron-star mass, i.e., 
$m_0=M_{\rm BH,0}+M_{\rm NS}=1.35(Q+1)M_\odot$.
In this initial setup, the binary merges after about three orbits. 
We note that the binary parameter for $Q=4$ is the same as that employed for the 
DD2 EOS in our previous paper~\cite{kyutoku2018jan}. 

\begin{table*}[]
  \centering

  \caption{
   Key parameters and quantities for the initial conditions together with the parameters of grid setup for our numerical simulations. $M_{\rm BH,0}$: the initial black-hole mass, $b_{\rm 0,max}$: the initial maximum magnetic-field strength, $\Omega_0$: the initial orbital angular velocity,  
   $M_{\rm ADM,0}$: the initial ADM mass, $\Delta x_{i_{\rm max}}$: the grid spacing for the finest refinement level, $L_1$: the location of the outer boundaries along each axis, and the values of $N$ and $i_{\rm max}$. 
   For all the models, the neutron-star mass 
   is $1.35M_\odot$ and the initial dimensionless black-hole spin is 0.75. 
   Note that $M_{\rm ADM,0}$ is by $\sim 1\%$ smaller than $m_0=(6.75$ and $9.45M_\odot$ 
   for $Q=4$ and 6) because of the presence of the gravitational binding energy. 
  }
  \label{tab:init_cond}
  
  \begingroup
  \setlength{\tabcolsep}{4pt} 
  \renewcommand{\arraystretch}{1.2} 
  
  \begin{tabular}{ccccccccc}
    \hline
    \hline
    model name           & $M_{\rm BH,0} ~[M_{\odot}]$ & $b_{\rm 0,max} ~[{\rm G}]$ & $m_0 \Omega_0$   &$M_{\rm ADM,0} ~[M_{\odot}]$& $\Delta x_{i_{\rm max}} ~[{\rm m}]$  & $L_1 ~[{\rm km}]$ & ~~~$N$~~~   & ~~$i_{\rm max}~~$\\
    \hline
Q4B5H  & 5.400   & $5\times 10^{16}$  & 0.056 & 6.679 & 270 & $1.62\times 10^4$ & 234 & 9\\
Q4B5L  & 5.400   & $5\times 10^{16}$  & 0.056 & 6.679 & 400 & $1.74\times 10^4$ & 170 & 9\\
Q4B3L  & 5.400   & $3\times 10^{16}$  & 0.056 & 6.679 & 400 & $1.74\times 10^4$ & 170 & 9\\
Q6B5H  & 8.100   & $5\times 10^{16}$  & 0.064 & 9.368 & 270 & $3.90\times 10^4$ & 282 & 10\\
Q6B5L  & 8.100   & $5\times 10^{16}$  & 0.064 & 9.368 & 400 & $3.97\times 10^4$ & 194 & 10\\
Q6B3H  & 8.100   & $3\times 10^{16}$  & 0.064 & 9.368 & 270 & $3.90\times 10^4$ & 282 & 10\\
Q6B3L  & 8.100   & $3\times 10^{16}$  & 0.064 & 9.368 & 400 & $3.97\times 10^4$ & 194 & 10\\
    \hline
  \end{tabular}

  \endgroup

\end{table*}

We initially superimpose a poloidal magnetic field confined in the neutron star.
Following our previous work~\cite{kiuchi2015sep}, 
the poloidal field is given in terms of the vector potential as
\begin{eqnarray} \label{init_b_field}
  A_j &= & \{ -(y - y_{\mathrm{NS}}) \delta_j^{~x} + (x - x_{\mathrm{NS}} ) \delta_j^{~y} \} \nonumber \\
  && \times  A_b \max(P/P_{\mathrm{max}} - 10^{-3} , 0)^2,
\end{eqnarray}
where $(x_{\mathrm{NS}},y_{\mathrm{NS}})$ is the coordinate position of the 
neutron-star center (location of the maximum rest-mass density) on the orbital plane,
$P$ is the pressure, $P_{\mathrm{max}}$ is the maximum pressure,
and $j = x, y$, and $z$.
$A_b$ is a constant and is chosen so that the initial maximum magnetic-field strength $b_{\mathrm{0,max}}$ is
$3 \times 10^{16}~\mathrm{G}$ or $5 \times 10^{16}~\mathrm{G}$.
These values are chosen to obtain a strong magnetic field in the remnant disk formed 
after tidal disruption of the neutron star in a short timescale after the merger. 
The strong magnetic field is required to resolve 
the fastest growing mode of the magnetorotational instability (MRI)~\cite{balbus1991,balbus1998} in the accretion disk with the limited 
grid resolution, because its wavelength is
proportional to the magnetic-field strength. 
Although such strong fields are not realistic in orbiting neutron stars, 
the resulting turbulent state in the accretion disk 
established by the MRI is not likely to 
depend strongly on the initial magnetic-field strength.\footnote{\addms{That is, we implicitly assume that the magnetic-field strength would be increased by the MRI and a turbulent state would be eventually established even if we started a simulation from low magnetic-field strengths (as is often done in this research field). This is just an assumption, and confirming this by better-resolved simulations remains as an issue for the future.}} 
Thus, it \addms{would be reasonable to suppose} that the resulting strong magnetic field and turbulent state will be established even for the case that 
we start a simulation from a much weaker magnetic-field strength in 
the presence of a sufficient grid resolution. We also note that even with 
$b_{\mathrm{0,max}}=5\times 10^{16}$\,G, the electromagnetic energy 
(of order $10^{49}$\,erg) is much smaller than the internal energy and gravitational 
potential energy (of order $10^{53}$\,erg) of the neutron star. 
We do not consider the effect of the neutrino viscosity to the MRI supposing that 
the magnetic-field strength could be enhanced to be $\agt 10^{14}$\,G due to 
the rapid winding in the main region of the accretion disk (see Sec.~\ref{sec:results-disk}) even if the early growth of the MRI is suppressed~\cite{masada2008,guilet2016}. 

We perform 7 simulations changing the black-hole mass, value of $b_{\mathrm{0,max}}$,  and grid resolution. The parameters and quantities for the 7 models employed in this study 
are summarized in Table~\ref{tab:init_cond}. 
Numerical simulations with the low-resolution setting are 
always performed for the duration of $\geq 1$\,s. In particular for $Q=4$ models, 
the low-resolution simulations are performed for $\agt 2$\,s. On the other hand, 
the high-resolution simulations are performed only for $\alt 1$\,s  
because such simulations require an extremely high computational cost. However, as we  show below, the results for the low-resolution runs are quantitatively similar to 
those for the corresponding high-resolution runs, and hence, we consider that 
a fair convergence is achieved even with the low-resolution runs. 
The computational time with the low-resolution setting for 2\,s is 
about 1400 hours using 64 nodes of our Sakura cluster in which 1 node has 
2 Intel Xeon Gold 6248 CPUs (1 node has 40 cores). 

\section{Results} \label{sec:results}

\subsection{Overview of the evolution} \label{sec:results-overview}
\begin{figure*}[]
  \begin{tabular}{c}

    \begin{minipage}{1.0\hsize}
      \begin{center}
        \includegraphics[scale=0.09]{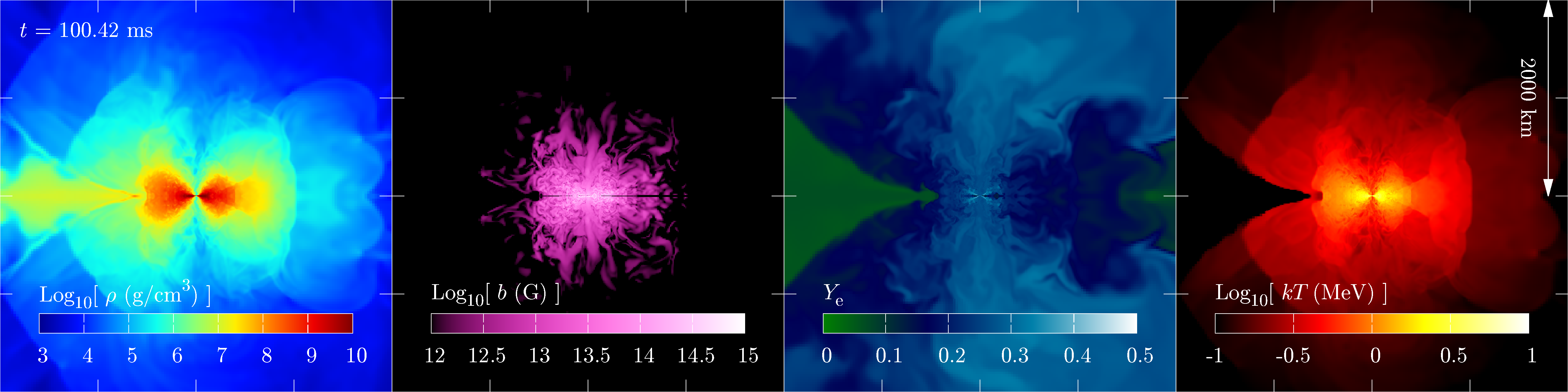}
        \includegraphics[scale=0.09]{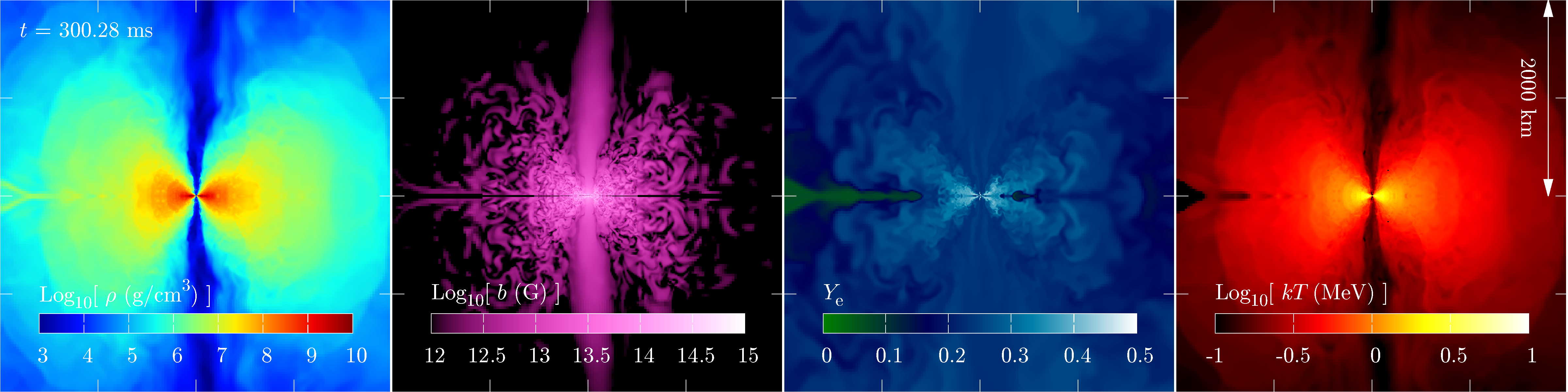}
        \includegraphics[scale=0.09]{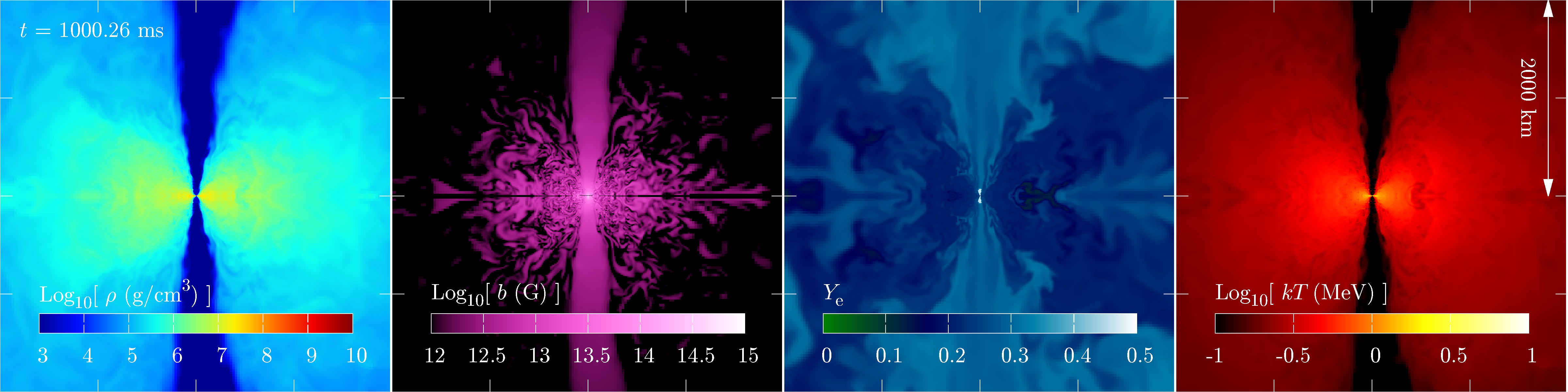}
        \includegraphics[scale=0.09]{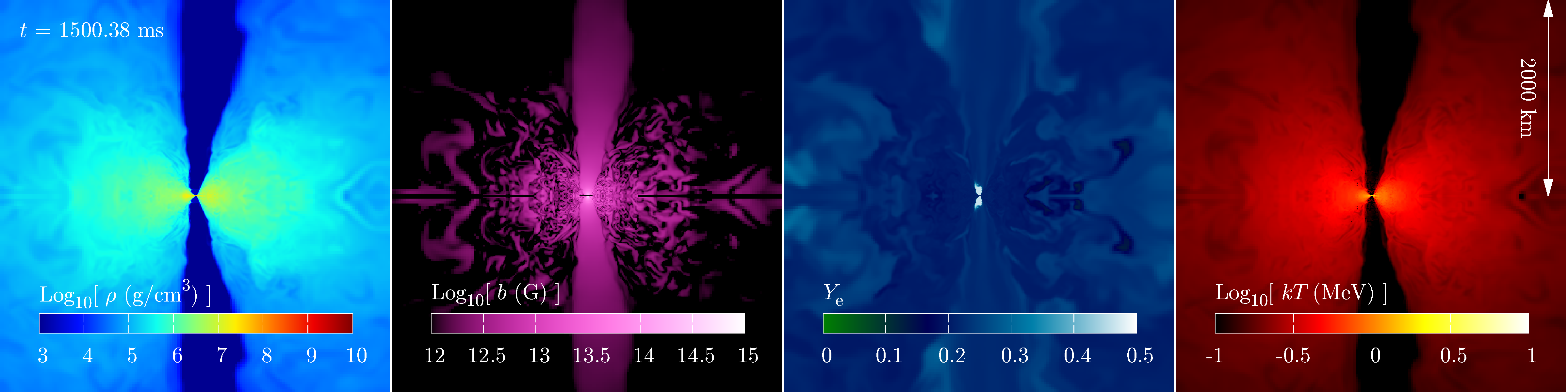}
        \includegraphics[scale=0.09]{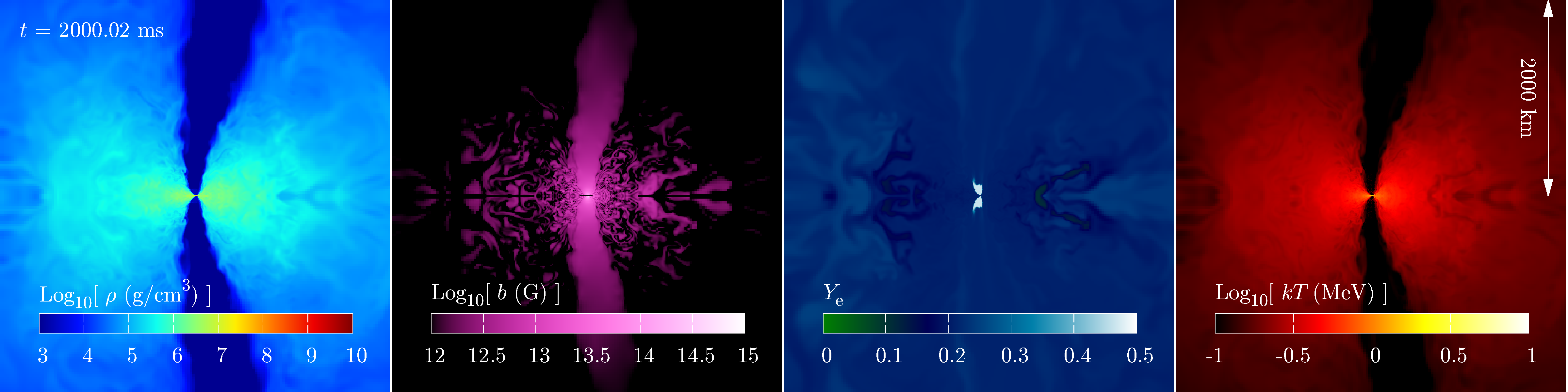}

        \caption{The snapshot for the rest-mass density $\rho~({\rm g/cm^3})$, 
        magnetic-field strength $b=\sqrt{b^{\mu}b_{\mu}}~({\rm G})$,
          electron fraction $Y_{\rm e}$, and temperature $T$ ($kT$ in units of MeV) on the $x$-$z$ plane with $[-2000~{\rm km}:2000~{\rm km}]$ for both $x$ and $z$ at $t\approx 0.1$, $0.3$, $1.0$, $1.5$, and $2.0$\,s for model Q4B5L. 
    Note that the green region in $Y_{\rm e}$ found in the left side at the first and second rows shows the dynamical ejecta and fall-back matter. 
    See also an animation: \url{https://www2.yukawa.kyoto-u.ac.jp/~kota.hayashi/Q4B5L-2000a.mp4 }.
        }
        \label{fig:xz-snap}
      \end{center}
    \end{minipage}

  \end{tabular}
\end{figure*}

First, we summarize the entire merger process found in a seconds-long simulation 
presenting the result for model Q4B5L for which the system was 
evolved up to $\sim 2.1$\,s.
Figure~\ref{fig:xz-snap} displays the snapshot for the rest-mass density, 
absolute value of the magnetic-field strength, 
electron fraction  $Y_{\rm e}$, and temperature $T$, respectively, on the $x$-$z$ plane. 
The magnetic-field strength is defined by $b=(b_\mu b^\mu)^{1/2}$ where $b^\mu$ is the magnetic field in the frame comoving with fluid and the temperature is shown by multiplying the Boltzmann's constant $k$ and in units of MeV. 

\begin{table}[]
  \centering

  \caption{
   The mass $M_{\rm BH}$ and the dimensionless spin parameter $\chi_{\rm BH}$ of the remnant black hole evaluated at $t=100~{\rm ms}$ together with the gravitational-wave and neutrino energy emitted before $t=100$\,ms, $E_{\rm GW}$ and $E_\nu$, and the rest mass of the matter located outside the black hole at $t=100$\,ms, $M_{>{\rm AH,0.1}}$. 
   All the quantities related to the mass or energy are described in units of $M_{\odot}$. 
  }
  \label{tab:bh_mass_spin}
  \begingroup
  \setlength{\tabcolsep}{4pt} 
  \renewcommand{\arraystretch}{1.2} 
  
  \begin{tabular}{cccccc}
    \hline
    \hline
    model   &~$M_{\rm BH}$~ & ~$\chi_{\rm BH}$~ & ~$E_{\rm GW}$~  & ~$E_{\nu}$~ 
    & ~$M_{\rm >AH,0.1}$~\\
    \hline
    Q4B5H & 6.466 & 0.856 & 0.069 & 0.008 & 0.129\\
    Q4B5L & 6.400 & 0.838 & 0.066 & 0.008 & 0.135\\
    Q4B3L & 6.396 & 0.838 & 0.066 & 0.008 & 0.138\\
    Q6B5H & 9.145 & 0.837 & 0.117 & 0.007 & 0.097\\
    Q6B5L & 9.138 & 0.832 & 0.112 & 0.007 & 0.104\\
    Q6B3H & 9.145 & 0.838 & 0.117 & 0.007 & 0.097\\
    Q6B3L & 9.136 & 0.833 & 0.112 & 0.007 & 0.106\\
    \hline
    \hline
  \end{tabular}
  \endgroup

\end{table}

\begin{figure*}[th]
  \begin{tabular}{c}
        \includegraphics[scale=0.6]{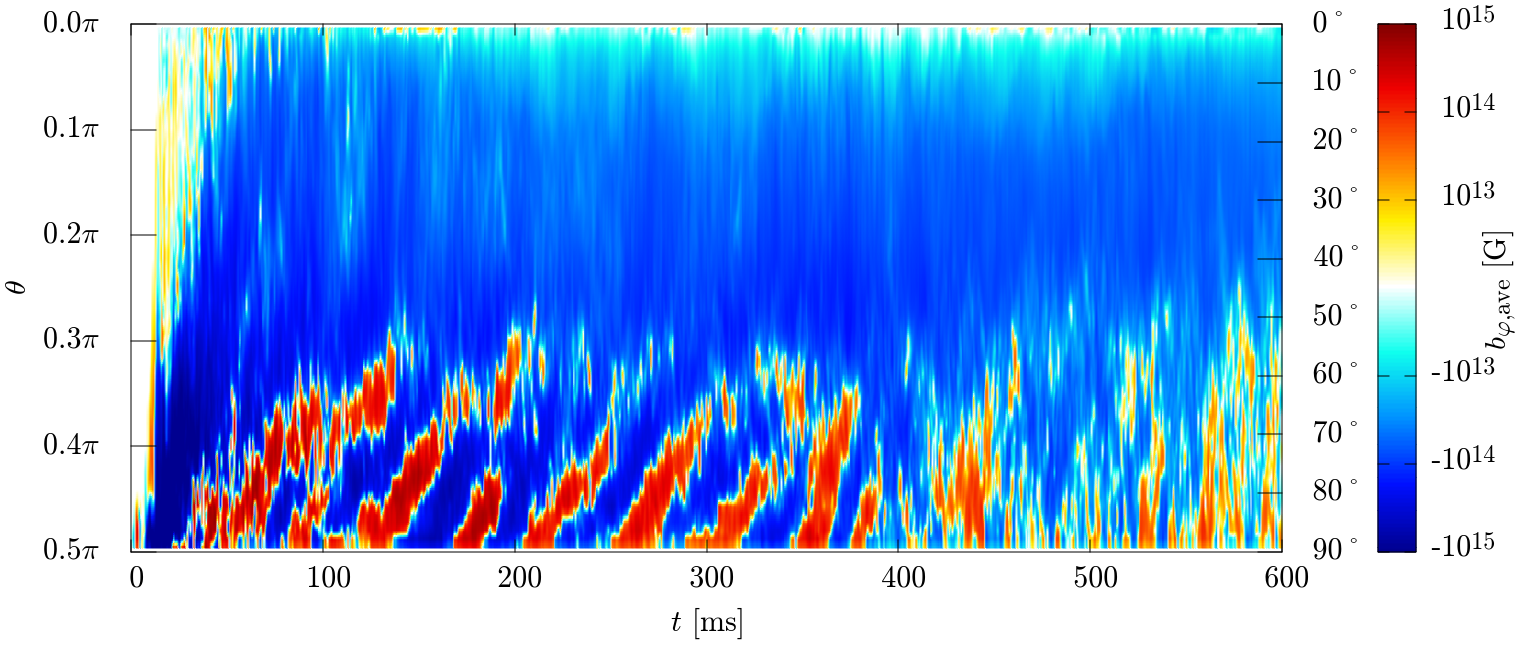}
        \\
        \\
        \includegraphics[scale=0.6]{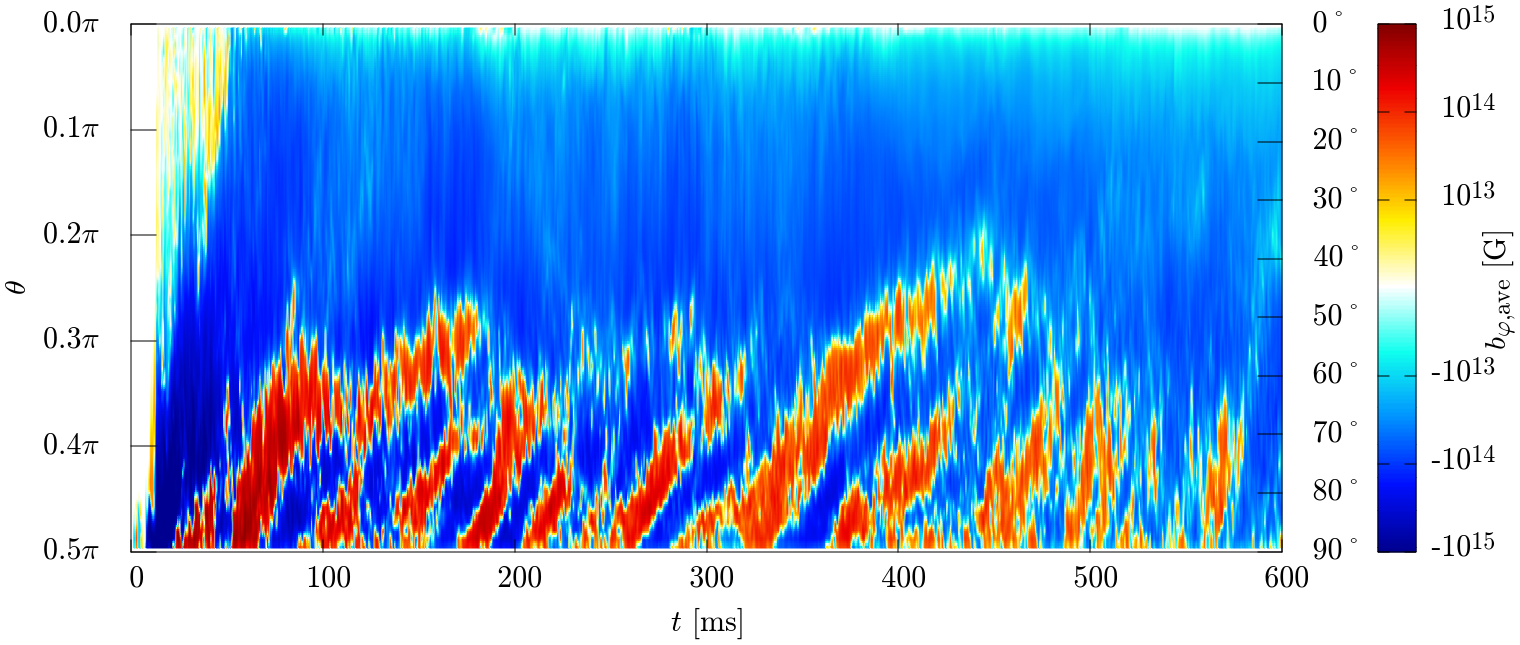}
  \end{tabular}
  \caption{The profile of the average toroidal magnetic field along the polar
  direction ($\theta$) at $r \approx 50~{\rm km}$ as a function of time for models Q4B5L (top panel) 
        and Q4B5H (bottom panel).
        }
    \label{fig:BUT_Q4_B5_low}
\end{figure*}

In the present choice of the dimensionless spin parameter for the black hole and 
the fairly large radius of the neutron star,  
the neutron star is tidally disrupted by the black hole before the binary reaches 
the innermost stable circular orbit both for $Q=4$ and 6. During the tidal disruption  process, the neutron-star matter located in the black-hole side falls into 
the black hole. Specifically, 
$\sim80$\% of the neutron-star matter falls into the black hole 
in a short timescale of a few ms. On the other hand, the neutron-star matter 
located distant from the black hole forms a one-armed spiral structure. Due to the subsequent angular-momentum transport inside the spiral arm and the dynamical evolution of the black-hole spacetime resulting from the matter infall into it, a part of the matter in the outer part of the spiral arm gains specific energy and angular momentum. 
The matter which gains sufficient specific energy eventually becomes dynamical ejecta,
while the other part in the spiral arm which is bound to the remnant black hole 
forms an accretion disk. The timescale of this stage is $\alt 10$\,ms 
(see the first row of Fig.~\ref{fig:xz-snap} for the resulting state).
All these processes have been studied by a number of previous numerical-relativity 
work and our present result on the tidal disruption and disk formation processes is essentially the same as the previous findings.

The mass, $M_{\rm BH}$, and the dimensionless spin parameter, $\chi_{\rm BH}$, of the remnant black holes evaluated at $t=100~{\rm ms}$ are summarized in Table~\ref{tab:bh_mass_spin}. 
Irrespective of the runs, the black-hole mass and dimensionless spin are increased by 
$\approx 1.05M_\odot$ and $\sim 0.1$, respectively, due to the matter infall. 
The black-hole mass is by $\sim 0.3M_\odot$ smaller than the initial 
Arnowitt-Deser-Misner (ADM) mass. The reason for this is that a part of the 
neutron-star matter forms an accretion disk and ejecta, and in addition, 
gravitational waves and neutrinos 
carry away the energy (see Tables~\ref{tab:init_cond} and \ref{tab:bh_mass_spin}) 
in the inspiral and early merger stages. 

We also list the total gravitational-wave and neutrino energy emitted before $t=100$\,ms, $E_{\rm GW}$ and $E_\nu$, and 
the rest mass of the matter located outside the apparent horizon at $t=100$\,ms, 
$M_{>{\rm AH,0.1}}$, in Table~\ref{tab:bh_mass_spin}. By comparing 
$M_{\rm BH}$ and $M_{\rm ADM,0}-E_{\rm GW}-E_\nu-M_{>{\rm AH,0.1}}$, we 
can assess how good (or bad) the energy conservation is satisfied in our 
simulation. It is found that for $Q=4$, the energy conservation is satisfied 
with about 0.1\% and 1.1\% error for high- and low-resolution runs of $Q=4$ model, and 
with $\ll 0.1\%$ and about 0.1\% error for high- and low-resolution runs of $Q=6$ model, respectively. 
The reason that the accuracy depends strongly on the grid resolution for $Q=4$ 
(i.e.,~for the smaller black-hole mass)
is that the accuracy for resolving the black hole depends strongly on it. 
This is found by taking a look at the value of the black-hole mass for $Q=4$: 
For the low-resolution runs, the black-hole mass is underestimated. 
However, the error of $\alt 1\%$ at $t=100$\,ms is still in an acceptable level, 
indicating the reliability of the numerical results. 


After the spiral arm winds around the black hole, a compact accretion disk is formed. 
The orbital period at the innermost region of the accretion disk is 1--2~ms. 
During the tidal disruption process, the neutron-star matter which eventually 
forms an accretion disk experiences a strong differential rotation stage in the 
spiral arm, and then, a toroidal magnetic field is developed from the 
initially poloidal magnetic field by winding. After the formation of the accretion disk, 
the winding continues to enhance the toroidal magnetic-field strength, 
in particular in the innermost region of the accretion disk. After the sufficient 
amplification of the magnetic-field strength, an outward expansion of the matter 
is driven toward the polar direction due to the enhanced magnetic pressure, 
and as a result, poloidal fields for which the strength is comparable to 
that of the toroidal fields are also generated. 
With these strong magnetic fields, the wavelength for the fastest growing mode of the MRI becomes $\sim 10$\,km and can be numerically resolved \addms{(see Appendix~\ref{appendixC})}. 
Then, a turbulent state associated with the MRI is developed, and eventually, 
an MRI dynamo is activated in the accretion disk.
This can be also observed from a spacetime diagram of the toroidal-field strength. 
In Fig.~\ref{fig:BUT_Q4_B5_low}, we plot the average value of the toroidal field 
as a function of time and polar angle 
$\theta=\tan^{-1}(\sqrt{x^2+y^2}/z)$ for models Q4B5L and Q4B5H. 
Here, $x$, $y$, and $z$ are defined with respect to the black-hole center. 
The toroidal field is defined by $b_{\bar\varphi}=(xb_y-yb_x)/\sqrt{x^2+y^2}$.
The average is performed with respect to the azimuthal angle $\varphi=\tan^{-1}(y/x)$ 
at the selected radius of $r:=\sqrt{x^2+y^2+z^2} \approx 50$\,km. 
From Fig.~\ref{fig:BUT_Q4_B5_low}, we find the so-called butterfly structure~\cite{Brandenburg2005} irrespective of the grid resolution: 
The polarity of the toroidal magnetic field is reversed due to the turbulent motion 
in a periodic manner with the period of $\sim 20$ local orbital periods 
($\approx 2.5$\,ms).\footnote{\addms{After the post-merger mass ejection sets in at $t \sim 400$\,ms (cf. Sec.~\ref{sec:results-disk}), the periodic butterfly diagram is not observed. However, it is still seen that the polarity of the magnetic field changes with time due to the presence of the turbulent motion. The decrease of the toroidal magnetic-field strength is due to the disk expansion.}} It is also found that strong magnetic-field 
regions move from the accretion disk to the polar region in the early stage, 
producing a global magnetic-field structure (see also the magnetic-field strength  
in the second row of Fig.~\ref{fig:xz-snap}). 

During this turbulent stage, the angular momentum is transported from the inner to the outer region of the accretion disk due to the effective viscosity induced by the 
turbulence. In addition to this effective viscous process, 
magnetohydrodynamics effects such as the magneto-centrifugal effect~\cite{blandford1982} 
which results from a global magnetic field \addms{could} play an 
important role for expelling the matter from the central region. Due to 
these effects, the matter near the innermost stable circular orbit loses 
its angular momentum and falls into the black hole, 
while the matter in the outer part of the disk receives the angular momentum and expands gradually. As a result, the rest-mass density and the temperature in the disk decrease in the viscous timescale of order 100\,ms to $1$\,s (see the third to fifth rows of Fig.~\ref{fig:xz-snap}). 

In addition to the disk expansion toward the equatorial direction, the matter expands  toward the direction perpendicular to the orbital plane (see the entire panels of  Fig.~\ref{fig:xz-snap}). Our interpretation for 
this expansion is that the magnetic tower effect plays a role: During the evolution 
of the accretion disk, the toroidal magnetic-field strength is enhanced by the MRI and 
winding. As a result, the magnetic pressure is enhanced to be high enough for the accretion disk to expand toward the direction perpendicular to the orbital plane (and thus the disk becomes a torus), while the serious baryon contamination 
in the vicinity of the rotational axis is prevented by the centrifugal force of the matter. This effect produces a funnel structure around the rotational axis (see the 
second to fifth rows of Fig.~\ref{fig:xz-snap}).\footnote{\addms{In the late stage with $t \agt 1.5$\,s, the funnel has an asymmetric structure. This is caused by the fall-back of the matter in the tidal tail that is formed predominantly for the negative $x$ direction at tidal disruption. This fall-back also lowers the electron fraction near the black hole in the late phase. }} 

\addms{In spite of the enhanced magnetic-field strength, we do not find appreciable early-post-merger mass ejection (which might occur within 100--200\,ms after the onset of the merger) associated with this enhancement. The absence of the clear early post-merger mass ejection agrees with some of the results found in Ref.~\cite{christie2019sep} in which the initial magnetic-field profile is chosen to be toroidal or weakly poloidal. Only in several previous magnetohydrodynamics studies~\cite{daniel2018may,fernandez2018oct,miller2019jul,christie2019sep} in which a strong poloidal magnetic field is given, the early post-merger mass ejection was found. In our simulations, the magnetic-field profile in the early stage of the post-merger evolution is primarily toroidal. Thus, we consider that the early post-merger mass ejection takes place only for the case that a strong poloidal field is present in the disk at the formation of the remnant disk, although our result indicates that such strong poloidal fields are not likely to be formed soon after the merger of black hole-neutron star binaries.}

Not only the magnetohydrodynamics effect but also the neutrino cooling plays an 
important role for the evolution of the accretion disk~\cite{fernamdez2013aug}. In the early stage of the accretion disk, the maximum density is $\agt 10^{12}\,{\rm g/cm^3}$ and the maximum temperature is several MeV.
In addition to the high density and high temperature, 
the disk is massive with the mass $\agt 0.1M_\odot$ in the early stage. 
In such a stage, neutrino luminosity becomes higher 
than $10^{53}\,{\rm erg/s}$ which is comparable to or higher than 
the viscous heating rate for a 
compact disk with a high viscous parameter~\cite{fujibayashi2020apr}. 
During the stage that the neutrino luminosity is as high as the rate of the viscous 
heating \addms{(and the shock heating associated with the magnetohydrodynamical activity in the present context)}, the matter in the accretion disk is not affected significantly by the heating effect, although the accretion disk gradually expands due to 
the viscous/magnetohydrodynamics angular-momentum transport and magnetic pressure 
resulting from the enhanced magnetic-field strength. However, with the expansion, 
the density and temperature of the accretion disk decrease, and consequently, 
the neutrino luminosity sharply decreases because the neutrino emissivity 
is approximately proportional to $T^6$~\cite{fuller1985jun}. 
As the neutrino luminosity drops below the heating rate due to the viscous and magnetohydrodynamics activities, neutrinos 
cannot efficiently carry away the thermal energy from the accretion disk and the 
thermal energy generated by the viscous/magnetohydrodynamics effect influences the evolution of the 
accretion disk. Specifically, convective motion of the matter at the 
innermost region of the disk, in which the viscous heating and shock heating are most efficient, is excited and blobs of the matter heated in the vicinity of the 
black hole are moved toward the outer region of the disk \addms{along the surface of the disk}.\footnote{See the following animation for the entropy per baryon ($s/k$) and for the convective activity: \\ 
\url{https://www2.yukawa.kyoto-u.ac.jp/~kota.hayashi/Q4B5L-2000a.mp4},
\addkh{and \url{https://www2.yukawa.kyoto-u.ac.jp/~kota.hayashi/Q4B5L_sent.mp4}.}}
As a result, 
the matter in the outer part of the disk obtains the thermal energy and 
the heated matter eventually becomes unbound from the system to be the post-merger ejecta 
(cf.~the second and third rows of Fig.~\ref{fig:xz-snap}). This mechanism is 
the same as that found in \addms{the previous viscous hydrodynamics  simulations~\cite{fernamdez2013aug,metzger2014may,just2015feb,fujibayashi2020apr} (see also Ref.~\cite{Metzger2008})}. 
This post-merger mass ejection continues from $0.2$--0.3\,s to $\sim 1$\,s after the merger (i.e., after the formation of the accretion disk). We note that in addition to this \addms{convective} effect, purely magnetohydrodynamical effects such as magneto-centrifugal effect~\cite{blandford1982} could also play a role for the mass ejection. 


In parallel with the accretion-disk evolution, a magnetosphere is 
developed in the low-density region near the rotational axis (see Fig.~\ref{fig:xz-snap}). 
For the merger of black hole-neutron star binaries that experience 
tidal disruption, such a low-density region is naturally developed because the 
matter is primarily ejected toward the  equatorial direction. 
During the magnetohydrodynamics evolution of the accretion disk, a mass outflow  
toward the direction perpendicular to the equatorial plane is driven by the 
activity of the accretion disk. However, 
the density in the vicinity of the rotation axis is still preserved to 
be low because of the presence of the 
centrifugal force on the injected matter. Thus the accretion of the matter into 
the black hole proceeds primarily from the innermost region of the disk. 
In ideal magnetohydrodynamics, 
the accretion of the matter accompanies the infall of the magnetic flux 
into the black hole. Although the magnetic field comoving with the infalling matter 
falls together into the black hole, the magnetic-field line located outside 
the black hole can expand toward the outer direction in particular along the 
rotational axis which has low matter density and low gas pressure (see Sec.~\ref{sec:results-outflow}). Such magnetic 
fields eventually develop a magnetosphere for which the 
magnetic-field lines are nearly aligned with the rotational axis 
(except for the vicinity of the black hole).\footnote{
Although we find it in our present simulation, it is not conclusive 
whether an aligned magnetic field with constant polarity is always formed or not: 
see, e.g., Refs.~\cite{christie2019sep,liska2020,shibata2021sep} for related work.} 
The magnetic pressure in such a region is lower than the gas pressure of the surrounding 
thick torus which is formed after the activity of the accretion disk is enhanced (see the second to fifth rows of Fig.~\ref{fig:xz-snap}). In other word, the size of the magnetosphere 
is determined by the structure of the thick torus. 

The magnetic-field lines penetrate the black hole spinning 
rapidly with the dimensionless spin $\agt 0.8$, and thus, the system can be subject 
to the Blandford-Znajek mechanism~\cite{blandford1977} by which the 
rotational kinetic energy of the black hole is converted to the outgoing 
Poynting flux. In the presence of the matter for which the rest-mass energy 
density is comparable to or larger than the electromagnetic energy density, 
the Poynting flux cannot propagate away efficiently. However, the density in the polar region decreases with time because the matter in the vicinity of the  
black hole falls into the black hole and a part of the matter is expelled  
by the magnetic pressure. Hence, eventually, 
electromagnetic waves generated by the Blandford-Znajek effect can propagate away 
(cf.~the second to fifth rows of Fig.~\ref{fig:xz-snap}). 
If an efficient conversion of the electromagnetic energy to the 
kinetic energy of the matter occurs during the subsequent propagation, 
a gamma-ray burst jet may be launched. 
Since the magnetic field has a collimated structure, the electromagnetic 
emission is also collimated. 
This collimated emission continues as far as the gas pressure of the 
thick and dense torus confines the magnetosphere (see Sec.~\ref{sec:results-outflow}). 

We note that the evolution processes described above are qualitatively universal  
irrespective of the black-hole mass, initial magnetic-field strength, and grid resolution employed in this paper.
In the following subsections, we describe 
the quantitative details about the accretion disk evolution, mass ejection, 
and generation of strong Poynting flux in the magnetosphere separately. 

\subsection{The evolution of the accretion disk and post-merger mass ejection} \label{sec:results-disk}

\subsubsection{Disk evolution and ejecta}

\begin{figure*}[th]
      \begin{center}
        \includegraphics[scale=0.38]{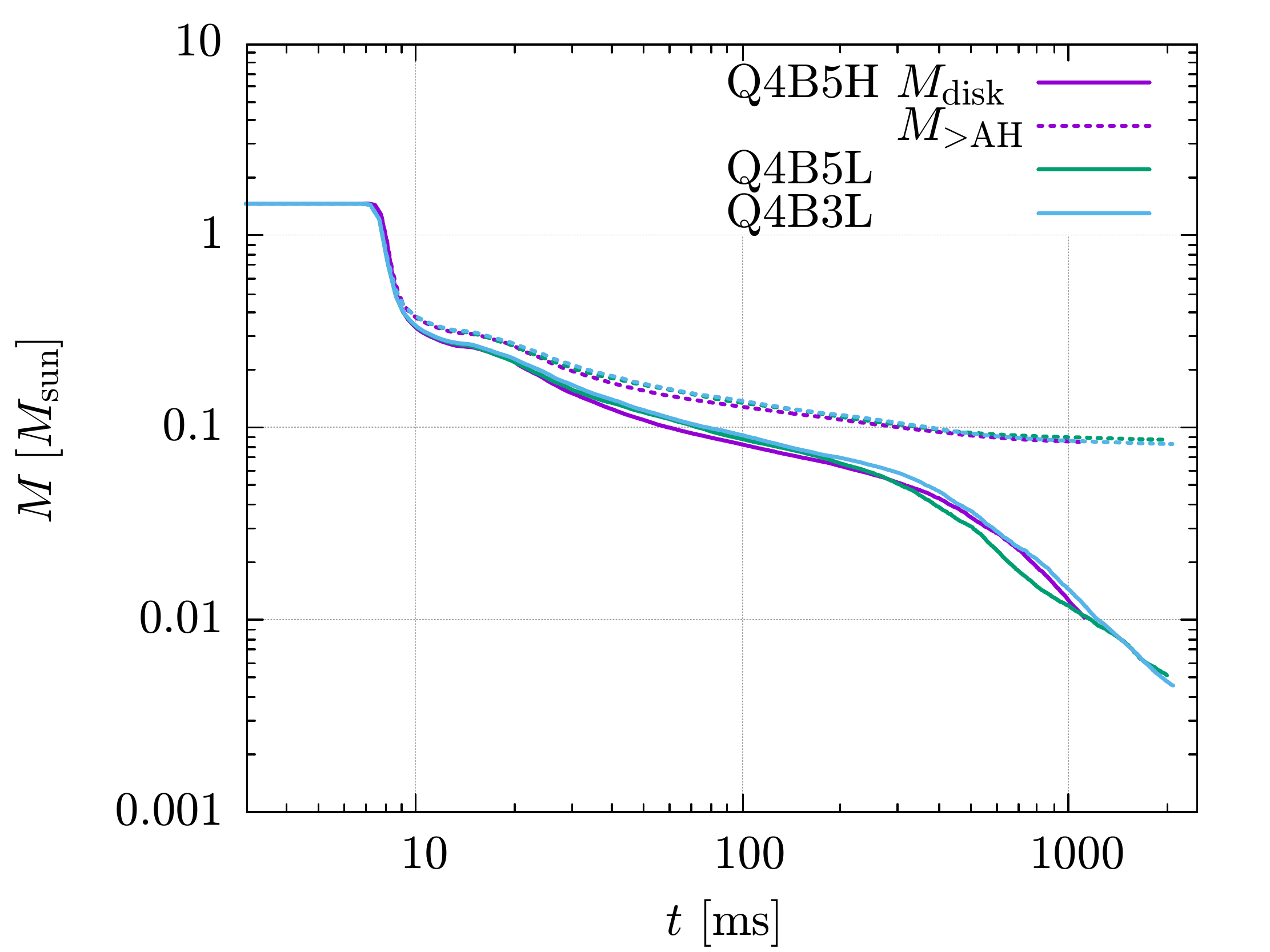}~~~~~
        \includegraphics[scale=0.38]{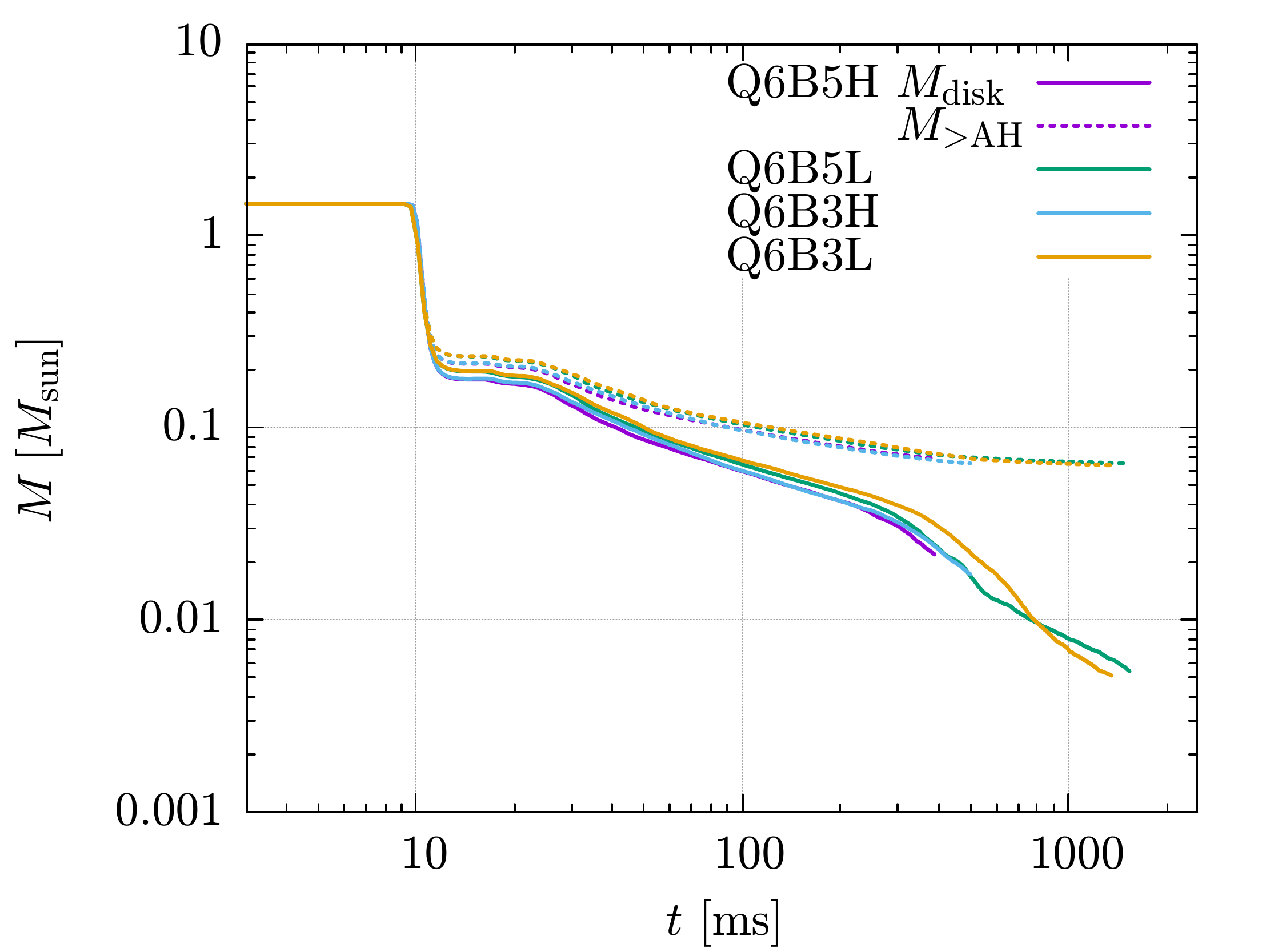}
        \caption{The time evolution of the rest mass of the matter located outside the apparent horizon (dashed curves) and the accretion-disk mass (solid curves) 
        for all the runs with $Q=4$ (left panel) and $Q=6$ (right panel). 
	}
        \label{fig:mrem}
      \end{center}
\end{figure*}

\begin{figure*}[th]

      \begin{center}
        \includegraphics[scale=0.38]{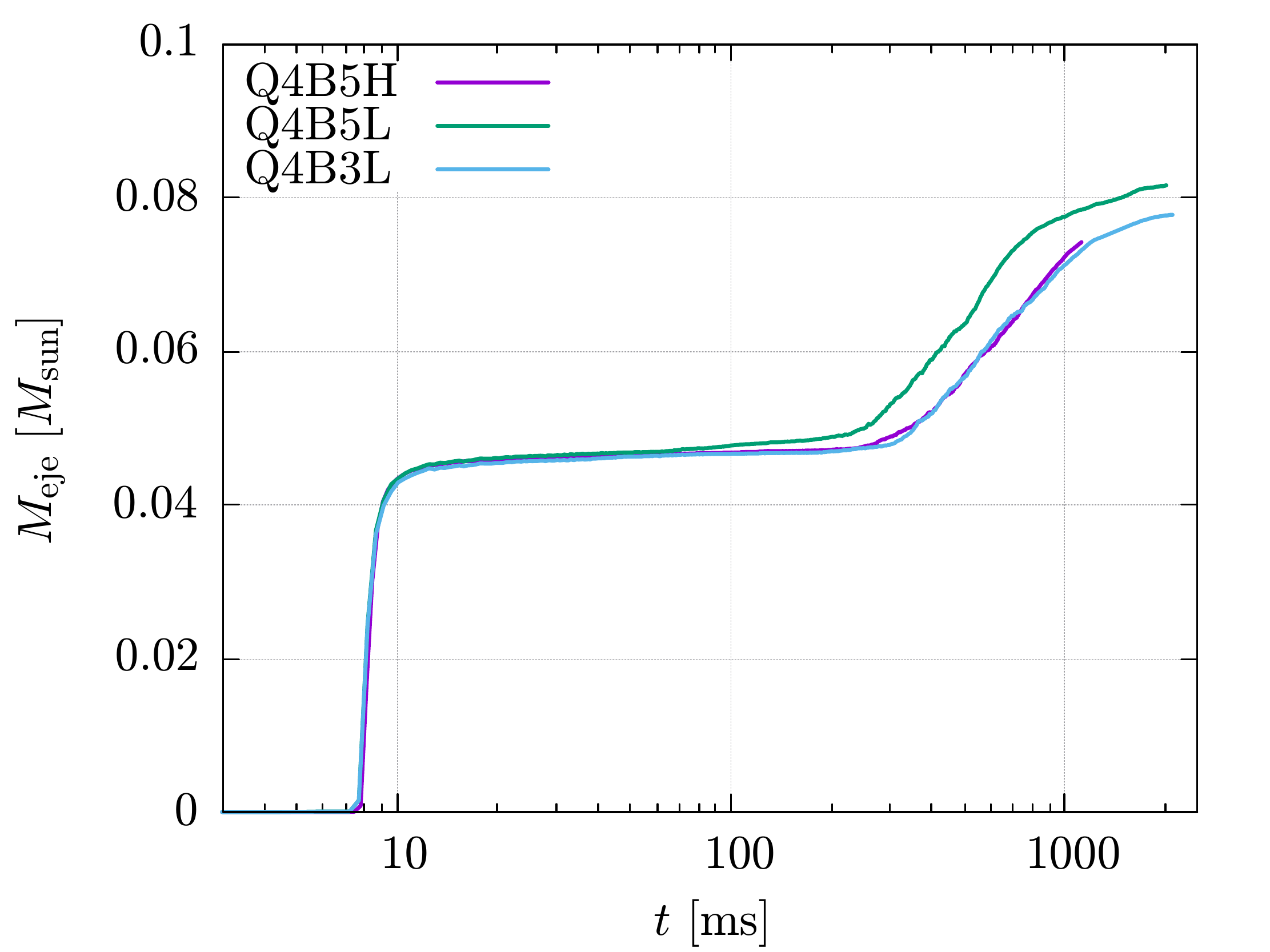}~~~~~
        \includegraphics[scale=0.38]{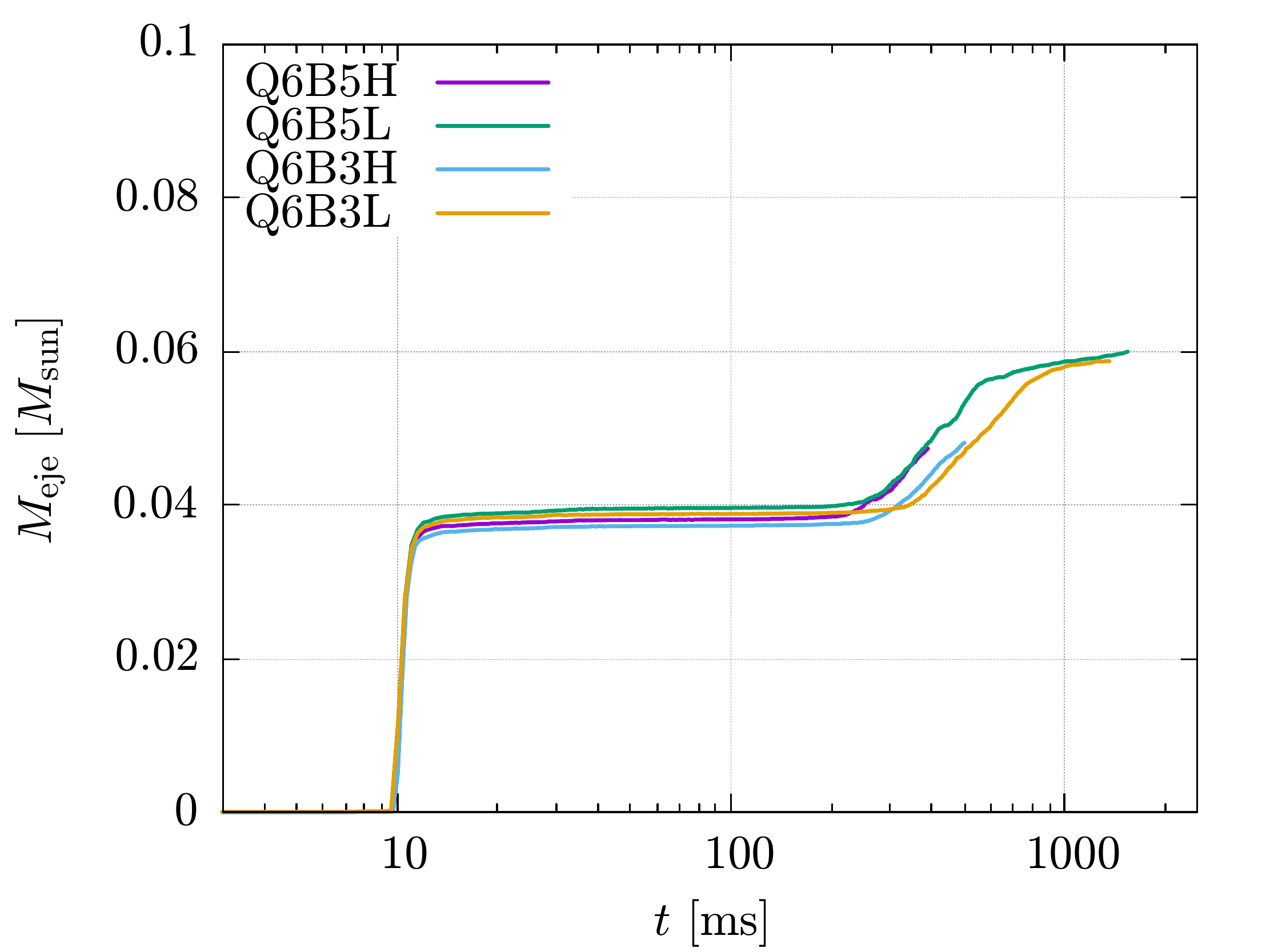}
        \caption{The time evolution of the rest mass of the unbound matter (ejecta) for all the runs with $Q=4$ (left panel) and $Q=6$ (right panel). 
	}
        \label{fig:meje}
      \end{center}
\end{figure*}

In this subsection, we present the quantitative details on the evolution of the accretion disk and on the mass ejection. Figure~\ref{fig:mrem} shows 
the rest mass of the matter located outside the apparent horizon $M_{>\mathrm{AH}}$ 
(dashed curves) and the accretion disk mass $M_{\rm disk}$ (solid curves) as functions of time. Figure~\ref{fig:meje} shows the rest mass of the unbound matter (ejecta) $M_{\mathrm{eje}}$ as a function of time. These quantities are defined by
\begin{eqnarray}
  M_{> {\rm AH}}&:=&\int_{r>r_{\rm AH}}\rho_*d^3x + M_{\rm esc}, \\
  M_{\rm eje}&:=&\int_{-hu_{t}>h_{\rm min},r>r_{\rm AH}}\rho_*d^3x + M_{\rm esc}, \\
  M_{\rm disk}&:=&M_{> {\rm AH}}-M_{\rm eje},
\end{eqnarray}
where $\rho_*:=\rho \sqrt{-g} u^t$ with $g$ the determinant of the 
spacetime metric, $g_{\mu\nu}$, $u^t$ the 
time component of the four velocity, $u^\mu$, and $r_{\rm AH}$ denotes the 
coordinate radius of the apparent horizon with the respect to the 
black-hole puncture.  
$M_{\rm esc}$ denotes the rest mass escaping from the computational domain, 
which is calculated from
\begin{eqnarray}
  \dot{M}_{\rm esc}&:=&\oint \rho \sqrt{-g} u^{i} d S_{i}, \\
  M_{\rm esc}&:=&\int^{t} \dot{M}_{\rm esc} d t.
\end{eqnarray}
The surface integral is performed near the outer boundaries of the computational domain.

\begin{figure*}[t]

      \begin{center}
        \includegraphics[scale=0.38]{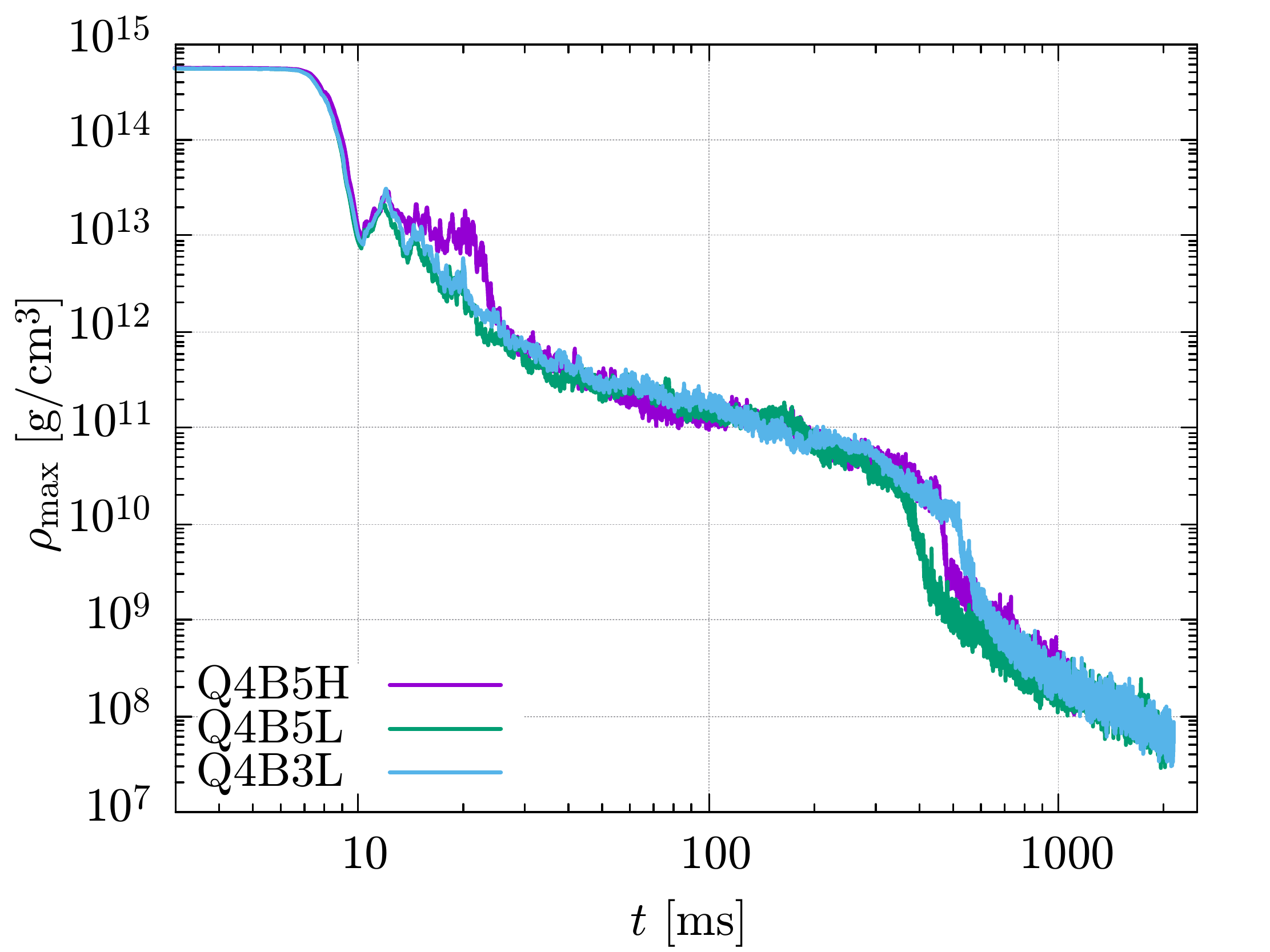}~~~~~
        \includegraphics[scale=0.38]{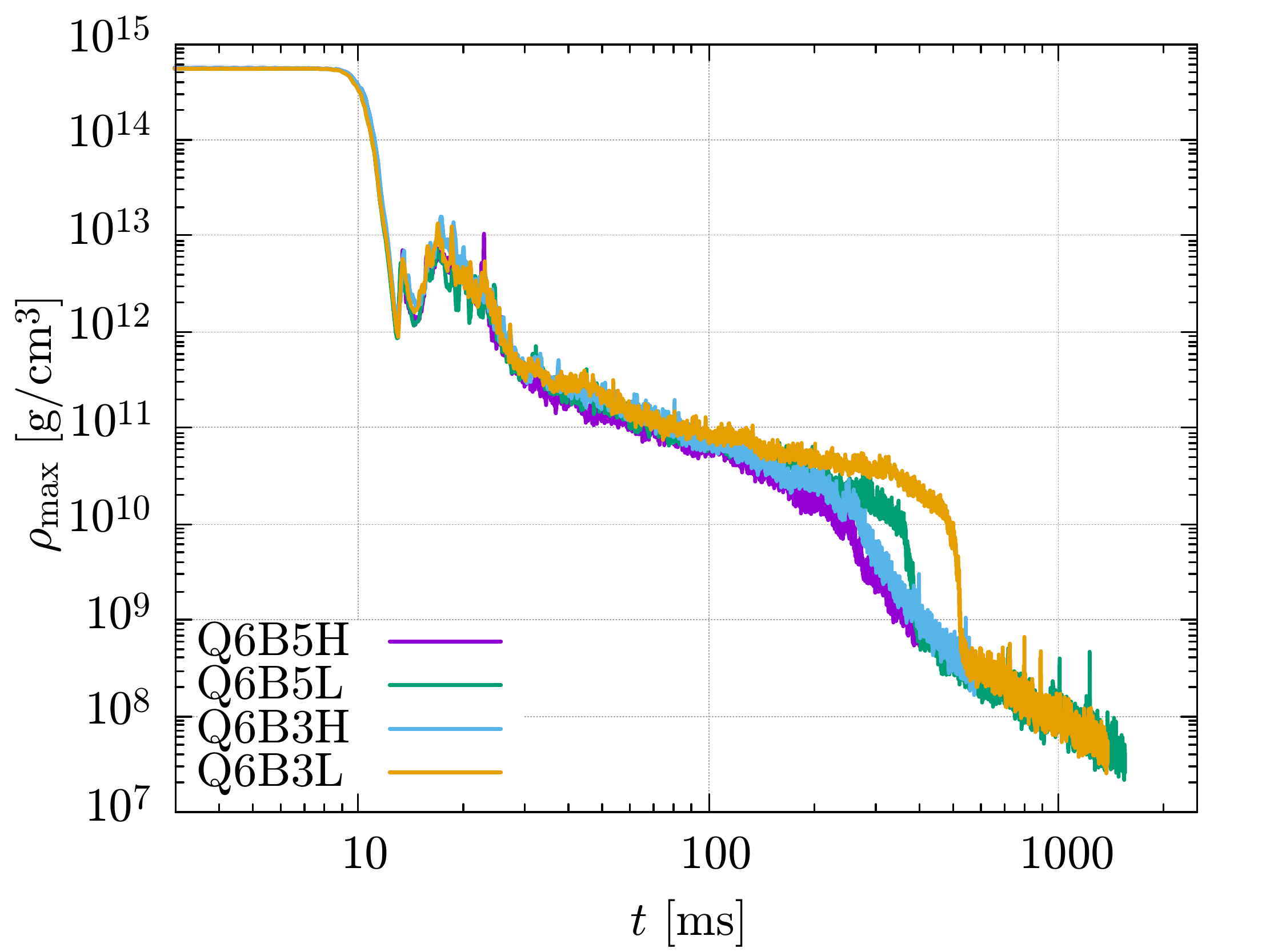}
        \caption{The time evolution of the maximum rest-mass density of the bound matter located outside the apparent horizon for all the runs with $Q=4$ (left panel) and $Q=6$ (right panel).
	}
        \label{fig:rhomax}
      \end{center}

\end{figure*}

\begin{figure*}[t]

      \begin{center}
        \includegraphics[scale=0.38]{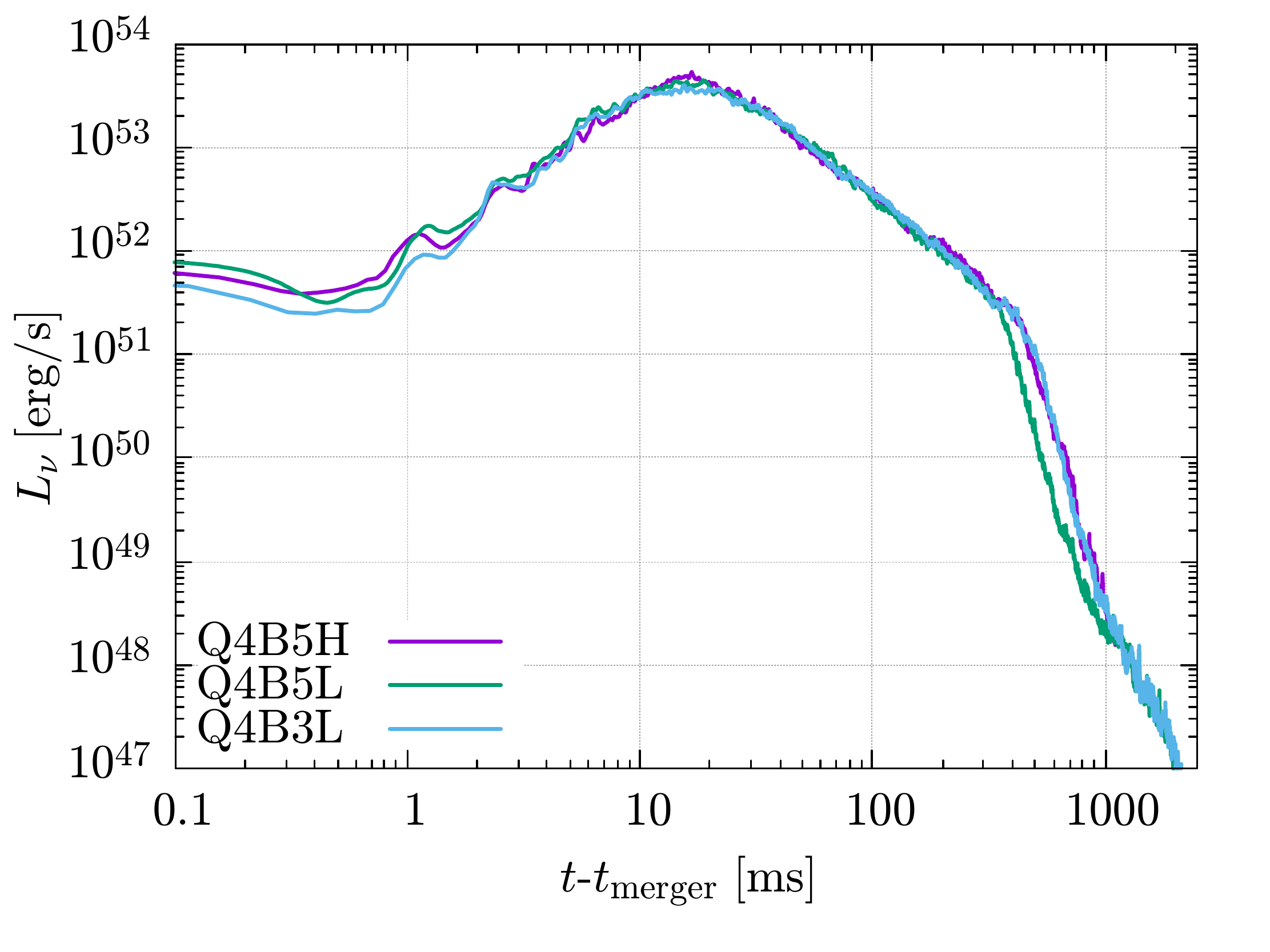}~~~~~
        \includegraphics[scale=0.38]{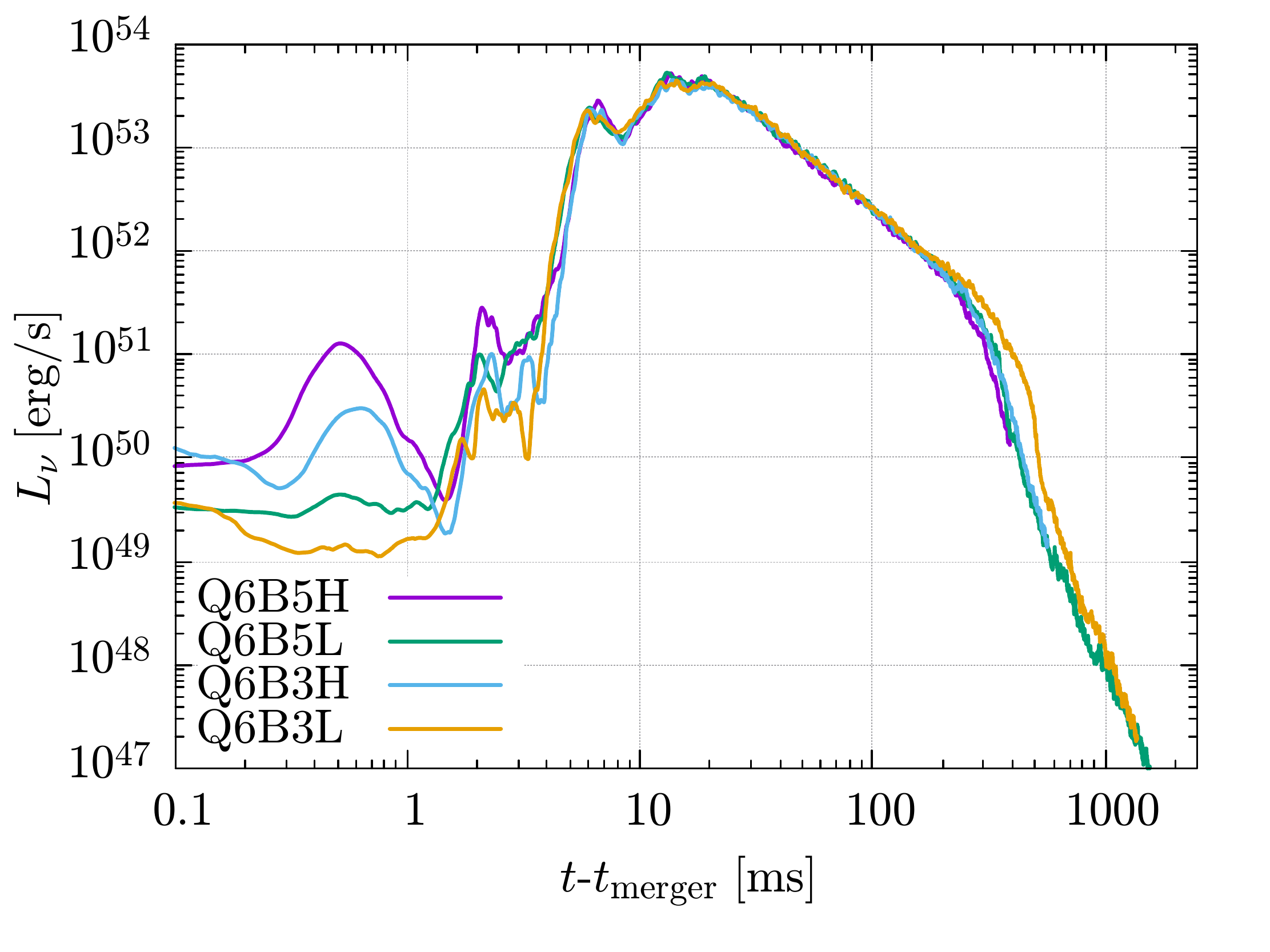}
        \caption{The time evolution of the total neutrino luminosity (sum of the 
        luminosity for all the neutrino species) for all the runs 
        with $Q=4$ (left panel) and $Q=6$ (right panel).
	The post-merger mass ejection sets in at $t-t_{\rm merger} \sim 300$--500\,ms at which $L_\nu \sim 10^{51.5}~{\rm erg/s}$. 
	}
        \label{fig:nlum}
      \end{center}
\end{figure*}

\begin{figure*}[t]

      \begin{center}
        \includegraphics[scale=0.38]{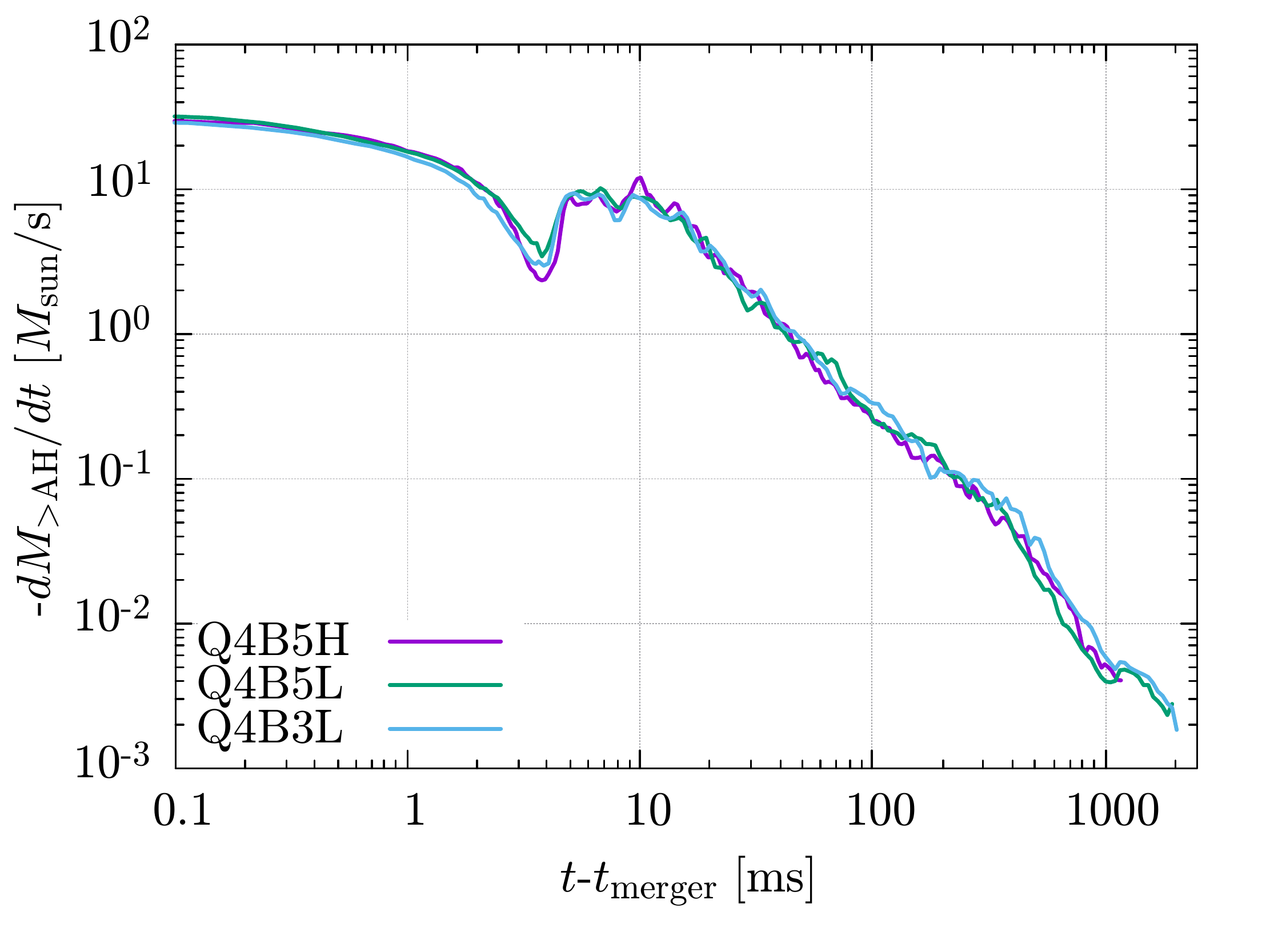}~~~~~
        \includegraphics[scale=0.38]{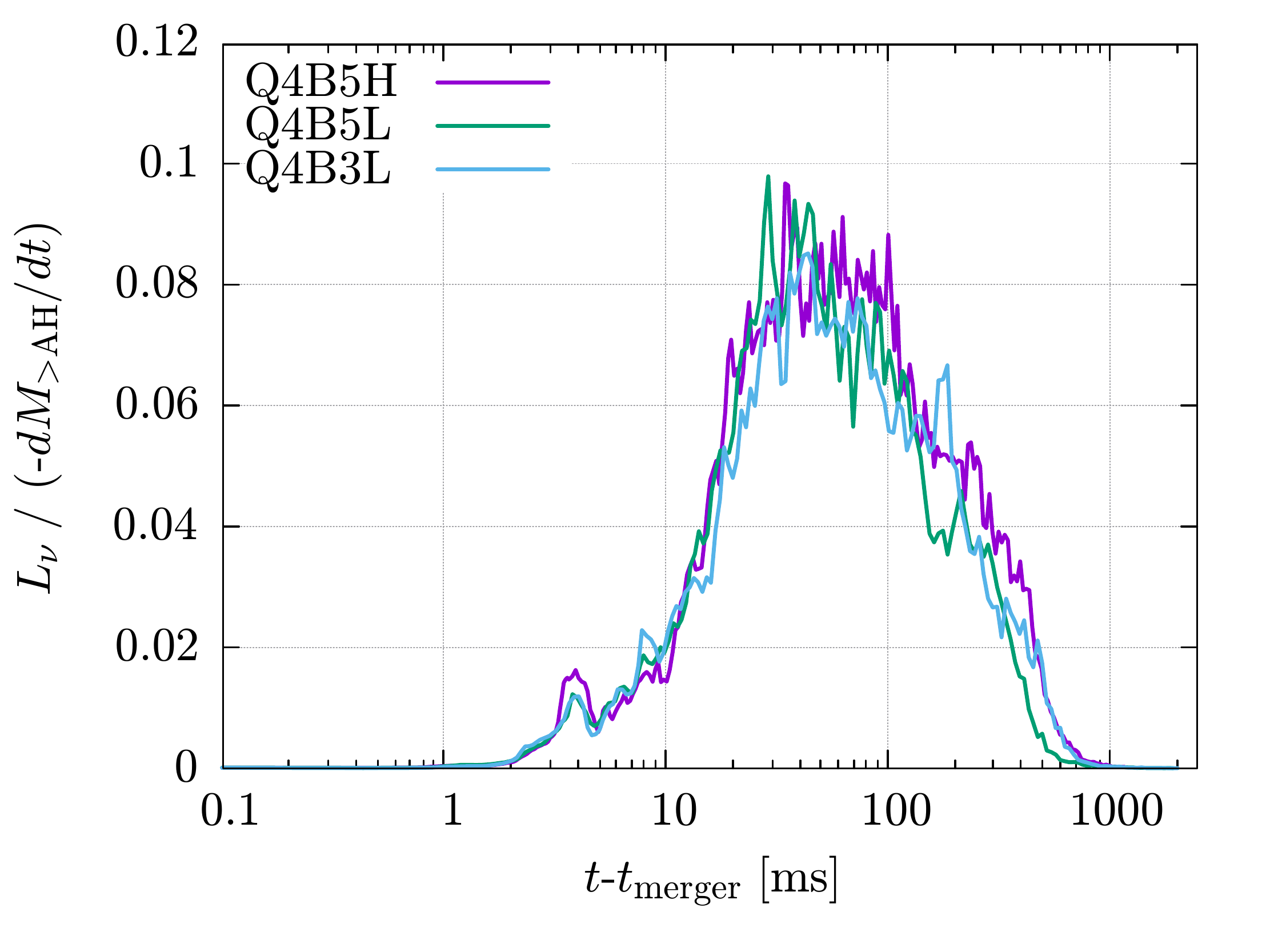}
        \caption{The time evolution of the rest-mass accretion rate calculated from $-dM_{\rm >AH}/dt$ (left panel), and the neutrino emission efficiency $L_{\nu}/(-dM_{\rm >AH}/dt)$ (right panel) for all the runs with $Q=4$.
	}
        \label{fig:mdot}
      \end{center}
\end{figure*}

The ejecta component is identified by considering the Bernoulli criterion; i.e., we 
regard the matter located outside the apparent horizon that satisfies 
$hu_t<-h_{\rm min}$ as the unbound component. Here, 
$u_t (<0)$ is the lower time component of the four velocity and $h$ is the specific enthalpy.
$h_{\rm min}$ is the minimum specific enthalpy for a given electron fraction $Y_{\rm e}$ and it is obtained from the tabulated EOS employed. 
The value of $M_{>\mathrm{AH}}$ for $t \alt 20$\,ms is in approximate agreement with 
that in our previous paper~\cite{kyutoku2018jan} in which magnetohydrodynamics 
and resulting viscous effects were absent. In the 
present simulation, by contrast to the one in Ref.~\cite{kyutoku2018jan}, 
for $t \agt \SI{20}{\ms}$, 
$M_{>\mathrm{AH}}$ continuously decreases due to the matter accretion onto 
the black hole induced by the angular-momentum transport resulting from the 
magnetohydrodynamics effects as already mentioned in Sec.~\ref{sec:results-overview}. 
We note that the curves of $M_{>\mathrm{AH}}$ depend only weakly on the initial 
magnetic-field strength and grid resolution. 

The value of $M_{\mathrm{eje}}$ steeply increases at two characteristic moments. 
The first increase is found right after the tidal disruption, and the steep increase continues only for a few ms, comparable to the dynamical timescale of the system. Thus, this mass ejection component is the dynamical ejecta. The rest mass for this component is $\approx 0.05M_\odot$ and $\approx 0.04M_\odot$ for models 
with $Q=4$ and 6, respectively. The result for $Q=4$ is in good agreement with our 
previous radiation-hydrodynamics result~\cite{kyutoku2018jan} \addms{because the magnetic-field strength is still weak at the tidal disruption, and hence, the magnetohydrodynamics effects play essentially no role in the dynamical mass ejection}. After the steep increase, the value of $M_{\mathrm{eje}}$ 
remains approximately constant for the next few hundreds ms, reflecting that 
an efficient mass ejection activity is quiescent during this time. In this quiescent 
stage, however, the accretion disk is actively evolved due to the MRI and associated 
turbulent motion, and the density and temperature of the disk decrease 
(see, e.g., Fig.~\ref{fig:rhomax} for the rest-mass density)
due to the expansion of the disk resulting from the angular-momentum transport 
process and enhanced magnetic pressure. As a result of the decrease in temperature,  
the neutrino luminosity eventually drops below the heating rate associated with the turbulent motion (cf.~Fig.~\ref{fig:nlum}), 
and then, the post-merger mass ejection driven by the \addms{heating associated with the MRI turbulence} sets in. 
Thus, the second steep increase of $M_{\mathrm{eje}}$ that starts 
at $t \sim 300$--$500\,{\rm ms}$ is triggered by the quick damping of the neutrino 
luminosity (see Fig.~\ref{fig:nlum}). 
We emphasize here that even in the presence of pure magnetohydrodynamics process (not effectively viscous process resulting from the MRI turbulence), the post-merger mass ejection appreciably occurs only after these onset time, and that, since the post-merger mass ejection continues for several hundred ms, simulations with the duration shorter than $\sim 500$\,ms cannot clarify this ejection process. 

The rest mass of the post-merger ejecta is $\approx 0.035M_\odot$ 
and $\approx 0.020M_\odot$ for models with $Q=4$ and 6, respectively, and 
these values are about 10\% of the disk mass at its formation (at $t \sim 10$\,ms). 
For both $Q=4$ and 6, the dynamical ejecta is the primary component of the ejecta in 
the present setting, and this tendency is stronger for the larger mass ratio, as
discussed, e.g., in Refs.~\cite{kyutoku2015aug,KST2021}. 
The onset time of $t \sim 300$--$500\,{\rm ms}$ for the post-merger mass 
ejection depends on the initial magnetic-field strength and grid resolution 
by 100--200\,ms. Our interpretation for this difference is that the magnetohydrodynamics 
turbulence is a stochastic process, and hence, the angular-momentum transport 
process can depend on the difference in the initial-field strength and grid resolution. 
However, the total ejecta mass and the properties of the 
post-merger ejecta do not depend strongly on them (see below for the electron fraction and velocity of the ejecta). 


Figures~\ref{fig:rhomax} and \ref{fig:nlum} display the time evolution of 
the maximum rest-mass density $\rho_{\rm max}$ and the total 
neutrino luminosity $L_{\nu}$, respectively. For generating Fig.~\ref{fig:nlum}, 
we define the merger time $t_{\rm merger}$ as the time at which 
the rest-mass density reaches its local minimum value for the first time; i.e., 
$t\approx 10$ and 13\,ms for $Q=4$ and 6, respectively.   
These figures indeed show that the density and neutrino luminosity steeply decrease 
at $t \approx 300$--500\,ms. This simultaneous decrease 
clearly elucidates that the evolution of  
the accretion disk and the timing of the post-merger mass ejection are controlled by 
the neutrino cooling. We also note that after the onset of the post-merger mass 
ejection, the accretion rate of the matter onto the black hole also decreases 
steeply with time: see the left panel of Fig.~\ref{fig:mdot}. 

\begin{figure*}[t]

      \begin{center}
        \includegraphics[scale=0.38]{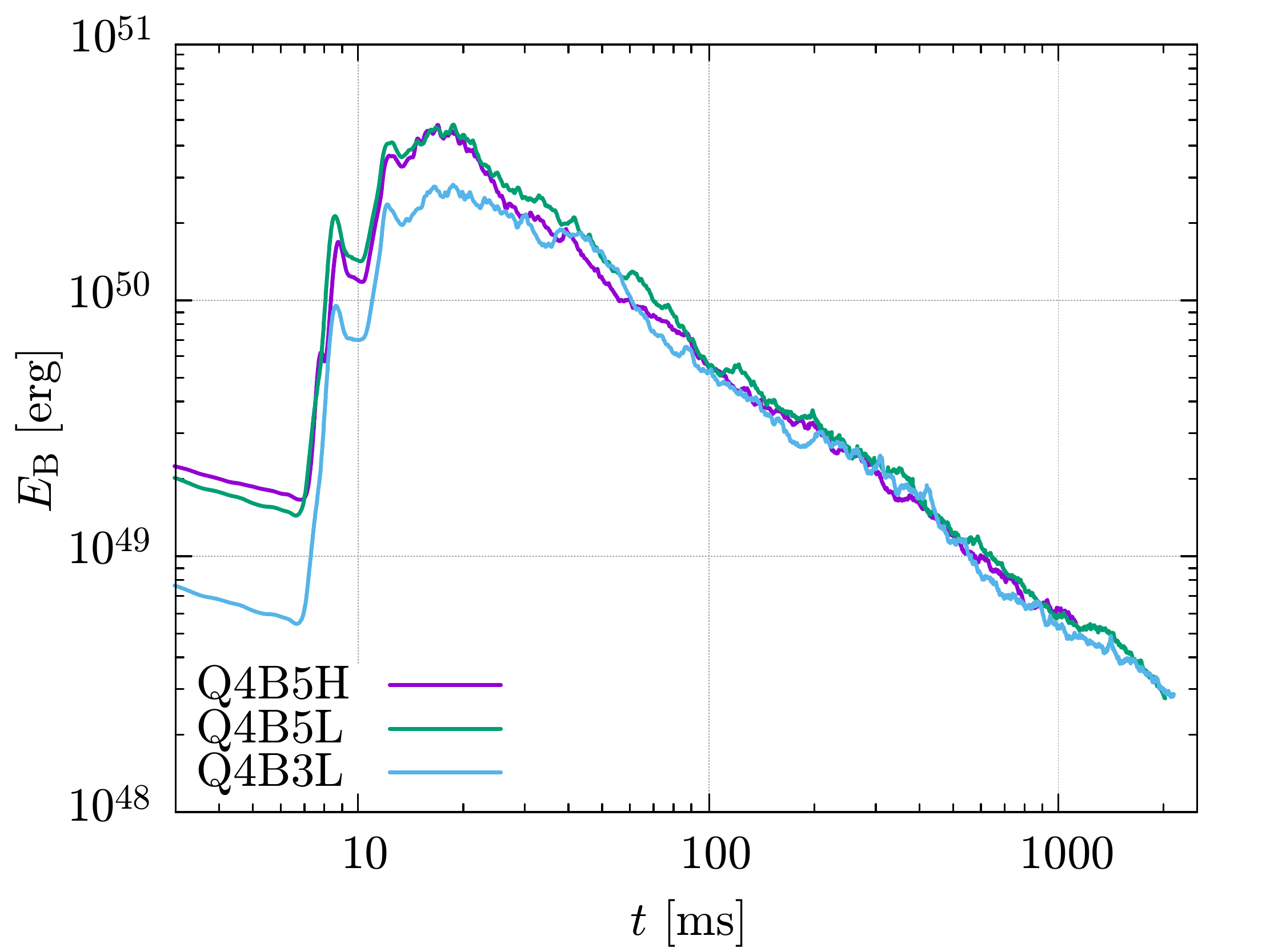}~~~~~
        \includegraphics[scale=0.38]{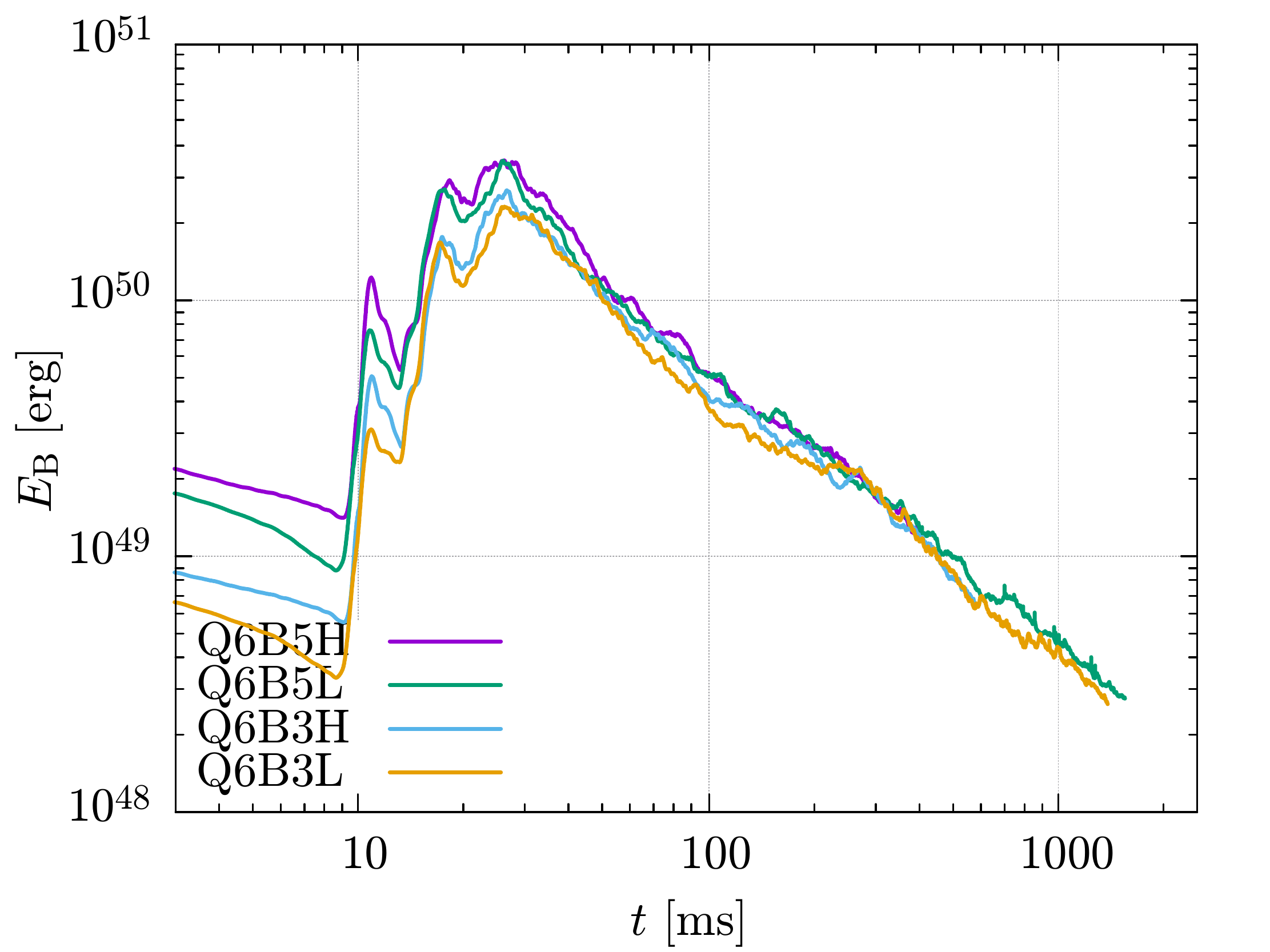}
        \caption{The time evolution of the electromagnetic energy evaluated for the 
        outside of the apparent horizon for all the runs with 
        $Q=4$ (left panel) and $Q=6$ (right panel)
	}
        \label{fig:eb}
      \end{center}
\end{figure*}

\begin{figure*}[th]

      \begin{center}
        \includegraphics[scale=0.6]{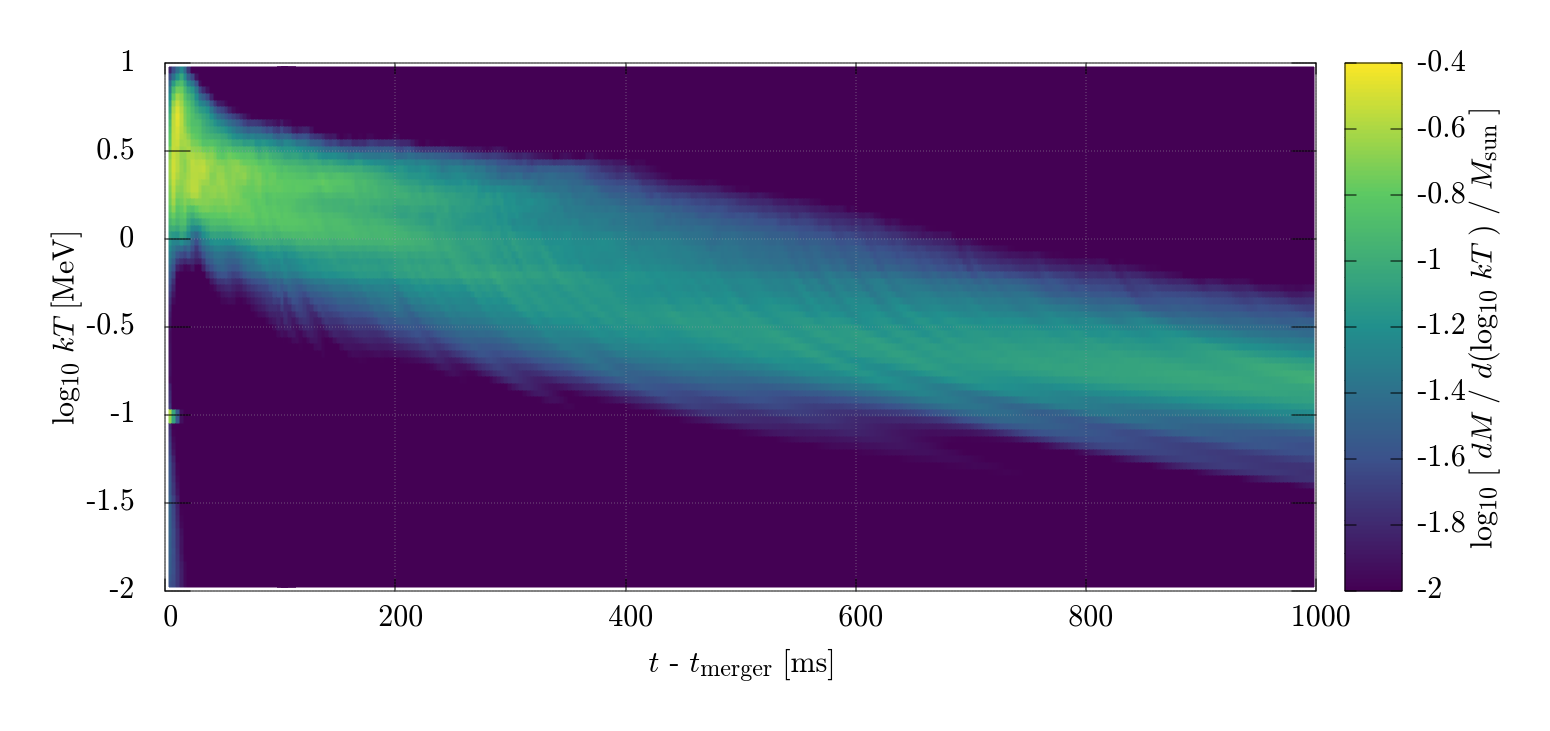}
        \includegraphics[scale=0.6]{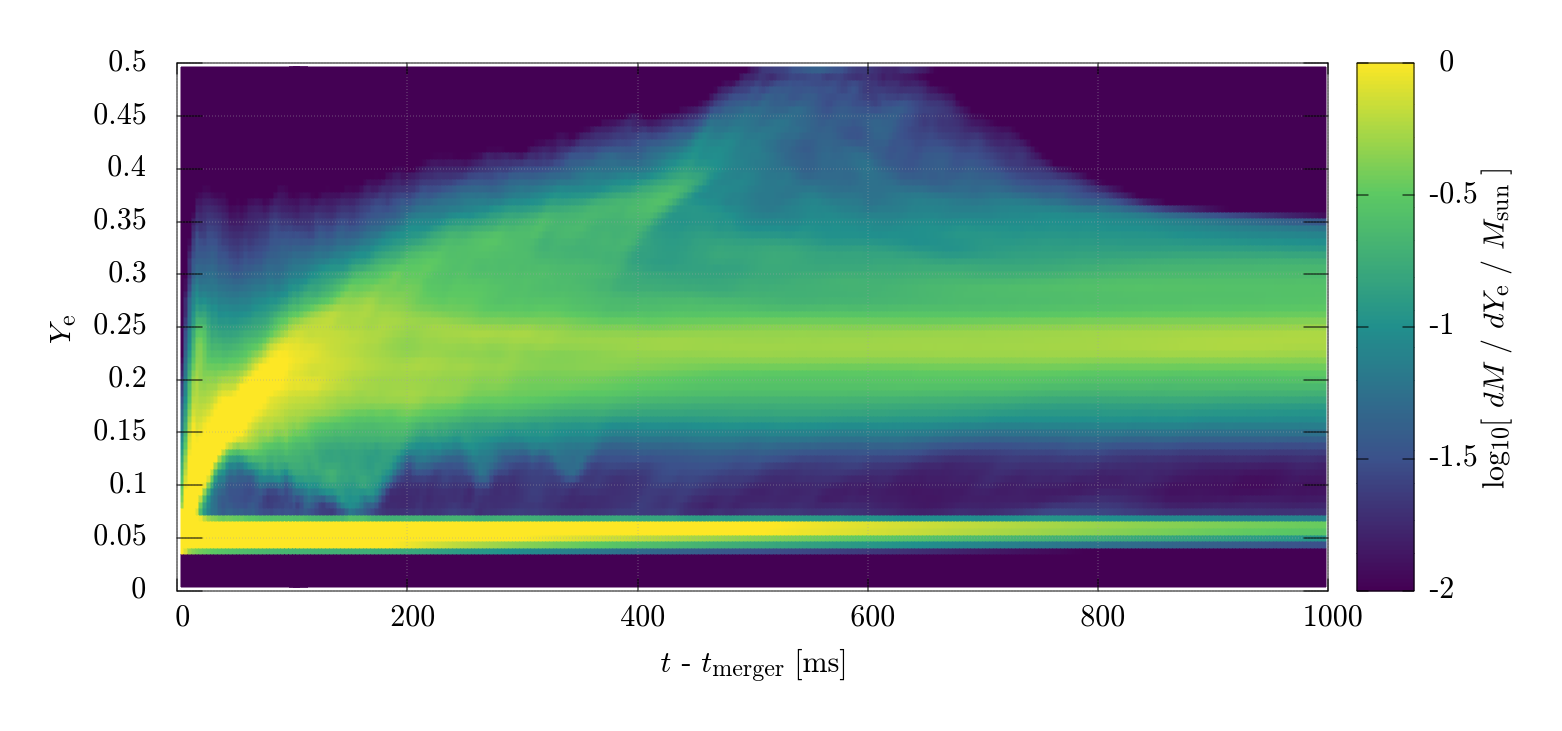}
        \caption{Time evolution of the mass histograms with respect to the temperature (upper panel) and electron fraction (lower panel) for model Q4B5H.
        The post-merger mass ejection sets in when the temperature for 
        most of the matter decreases below $3~{\rm MeV}$ at $t \sim 400$\,ms for this model. 
        Note that the matter only in the computational domain is taken into account for plotting this figure, and thus, the matter which has escaped from the computational domain is neglected in the late stages.
        Thus, for $t\gtrsim 300~{\rm ms}$, the dynamical ejecta mass decreases with time.
        }
        \label{fig:tem-spect}
      \end{center}

\end{figure*}

One interesting point is that the curve of $L_\nu$ well reflects the evolution of the accretion disk. From $t-t_{\rm merger} \approx \SI{1}{\ms}$ to $\sim 20$\,ms, 
$L_\nu$ increases by orders of magnitude both for $Q=4$ and 6. This reflects the 
temperature increase during the formation of the accretion disk 
(e.g., due to the compressional heating and shock heating) and the subsequent 
enhancement of the 
turbulent state in the accretion disk due to the MRI (see, e.g., Fig.~\ref{fig:eb}, 
which shows the increases of the electromagnetic energy in this stage). Subsequently, 
$L_{\nu}$ monotonically decreases for $t-t_{\rm merger} \gtrsim \SI{20}{\ms}$, because 
in this stage, the accretion disk expands due to the angular-momentum 
transport process and enhanced magnetic pressure, 
and the density and temperature decrease gradually. 
However, the thermal energy generated by the \addms{heating associated with the MRI turbulence} is consumed primarily by neutrino cooling prior to the onset of the post-merger mass ejection. 
Hence, the expansion of the accretion disk does not rapidly proceed, and thus, 
the mass ejection due to the thermally generated energy is suppressed. 
It is found that $L_\nu$ decreases approximately as $t^{-1.6}$ in this stage, 
and the decrease is fairly mild. 
However, after  $L_\nu$ decreases below $\approx  10^{51.5}$\,erg/s as a result of the disk expansion and resulting decrease of the temperature, the neutrino emission rate becomes smaller than the thermal energy generation rate due to the MRI turbulence. Then, the turbulent heating is used for the outward expansion of the disk efficiently, 
in particular through the convective motion from the inner to outer region
(see footnote 1), and the post-merger mass ejection is driven. 
(We note that the critical neutrino luminosity, which is $\sim 10^{51.5}$\,erg/s in 
the present case, should depend on the disk mass because the luminosity should 
be approximately proportional to it.) Subsequently, the neutrino luminosity 
exponentially drops at $t\approx 300$--500\,ms irrespective of the binary mass ratio and the initial choice of the magnetic-field strength. 
Specifically, this post-merger mass ejection sets in 
when the temperature for most of the disk matter decreases below $\sim 3\,{\rm MeV}$ 
(cf. the top panel of Fig.~\ref{fig:tem-spect} for a mass distribution with respect to the temperature as a function of time). This critical temperature at the onset of 
the post-merger mass ejection is quantitatively the same as that found in 
general relativistic neutrino-radiation viscous hydrodynamics simulations 
of black hole-torus systems~\cite{fujibayashi2020apr,fujibayashi2020dec}.
However, the time at the onset of the post-merger mass ejection is earlier than 
that in the viscous hydrodynamics result for the similar black-hole mass cases~\cite{fujibayashi2020dec}. As indicated in 
Refs.~\cite{christie2019sep,just2015feb,shibata2021sep}, the  
inherent magnetohydrodynamics effects such as magento-centrifugal effect~\cite{blandford1982} are likely to accelerate the mass ejection from the disk. The neutrino luminosity of 
$\approx 10^{51.5}\,{\rm erg/s}$ at the onset of the post-merger mass ejection which we find in this paper is indeed similar to that found in our recent magnetohydrodynamics study~\cite{shibata2021sep}. 

Figure~\ref{fig:mdot} plots the rest-mass accretion rate onto the black hole calculated by  $-dM_{>{\rm AH}}/dt$ and a neutrino emission efficiency defined by $L_\nu/(-dM_{>{\rm AH}}/dt)$. 
After the early matter infall associated with the onset of the merger, the 
mass accretion rate has a peak at $t-t_{\rm merger} \sim 10$\,ms. This is due to 
the fact that the magnetic-field strength is amplified in the accretion disk and the 
mass accretion rate is enhanced (cf. Fig.~\ref{fig:eb}). After the peak, the mass accretion rate decreases monotonically with time approximately as $\propto t^{-2}$ for 
$t-t_{\rm merger} \alt 50$\,ms and as $\propto t^{-1}$ in the subsequent stage 
before the onset of the post-merger mass ejection. After the onset of the post-merger 
mass ejection, the mass accretion rate drops steeply. 
Broadly speaking, the curve of the neutrino emission efficiency reflects that of $L_\nu$. 
However, the peak comes at $t-t_{\rm merger} \sim 40$--50\,ms, which is slightly 
later than the peak time of the neutrino luminosity and mass accretion rate. 
The reason for this is that 
$L_\nu \propto t^{-1.6}$ while $-dM_{>{\rm AH}}/dt \propto t^{-2}$ for $t-t_{\rm merger} \alt 50$\,ms and subsequently $-dM_{>{\rm AH}}/dt \propto t^{-1}$, and thus, the peak is shifted 
at $t-t_{\rm merger} \sim 50$\,ms. 
The maximum neutrino emission efficiency is $\sim 8$\%--10\%. 
Keeping the difference in the disk mass in mind, 
this value agrees broadly with those found in our viscous hydrodynamics simulations 
for similar black-hole mass ($M_{\rm BH}=6M_\odot$) and dimensionless spin ($\chi_{\rm BH}=0.8$)~\cite{fujibayashi2020dec}.

\begin{figure*}[th]

      \begin{center}
        \includegraphics[scale=0.38]{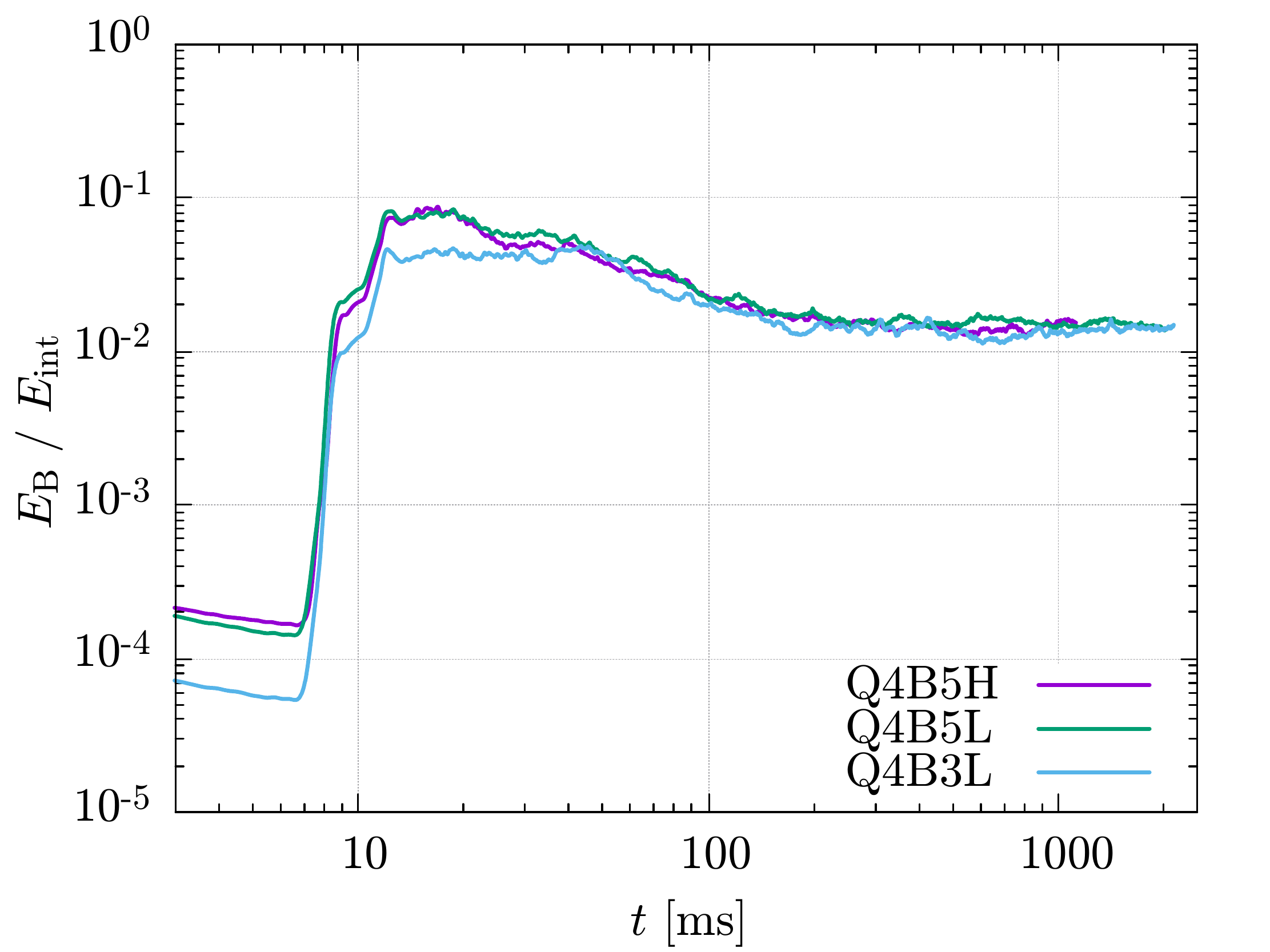}~~~~~
        \includegraphics[scale=0.38]{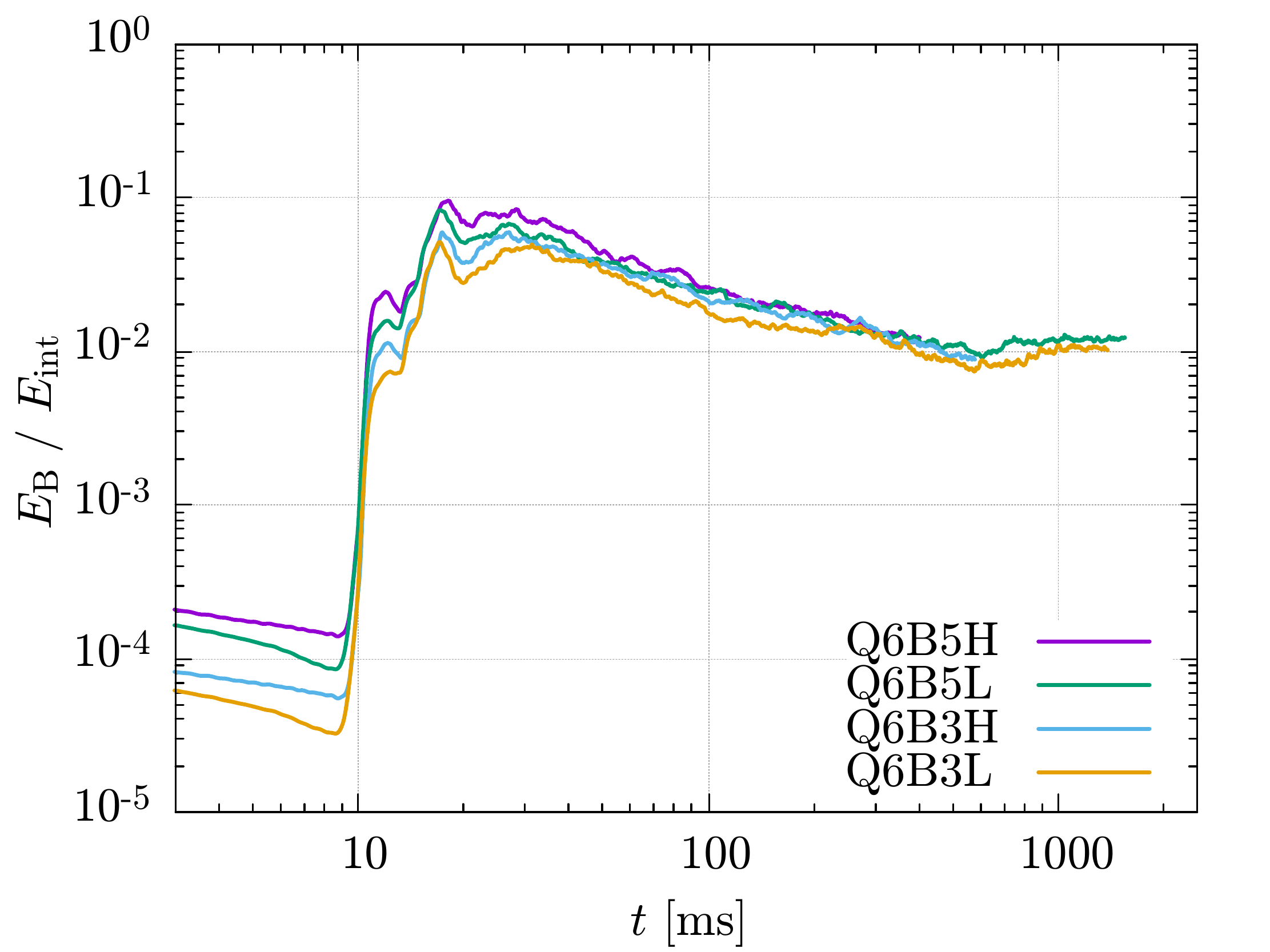}
        \caption{The time evolution of the ratio of the electromagnetic energy 
        to the internal energy evaluated for the outside of the apparent horizon for 
        all the runs with $Q=4$ (left panel) and $Q=6$ (right panel).
	}
        \label{fig:eb_eint}
      \end{center}

\end{figure*}

\subsubsection{Magnetic-field evolution}

Figures \ref{fig:eb} and \ref{fig:eb_eint} show the time evolution of 
the electromagnetic energy, $E_{\rm B}$, and the ratio of the electromagnetic 
energy to the internal energy, $E_{\rm int}$, respectively.
Here, $E_{\rm B}$ and $E_{\rm int}$ are defined, respectively, by
\begin{eqnarray}
  E_{\rm B}&:=&\frac{1}{8\pi}\int_{r>r_{\rm AH}}u^t \sqrt{-g}\,b_{\mu}b^{\mu}d^3x, \\
  E_{\rm int}&:=&\int_{r>r_{\rm AH}}\rho_* \varepsilon d^3x,
\end{eqnarray}
and $\varepsilon$ denotes the specific internal energy. 
Here we note that the energy-momentum tensor in the ideal magnetohydrodynamics 
is written as
\begin{eqnarray}
    T_{\mu\nu}&=&\rho h u_\mu u_\nu + P g_{\mu\nu} \nonumber \\
    &+&{1 \over 4\pi}\left(b^\alpha b_\alpha u_\mu u_\nu +{1 \over 2}b^\alpha  b_\alpha g_{\mu\nu}-b_\mu b_\nu\right),
\end{eqnarray}
and with $h=c^2+\varepsilon+P/\rho$, we have 
\begin{equation}
u^t \sqrt{-g} T_{\mu\nu}u^\mu u^\nu = \rho_* (c^2 + \varepsilon) +
{1 \over 8\pi} u^t \sqrt{-g}\, b^\mu b_\mu.\label{eq:12}
\end{equation}
Here we recover $c$ to clarify the physical units. 
Thus, the choice of $E_{\rm int}$ and $E_{\rm B}$ stems from Eq.~(\ref{eq:12}). 

During the merger stage, the magnetic-field strength \addms{in the accretion disk} is amplified quickly in a short timescale of a few ms. This is initially induced by the magnetic winding associated with the differential rotation in the accretion disk. 
In the Keplerian disk with the presence of the poloidal magnetic field of the 
cylindrically radial component $B^\varpi$, the strength of the toriodal magnetic field 
$B^T$ increases approximately linearly with time until a saturation as (e.g., Ref.~\cite{shibata2006}) 
\begin{equation}
B^T \approx {3 \over 2} B^\varpi \Omega t,
\end{equation}
where $\Omega$ denotes the local angular velocity. For a black hole with 
the dimensionless spin of $0.8$, the angular velocity at the innermost stable 
circular orbit of the black hole is 
$\Omega_{\rm ISCO} \approx 0.174 M_{\rm BH}^{-1}\approx 5.43\times 10^3(M_{\rm BH}/6.5M_\odot)^{-1}$\,rad/s~\cite{bardeen1972dec}. Thus for the models of $Q=4$ and $Q=6$, 
the matter near the innermost stable circular orbit rotates with the orbital 
period of $\approx 1.2$ and 1.6\,ms, respectively. This implies that 
in the first $\sim 10$\,ms, the toroidal field strength can be $\sim 60$--80 
times of $B^\varpi$, the maximum of which is $\sim 10^{14}$\,G at the formation of the accretion disk \addms{(i.e., much weaker than the field strength in the neutron star initially given)} in the present simulations. This is the reason that the initial steep amplification to $E_{\rm B} > 10^{50}$\,erg is found in our present simulations. 
Because the winding timescale is quite short, the magnetic-field amplification by 
$\sim 3$ orders of magnitude in $\alt 100$\,ms is possible even in the absence of 
other instabilities such as MRI: Even for the initial value of $B^\varpi=10^{12}$\,G, 
the toroidal field can be amplified to $\sim 10^{15}$\,G in $\sim 100$\,ms. 
After the sufficient amplification of the toroidal magnetic field, an 
outward expansion of the accretion disk is driven toward the polar direction due to the 
enhanced magnetic pressure and a poloidal field with its strength  
comparable to that of the toroidal field is also generated. 
We note that the Kelvin-Helmholtz instability 
that takes place during the winding of the spiral arm around the black hole 
and collision between different parts of the spiral arm may also 
contribute partly to the magnetic-field amplification. 

After the initial amplification of the magnetic-field strength, the ratio of 
$E_{\rm B}/E_{\rm int}$ reaches $\sim 0.05$--0.1. Then, the magnetic-field 
growth is saturated. The electromagnetic energy at the saturation, $E_{\rm B,sat}$, 
is smaller for the smaller value of the initial magnetic-field strength. However, 
the relative difference in the saturated electromagnetic energy between models with 
different initial magnetic-field strengths 
is not as large as that in the initial electromagnetic
energy. Furthermore, the electromagnetic energy for $t \agt 30$\,ms depends 
only weakly on the initial condition (as well as on the grid resolution). 
Thus, we infer that the amplification and saturation of the magnetic-field strength 
take place in a universal manner irrespective of the initial magnetic-field strength. 

When reaching the saturation, the typical magnetic-field strength is 
$10^{15}$\,G (cf.~Fig.~\ref{fig:BUT_Q4_B5_low}) and the maximum rest-mass density 
is $\sim 10^{11}$--$10^{12}~{\rm g/cm^3}$ in the innermost region. 
Thus the Alfv\'en velocity is $\approx b/\sqrt{4\pi\rho}\approx 9\times 10^8$\,cm/s$\,(b/10^{15}\,{\rm G})(\rho/10^{11}\,{\rm g\,cm^{-3}})^{-1/2}$ and 
the wavelength of the fastest growing mode of the MRI is typically 
$\sim 10$\,km~\cite{balbus1998}. 
As a result, the wavelength of this unstable mode is covered by 
tens of grid points in our setting \addms{(see Appendix~\ref{appendixC})}, and hence, the effect of the MRI comes into 
play subsequently. With the evolution of the disk, the typical 
magnetic-field strength and rest-mass density decrease, but in 
the equipartition stage (see below), the Alfv{\'e}n velocity is always of 
order $\sqrt{E_{\rm B}/E_{\rm int}} (\sim 10\%)$ of the sound speed, 
which changes weakly with time. Thus, the wavelength of the 
fastest growing mode of the MRI is always covered by tens of grid points 
in the present setting. Indeed, our numerical analysis shows that the 
wavelength is covered by $\sim 10$ grid points for the 
region with $\rho=10^{11}\,{\rm g/cm^3}$, and more (several tens of) 
grid points for lower density region. 

After the MRI starts playing a role, a turbulent state is developed in the accretion 
disk and an effective viscosity is induced. We evaluate the following 
ratio of the anisotropic stress to the pressure
\begin{eqnarray}
    \addkh{
    \alpha_{ij}:=\left\langle \left|  {1 \over P}
    \left(\rho h \hat u_i \hat u_j -{1 \over 4\pi}b_i b_j \right)  \right|\right\rangle_{\rm ave}, 
    }
\end{eqnarray}
where $i\not=j$ ($i$, $j=x, y, z$) and 
$\langle \cdots \rangle_{\rm ave}$ denotes the spatial average with the weight of the rest-mass density ($\rho_*$) for the region with $z \geq 0$ and $\rho \geq 10^7\,{\rm g/cm^3}$. 
$\hat u_i$ is defined by $u_i-\langle u_i \rangle_{t,{\rm ave}}$ 
where $\langle u_i \rangle_{t,{\rm ave}}$ denotes the local time average of $u_i$. 
The time average needs to be subtracted from $u_i$ 
to eliminate the contribution of coherent motion 
(not random motion) for evaluating the anisotropic stress associated with the turbulent motion.
We find that all the components of $\alpha_{ij}$ are between $\approx 0.02$ and $0.1$ at the onset of the post-merger mass ejection depending only weakly on the initial magnetic-field strength and grid resolution. The values of $\alpha_{ij}$ are comparable to the viscous alpha parameter often employed in the viscous hydrodynamics simulations (e.g., 
Refs.~\cite{fernandez2018oct,fujibayashi2020apr,fujibayashi2020dec}). 
\addkh{The value is slightly larger than the result of previous magnetohydrodynamics simulations of black-hole accretion disks~\cite{de2021nov} but this difference is likely to come from the difference in the definition of $\alpha_{ij}$.}
In this stage, the ratio of 
$E_{\rm B}/E_{\rm int}$ is preserved to be of $O(10^{-2})$. 

As a result of the viscous angular-momentum transport, the matter in the 
inner region of the accretion disk falls into the black hole while the 
matter in the outer part expands outward. Because of the matter infall into the 
black hole, the rest mass of the accretion disk decreases (see Fig.~\ref{fig:mrem}), 
and associated with the decrease in the rest mass, the electromagnetic energy decreases 
with time although the ratio of $E_{\rm B}/E_{\rm int}=O(10^{-2})$ is preserved. 
Thus, for $t \agt 100$\,ms, the accretion disk is in a \addms{quasi-steady equipartition state; the magnetic-field energy relaxes to $\sim 1\%$ of the internal energy irrespective of the mass and internal energy of the accretion disk.}  
It is interesting to point out 
that the electromagnetic energy decreases approximately in proportion to $t^{-1}$. 
All these features are found both for the models of $Q=4$ and $Q=6$ irrespective 
of the initial magnetic-field strength and grid resolution. 

\begin{figure*}[th]
      \begin{center}
        \includegraphics[scale=0.35]{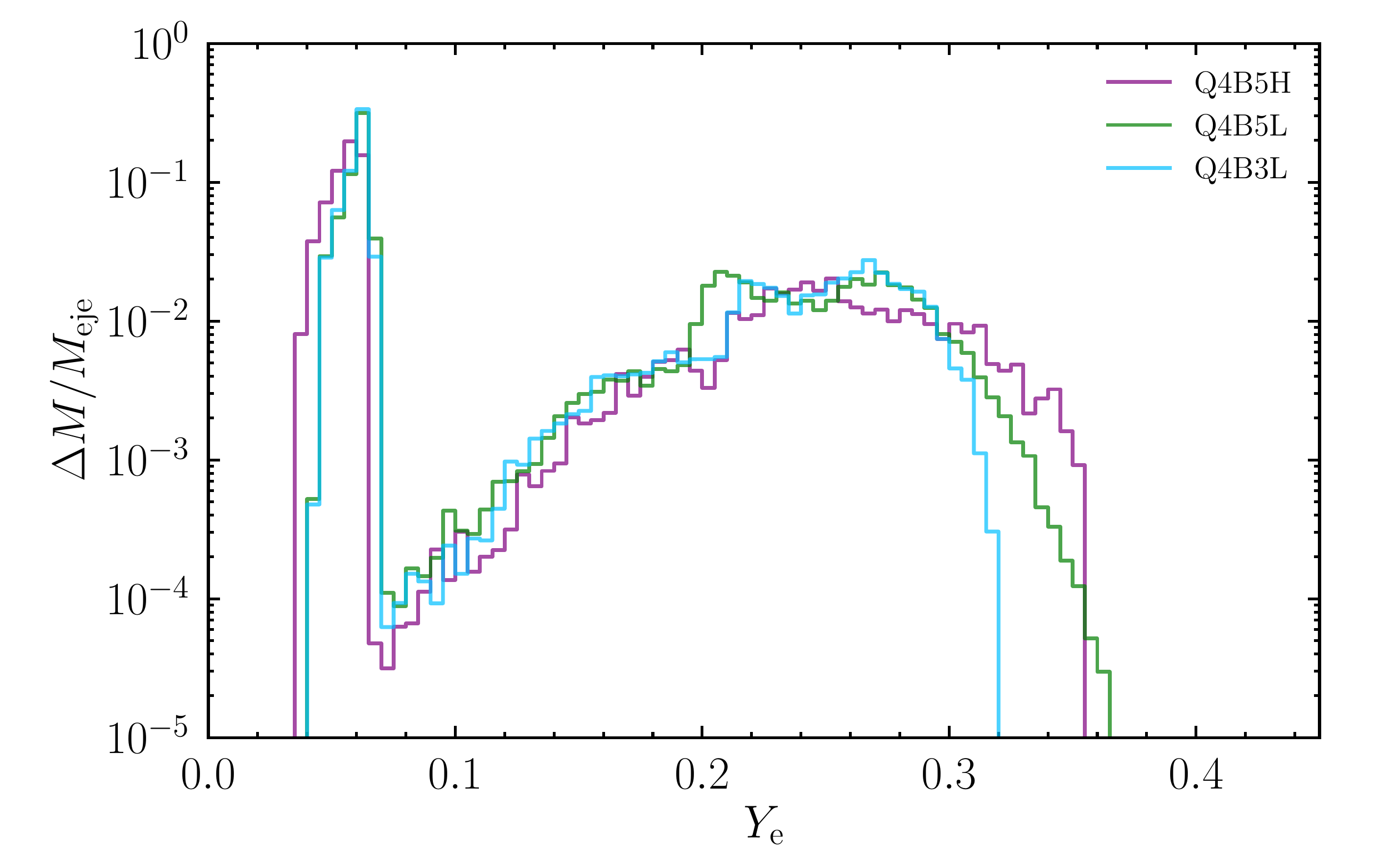}~~~
        \includegraphics[scale=0.35]{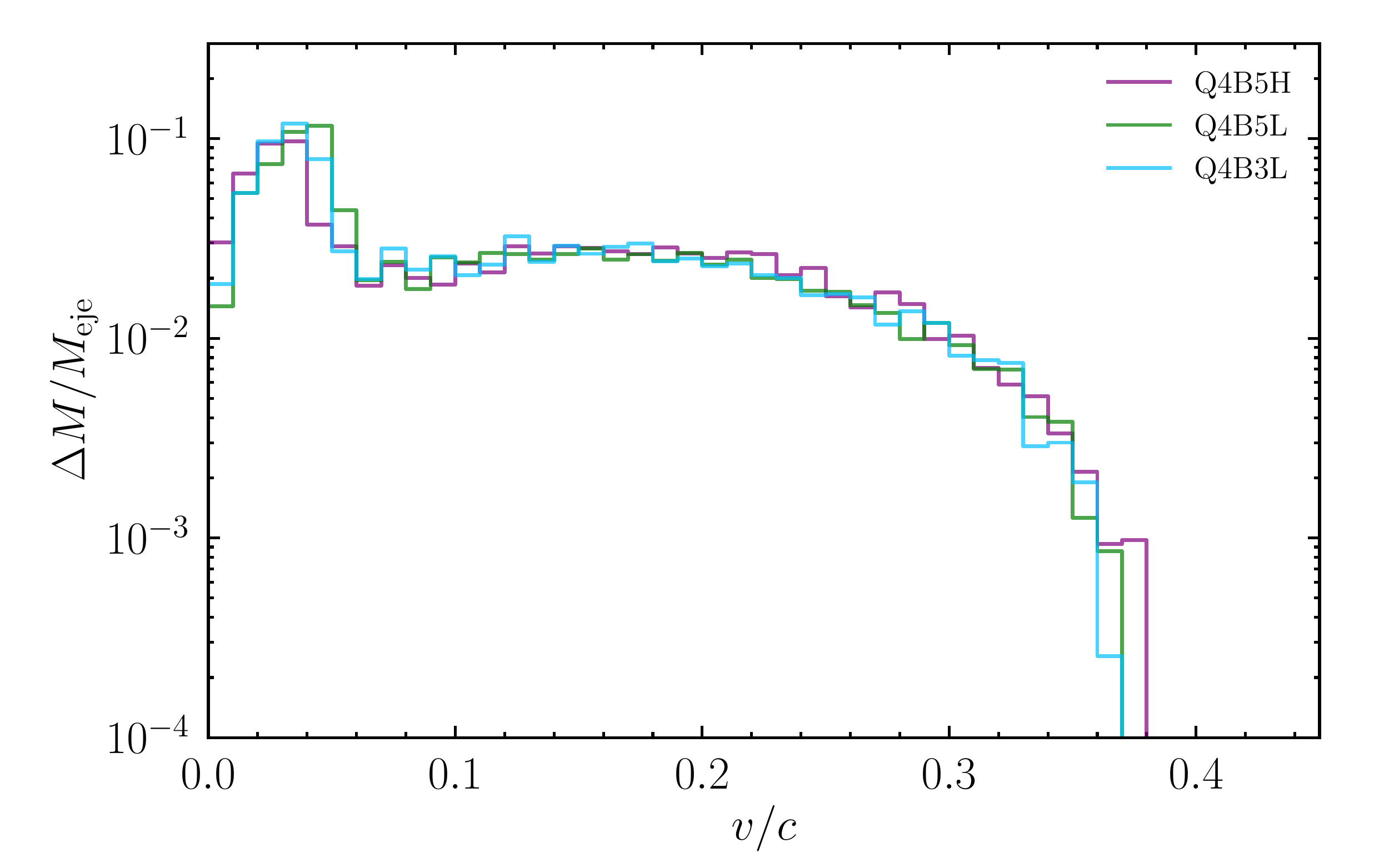}\\
        \includegraphics[scale=0.35]{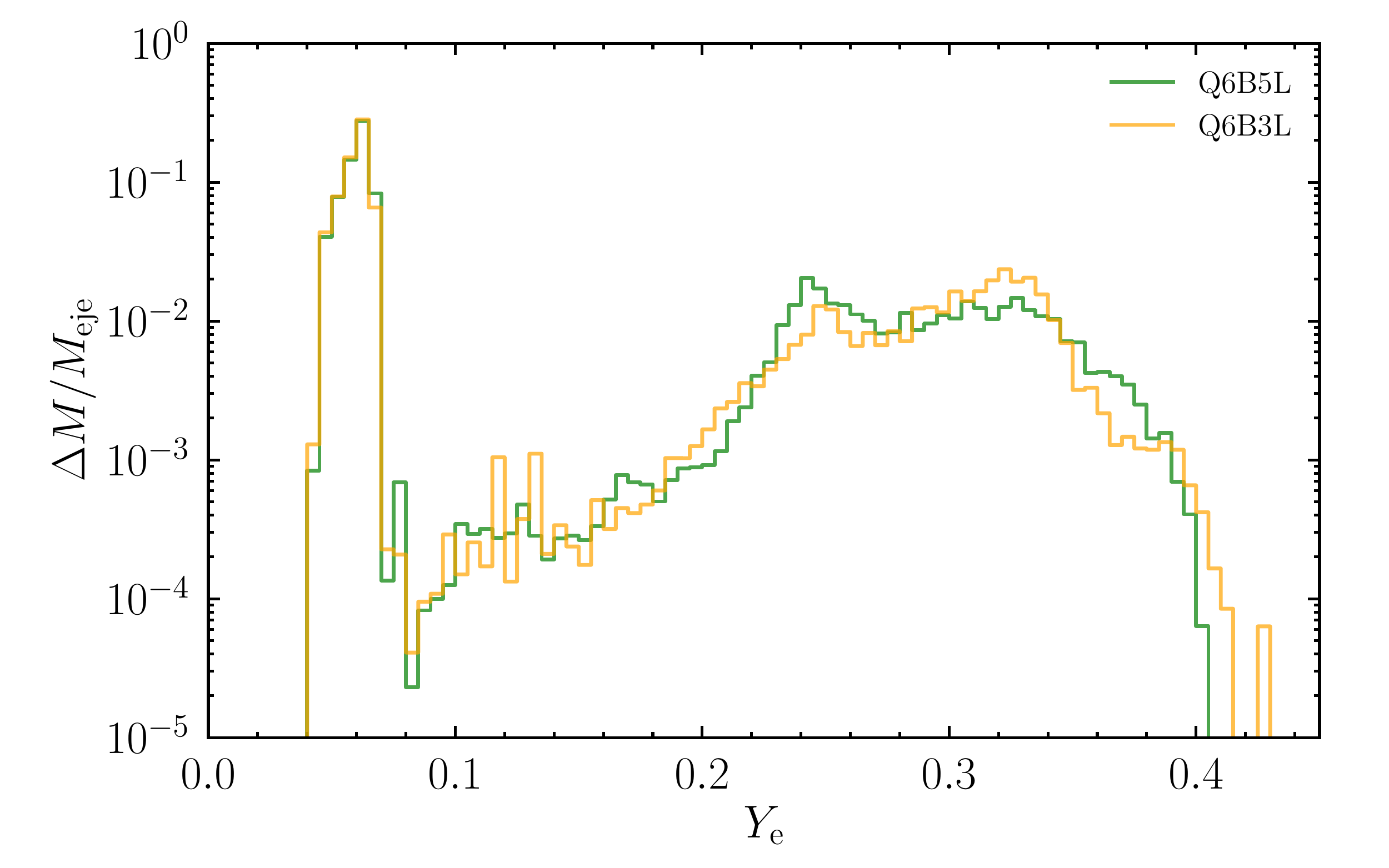}~~~
        \includegraphics[scale=0.35]{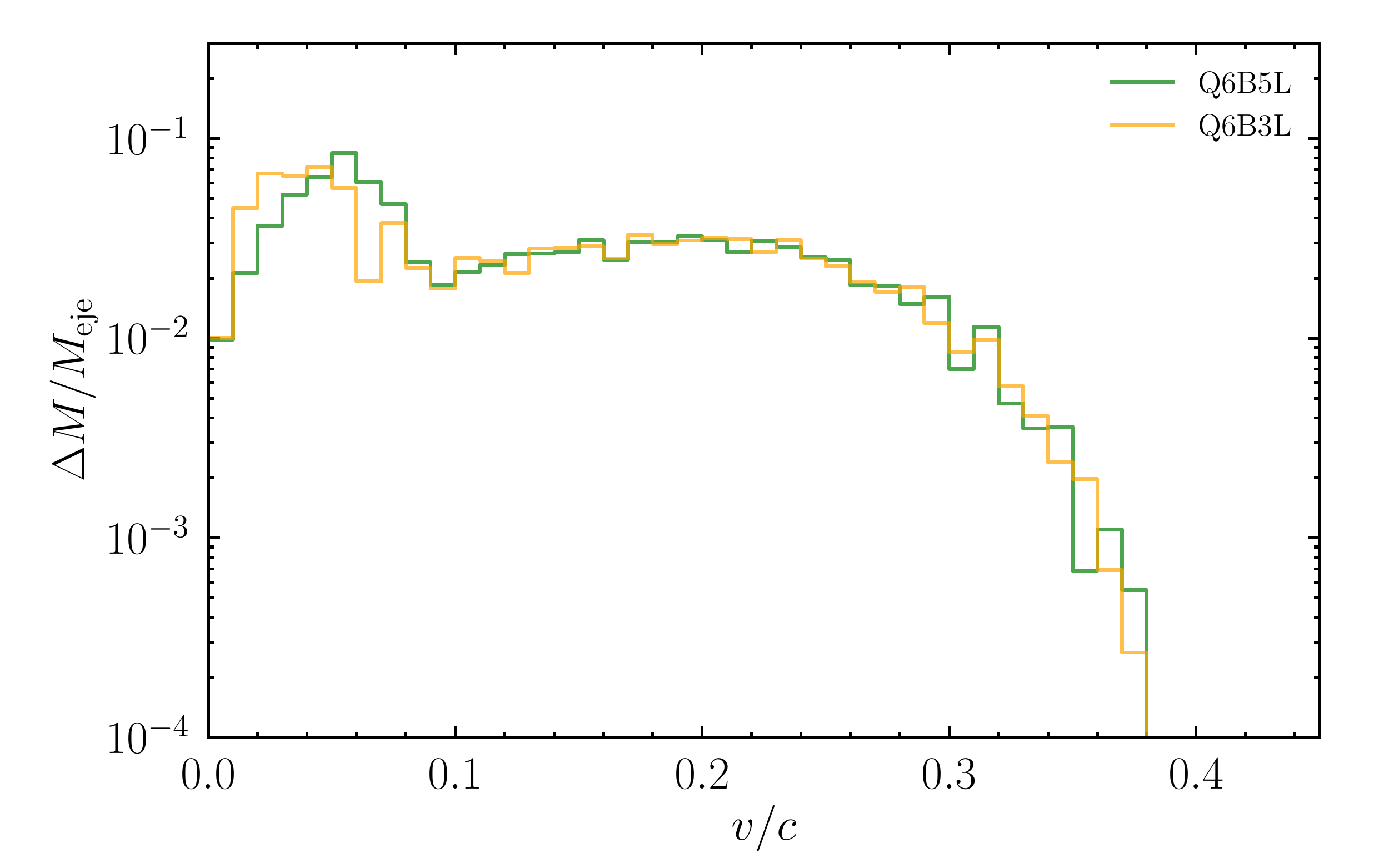}
        \caption{Mass histogram as functions of the electron fraction (left panels) and 
        the velocity (right panels) of ejecta 
        for the models with the simulation duration longer 
        than $1\,{\rm s}$ (models Q4B5H, Q4B5L, Q4B3L, Q6B5L, and Q6B3L).
        Models with $Q=4$ and $Q=6$ are displayed in the upper and lower panels, respectively.
        }
        \label{fig:hist}
      \end{center}

\end{figure*}

\subsubsection{Property of ejecta}

Now we turn our attention to the properties of the ejecta. The bottom panel of Fig.~\ref{fig:tem-spect} displays the mass distribution of the remnant matter 
with respect to the electron fraction $Y_{\rm e}$ for model Q4B5H. 
This shows that there are two characteristic 
peaks of $Y_{\rm e}$ at the regions around 0.05 and of 0.25--0.35, respectively. 
The former peak is associated primarily with the dynamical ejecta and the latter is 
with the accretion disk for $t \alt 400$\,ms and post-merger ejecta 
for $t \agt 400$\,ms. This figure clearly shows that the dynamical ejecta 
component with $Y_{\rm e}=0.03$--0.07 comes directly from  
the neutron star, because the values are unchanged from the beginning. 
That is, this dynamical ejecta component is not essentially affected by 
thermal or \addms{weak-interaction} processes in the merger and post-merger stages. 

By contrast, the electron fraction of the post-merger ejecta is found to be determined by the evolution process of the accretion disk, in which the typical electron fraction increases from $\sim 0.05$ to $\sim 0.25$ for $0 < t \alt 200$\,ms.
As already mentioned, in this stage, the accretion disk gradually expands due to the viscous \addms{and magnetohydrodynamical} angular-momentum transport and magnetic pressure by the amplified magnetic-field strength, and its rest-mass density and temperature monotonically decrease.
In the disk with its optical depth to neutrinos $\lesssim 1$, the electron fraction is determined predominantly by the reaction equilibrium between electron/positron capture reactions if the temperature is high enough (typically $kT\gtrsim 2$--3\,MeV; see Refs.~\cite{Beloborodov2003may,fujibayashi2020dec,just2022jan}) for their timescale to be shorter than that of the disk expansion.
Due to the disk expansion, the electron degeneracy becomes weak, and as a result, 
the electron fraction is shifted to higher values in the reaction equilibrium state.
With the decrease of the temperature, the neutrino luminosity decreases approximately in proportion to $T^6$.
As already mentioned, the post-merger mass ejection sets in when the neutrino luminosity drops below $\sim 10^{51.5}\,{\rm erg/s}$, which occurs for $t \agt 300$\,ms. The typical value of $Y_{\rm e}$ for the post-merger ejecta is determined around this timing, 
resulting in $Y_{\rm e}\approx 0.25 \pm 0.10$. 

Figure~\ref{fig:hist} displays the rest-mass histogram as functions of the electron fraction and velocity for the ejecta component for the models for which the simulation duration is longer than 1\,s. 
The mass histogram is derived for the ejecta component outgoing from the radius of $\approx 10^4$\,km. 
As described in the previous paragraphs, there are two distinct $Y_{\rm e}$ components for the ejecta, and this feature is clearly observed in Fig.~\ref{fig:hist}. 
The dynamical ejecta component always has $Y_{\rm e} \approx 0.03$--0.07 irrespective of the black-hole mass.
By contrast, the distribution of $Y_{\rm e}$ for the post-merger ejecta component depends 
on the black-hole mass in the present results.
Specifically, for larger black-hole mass, the value of $Y_{\rm e}$ tends to be larger. 
As a result, the peak of $Y_{\rm e}$ changes from $\sim 0.25$ for $Q=4$ to $\sim 0.31$ for $Q=6$. 
This is in agreement with our finding in our viscous hydrodynamics studies~\cite{fujibayashi2020dec} (see also Refs.~\cite{fernandez2020jul,just2022jan}), and the reason is as follows: In the condition 
that the disk mass has an approximately identical value, 
the density of the disk can be higher for the lower black-hole mass (the lower 
mass ratio, $Q$, in the present context), 
because the tidally disrupted matter can have a more compact orbit around the black hole 
due to the smaller radius of its innermost stable circular orbit.
Associated with this effect, the temperature is enhanced due to the compression and stronger shock heating, resulting in the higher neutrino emissivity and reducing the 
entropy per baryon of the matter in the accretion disk (cf. Fig.~\ref{fig:nlum}).
With the lower entropy per baryon, the degree of the electron degeneracy becomes higher 
and the neutron-richness is enhanced.
Therefore, for the lower black-hole mass, the electron fraction of the post-merger ejecta becomes slightly lower. Figure~\ref{fig:hist} shows that this effect is found irrespective 
of the initial-magnetic field strength and grid resolution (thus it is physical). 

The right panels of Fig.~\ref{fig:hist} present the rest-mass histogram as a 
function of the ejecta velocity. Again, there are two components. Here, the low-velocity 
component with $v/c \alt 0.08$ stems primarily from the post-merger ejecta, 
while the high-velocity component stems from the dynamical ejecta. 
We note that the velocity distribution for the dynamical ejecta is in 
good agreement with that in our previous study~\cite{kyutoku2018jan}, and 
the typical velocity of the post-merger ejecta agrees approximately with that 
found in viscous hydrodynamics simulations (e.g., Refs.~\cite{fujibayashi2020apr,fujibayashi2020dec}). 
As we reported in Ref.~\cite{kyutoku2015aug}, the velocity 
of the dynamical ejecta is at highest $\sim 0.4c$. This is in contrast 
to the case of binary neutron star mergers in which the maximum ejecta 
velocity can be $\agt 0.8c$~\cite{hotokezaka2013}. 

Our present results confirm that there are two distinct ejecta components, low-$Y_{\rm e}$ and 
high-velocity component, and relatively-high-$Y_{\rm e}$ and low-velocity component, 
as many previous numerical work have suggested.  
By our self-consistent simulations, the distinction of two components emerges 
clearly. The former (dynamical ejecta) is likely to synthesize heavy $r$-process elements, while the latter (post-merger ejecta) 
is likely to synthesize relatively light $r$-process elements as well as heavy ones (e.g., Refs.~\cite{wanajo2014,just2015feb}). Then, the former component is likely to shine as a 
red kilonova while the latter one is likely to contribute to a blue-kilonova component~\cite{metzger2014may}. However, the detailed light curve and spectrum are determined by a non-trivial radiation transfer effect~\cite{kawaguchi2020}. It is also likely 
that the light curve depends on the mass ratio $Q$. 
Thus, a nucleosynthesis calculation and radiation transfer simulation are topics to be explored as follow-up work.  


\subsection{Magnetic field in the funnel region and the relation to short gamma-ray bursts} \label{sec:results-outflow}



\begin{figure}[t]
      \begin{center}
        \includegraphics[scale=0.35]{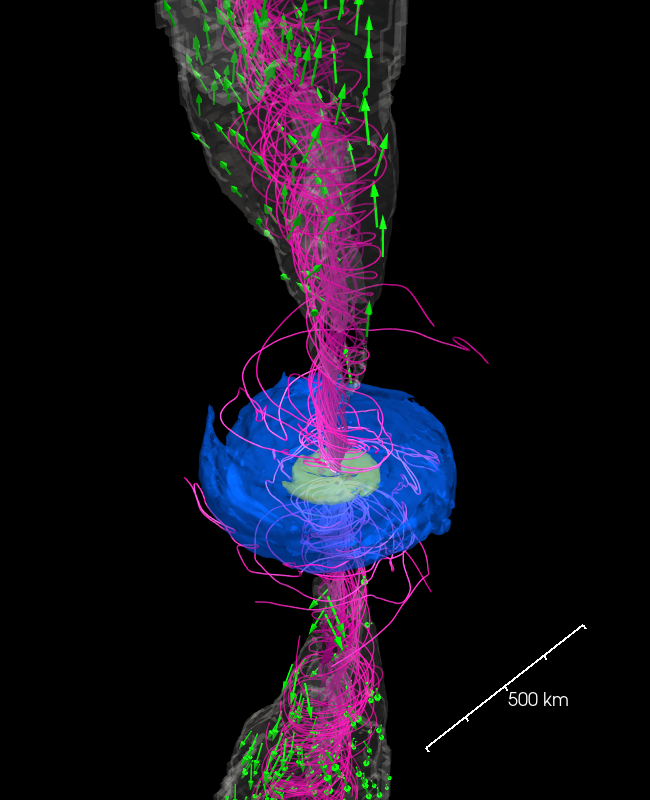}
        \caption{Snapshots of the rest-mass density profile (blue and green contours) 
        with the magnetic-field lines (pink curves),
          unbound matter (white color) and its velocity (green arrow) for model Q4B5L at $t=300$\,ms. Magnetic-field lines penetrating the black-hole horizon are displayed. 
          See also the following link for the time evolution:
          \url{https://www2.yukawa.kyoto-u.ac.jp/~kota.hayashi/Q4B5L-3D.mp4}
	}
        \label{fig:plum-ang}
      \end{center}

\end{figure}

In addition to aforementioned ejected matter (dynamical and post-merger ejecta),
we find a launch of an outflow of the matter and Poynting flux in the narrow 
funnel region established near the rotational axis of the black hole  (see Fig.~\ref{fig:plum-ang}). In particular, 
the isotropic Poynting luminosity estimated for most of the 
runs is comparable to the typical luminosity
of short-hard gamma-ray bursts~\cite{nakar2007apr,berger2014jun}. 
In this section, we discuss the quantitative details on this result. 

\begin{figure*}[th]
      \begin{center}
        \includegraphics[scale=0.38]{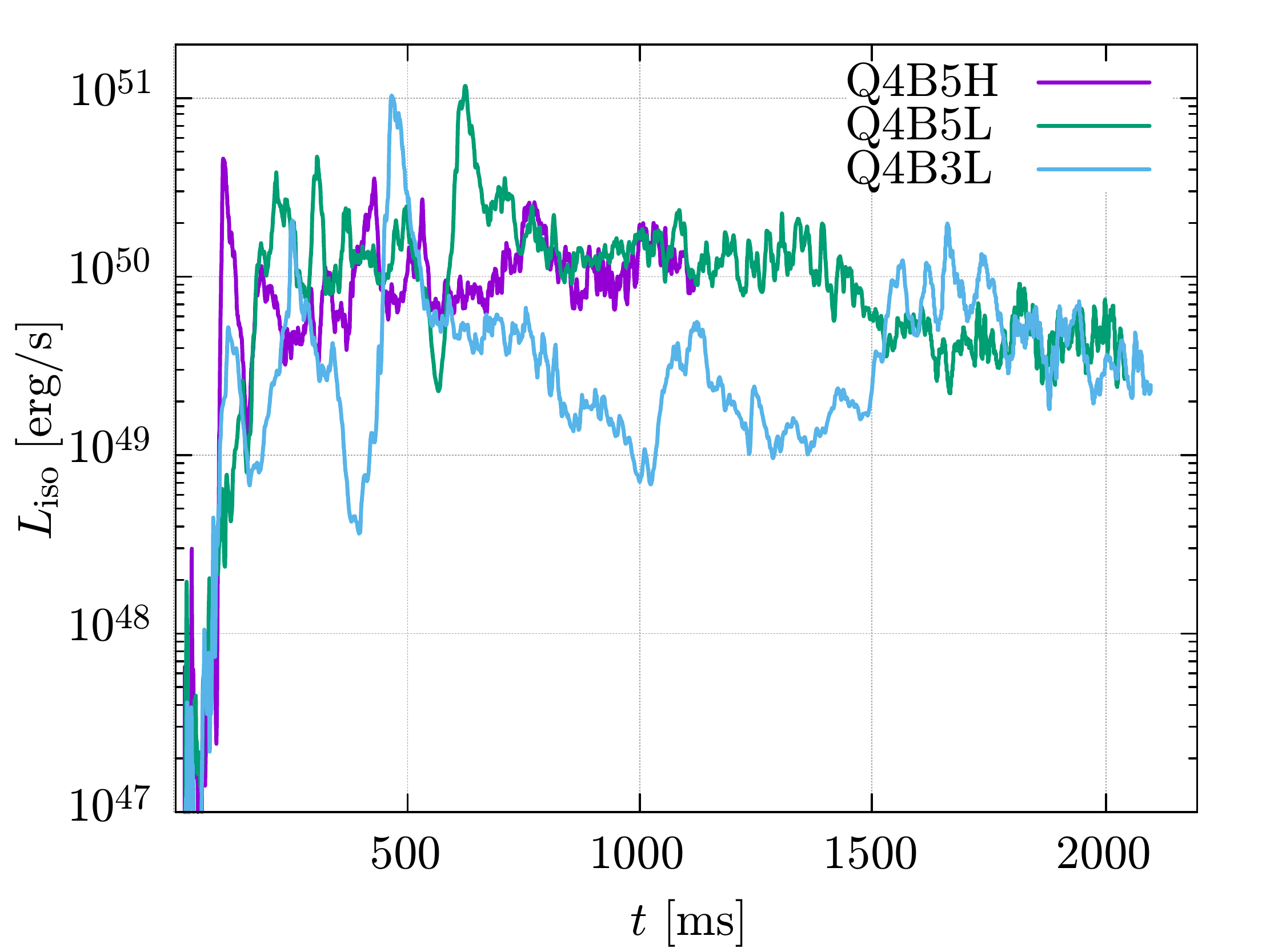}~~~~~
        \includegraphics[scale=0.38]{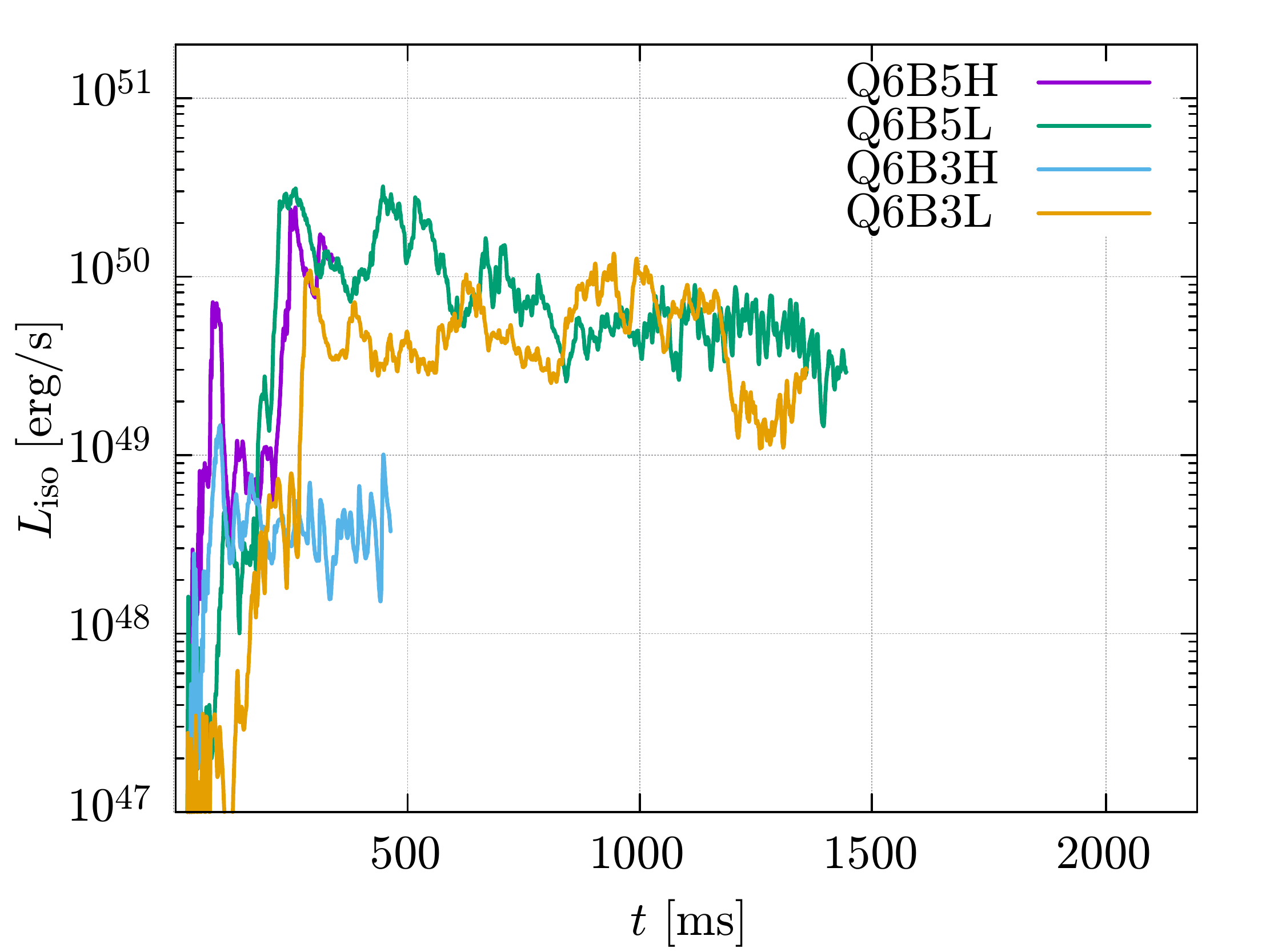}
        \caption{
          $L_{\rm iso}$ as a function of time for all the runs with $Q=4$ (left panel) and $6$ (right panel).
          The Poynting luminosity is evaluated at $r\approx 1500~{\rm km}$ for all the runs. 
	}
        \label{fig:plum}
      \end{center}
\end{figure*}

Irrespective of the black-hole mass, initial magnetic-field strength, 
and grid resolution, tidal disruption of the neutron star takes place 
in our present setting and a magnetized accretion disk is formed around 
the central black hole. As already mentioned in the previous subsections, 
the magnetic-field strength in the accretion disk is increased by the winding and MRI, and then, 
a turbulent state is established at $\sim 30$--40\,ms after the tidal disruption. 
Subsequently, the accretion disk evolves primarily due to the viscous effect 
stemming from the MRI turbulence. As already mentioned in 
the previous subsection, the magnetic-field strength is determined by 
an equipartition state, i.e., by the internal energy of the matter, which 
is typically $\rho c_{\rm s}^2$ where $c_{\rm s}$ is the sound speed of order 
$10^9$\,cm/s in the dense region of the disk. Since $E_{\rm B}/E_{\rm int}$ is 
of $O(10^{-2})$, the magnetic-field strength can be approximated as 
$\sim 0.1\sqrt{8\pi\rho c_{\rm s}^2}\sim 5 \times 10^{14}(\rho/10^{12}\,{\rm g\,cm^{-3}})^{1/2}(c_{\rm s}/10^9\,{\rm cm\,s^{-1}})$\,G 
near the inner edge of the accretion disk. 
The order of this field strength is indeed found in the accretion disk (see, e.g., Fig.~\ref{fig:BUT_Q4_B5_low}). 
By the angular-momentum transport, the matter in the innermost part 
of the accretion disk falls continuously into the black hole, and in this infall, 
the magnetic fluxes also fall in. As a result, the poloidal magnetic-field lines for which the field strength is $\agt 10^{14}$\,G at the horizon penetrate the black hole. 
Here, the infall magnetic fluxes do not have aligned polarity because the accretion process is determined by the turbulence in the accretion disk, and hence, the magnetic-field 
strength on the horizon does not monotonically increase. On the other hand, 
the poloidal magnetic fields in the polar region are twisted by the 
black-hole spin, and hence, the field strength could be larger than that for the 
accretion disk in the presence of a rapidly spinning black hole. Due to 
the twisting associated with the black-hole spin, the toroidal magnetic-field 
strength dominates over the poloidal one in the vicinity of the black hole 
(cf.~Fig.~\ref{fig:plum-ang}). 

However, such amplified magnetic fields do not immediately form a global 
magnetosphere. The reason for this is that at tidal disruption, 
a dense atmosphere ($\rho \sim 10^7~{\rm g/cm^3 }$) is formed in the polar region 
by the matter expelled by shocks generated during the winding and 
shock heating in the spiral arm. The matter also comes from the accretion disk 
due to its turbulent activity. Although a part of 
the matter in the polar region near the black hole 
eventually falls into the black hole, a certain fraction of the matter 
has to be expelled by the magnetic force to form a low-density magnetosphere. 
For this, the toroidal magnetic 
field amplified by the twisting due to the black-hole spin plays an important role, 
because a tower-like outflow is driven from the neighborhood of the black hole 
by this magnetic effect~\cite{ruiz2018dec}. Hence, eventually, the matter energy density
decreases below the magnetic energy density of $b^2/8\pi$ in the polar region 
of the black hole. This is satisfied 
for $\rho < b^2/8\pi c^2=4.4 \times 10^5 (b/10^{14}\,{\rm G})^2\,{\rm g/cm^3}$. 
Then, the magnetic pressure pushes the matter toward the outward direction along the 
rotation axis, establishing a low-density region near the rotational axis. 
During this process, the magnetic-field lines also expand outwardly, and a large-scale magnetosphere near the rotational axis is formed.  In this region, the poloidal field is 
dominant (see Fig.~\ref{fig:plum-ang}). 
As a result, the rest-mass density decreases in the black-hole polar region,
leading to the formation of the so-called funnel structure. At the 
funnel wall, the magnetic pressure is lower than 
the gas pressure of the surrounding thick torus and envelope, and hence, 
the magnetosphere is sustained by the surrounding matter. 

\begin{figure*}[t]
  \begin{tabular}{cc}
    \begin{minipage}[t]{1.0\hsize}
      \begin{center}
        \includegraphics[scale=0.25]{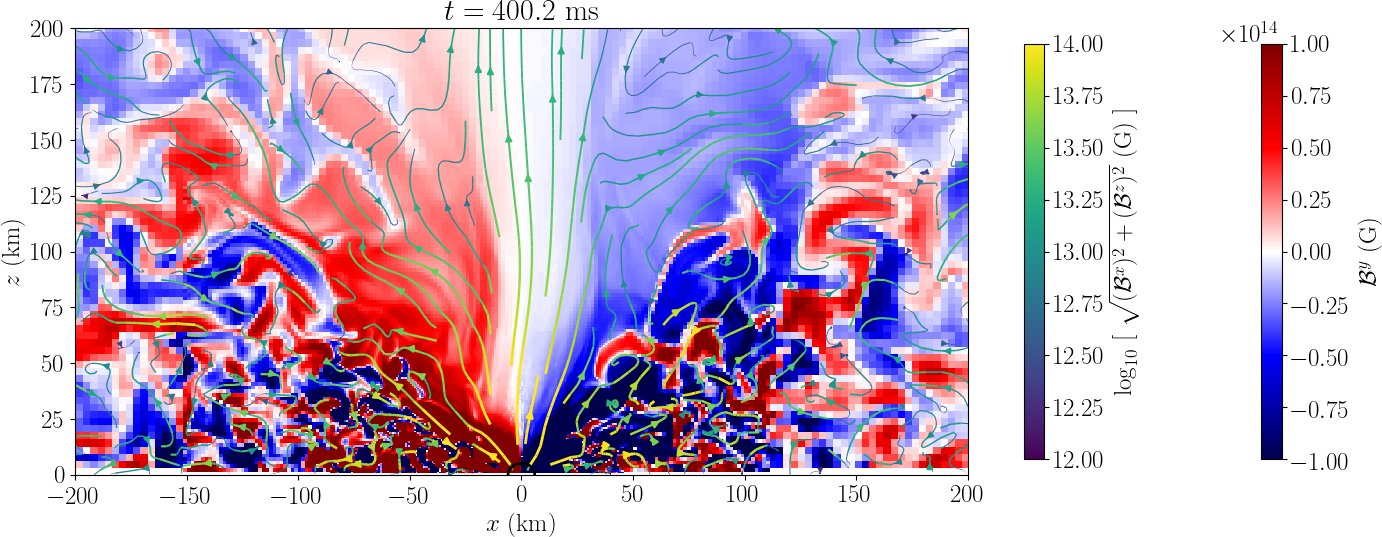}
        \includegraphics[scale=0.25]{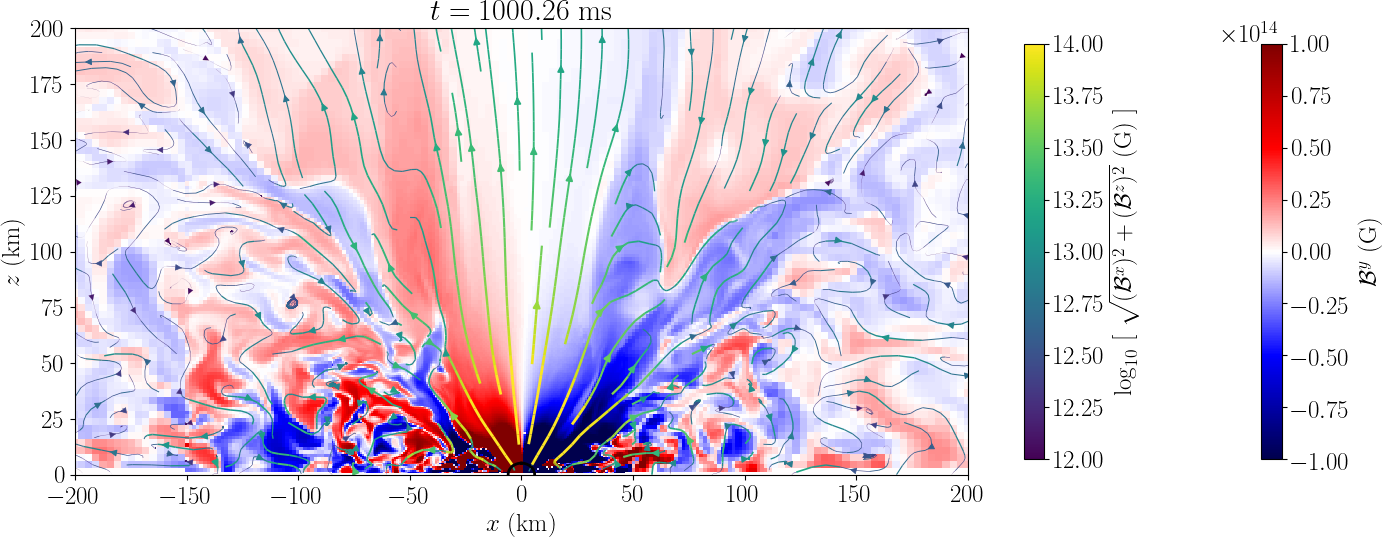}
        \includegraphics[scale=0.25]{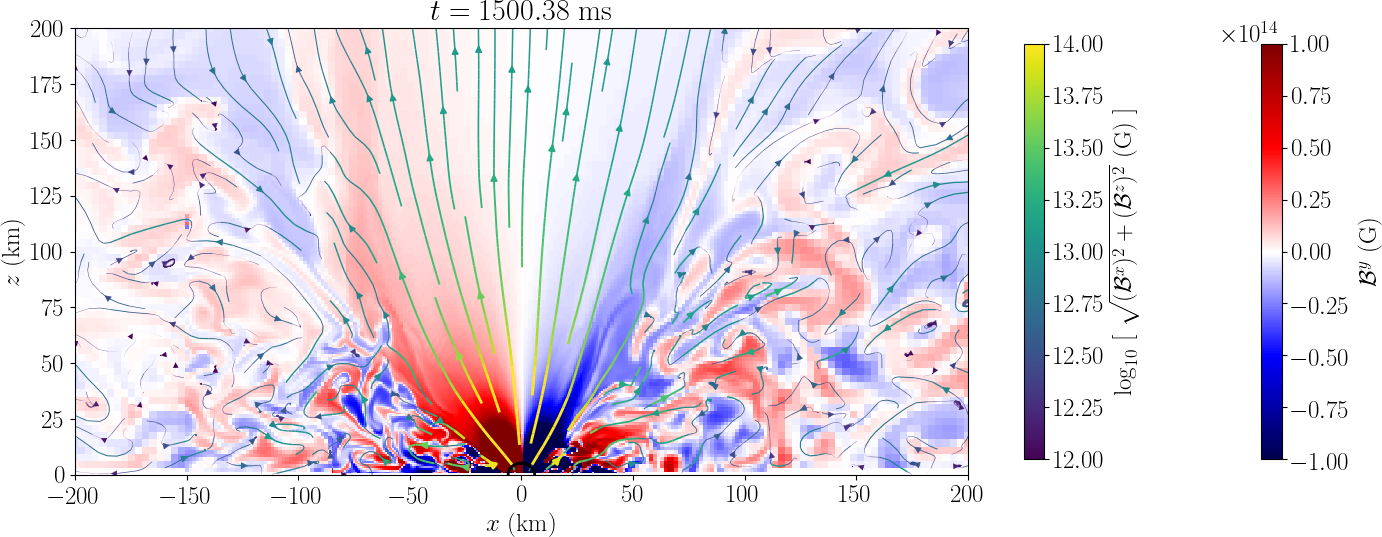}
        \includegraphics[scale=0.25]{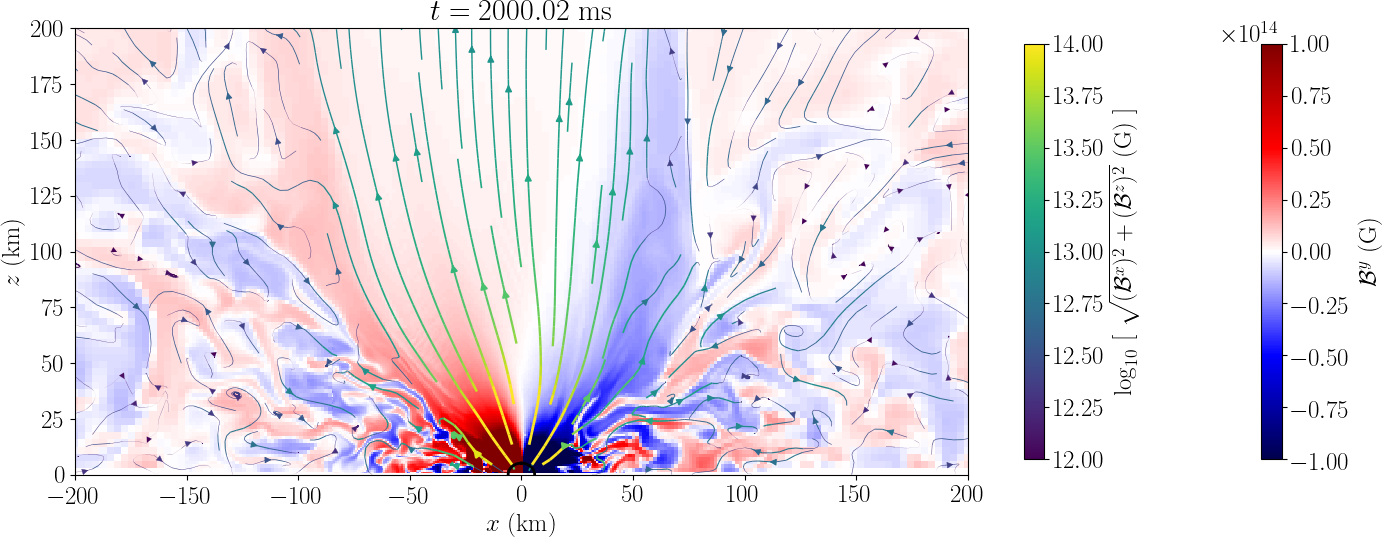}
      \end{center}
    \end{minipage}
  \end{tabular}
  \caption{The snapshot of the toroidal magnetic field (color profile) together 
  with the poloidal magnetic-field lines (curves) on the $x$-$z$ plane at selected time slices
  for model Q4B5L. See also the following link for an animation:
  \url{https://www2.yukawa.kyoto-u.ac.jp/~kota.hayashi/Q4B5L-mf.mp4}
  }
   \label{fig:snap_xz_mag}
\end{figure*}

\begin{figure*}[t]
  \begin{tabular}{cc}
    
    \begin{minipage}[t]{0.5\hsize}
      \begin{center}
        \includegraphics[scale=0.145]{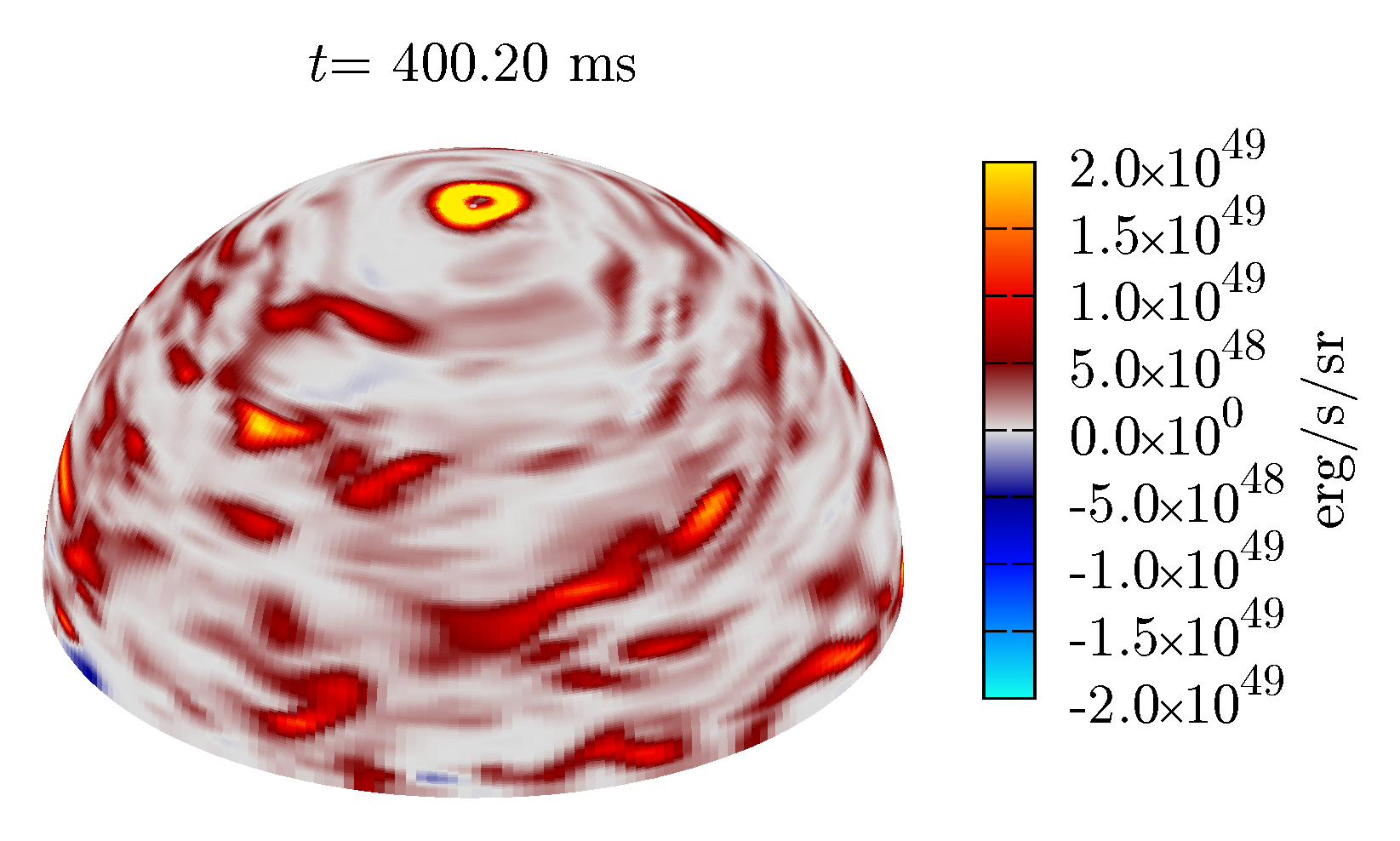}
      \end{center}
    \end{minipage}
    \begin{minipage}[t]{0.5\hsize}
      \begin{center}
        \includegraphics[scale=0.145]{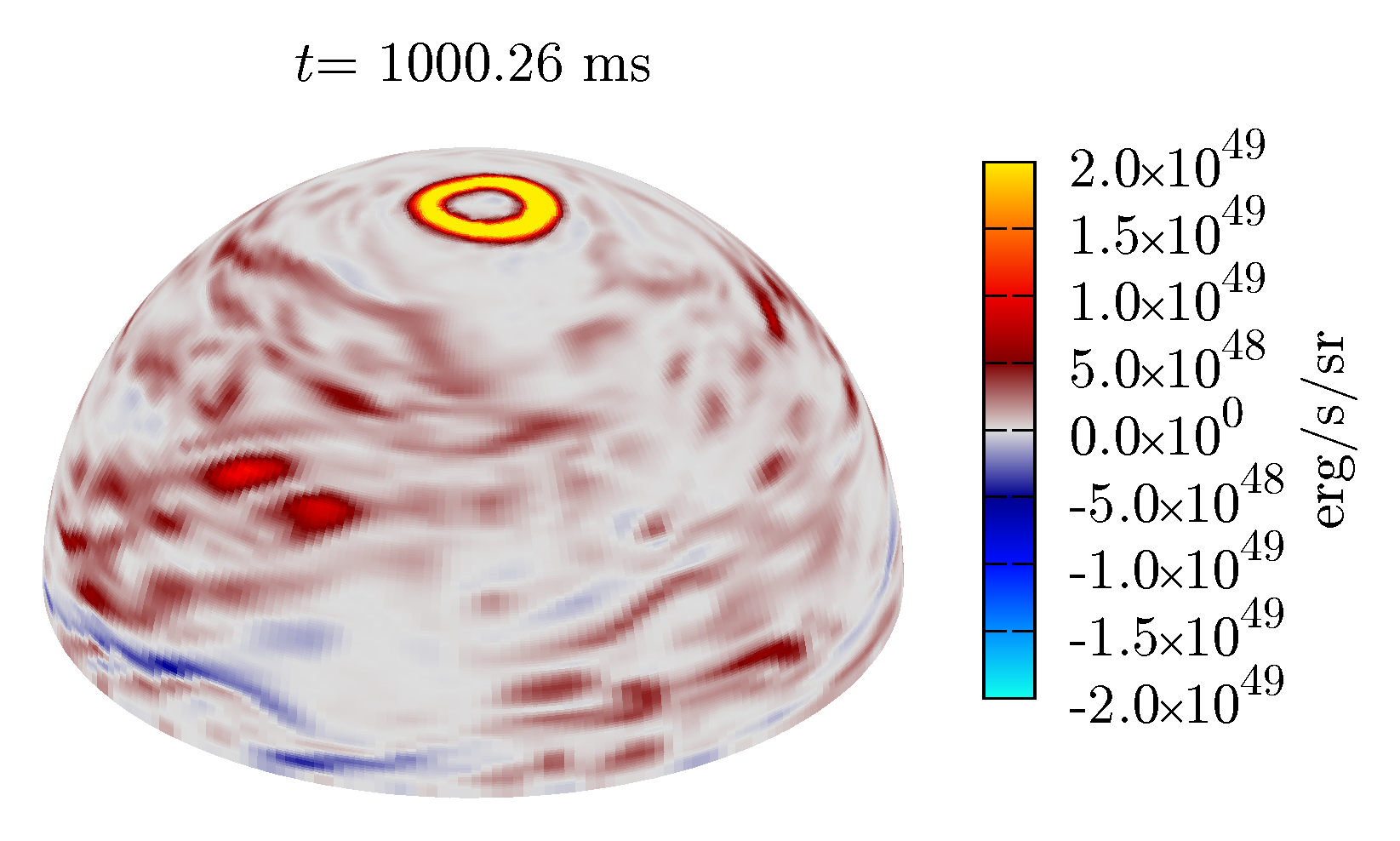}
      \end{center}
    \end{minipage}

  \end{tabular}
\begin{tabular}{cc}
   \begin{minipage}[t]{0.5\hsize}
      \begin{center}
        \includegraphics[scale=0.145]{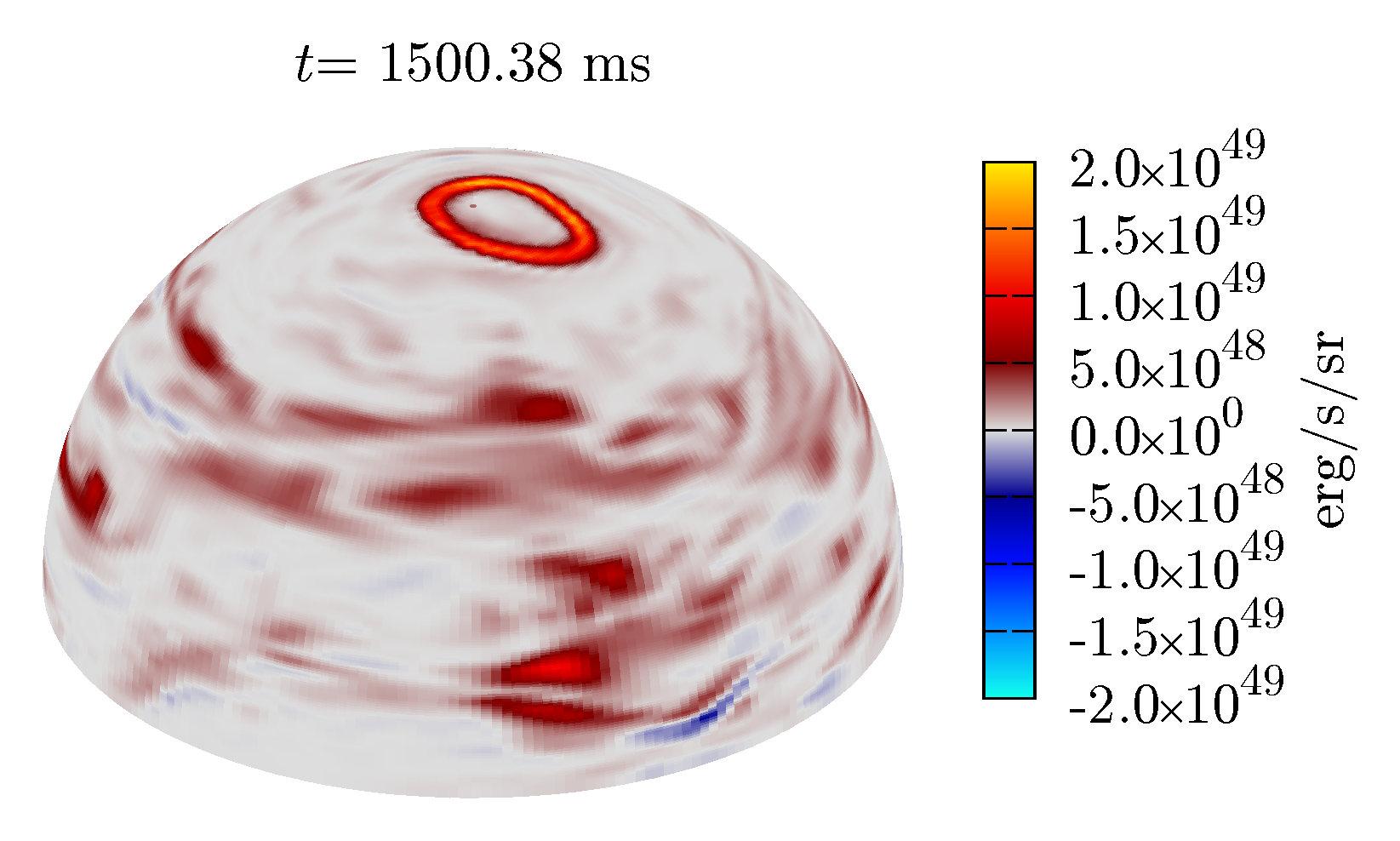}
      \end{center}
    \end{minipage}
    \begin{minipage}[t]{0.5\hsize}
      \begin{center}
        \includegraphics[scale=0.145]{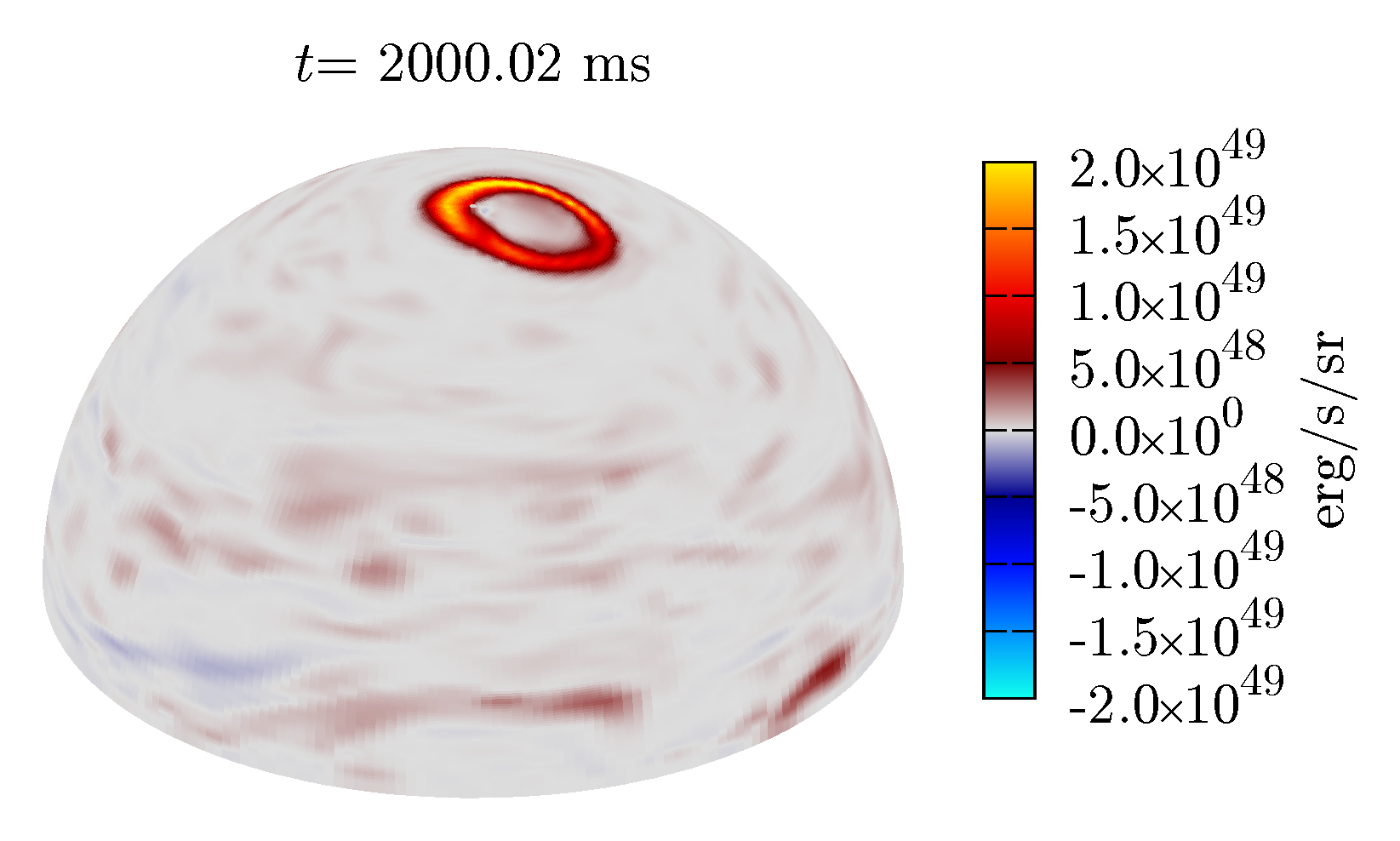}
      \end{center}
    \end{minipage}

  \end{tabular}

  \caption{The angular distribution of the Poynting flux per steradian on a sphere of 
  $r \approx 1500$\,km for model Q4B5L at selected time slices. The bright color displayed in the polar region stems from the Blandford-Znajek effect, while for other regions, the magnetic fields accompanying with the outflowing matter contribute mainly to the Poynting flux. 
 The opening angle of the Poynting flux in the polar region is shown to increase with time. See the following link for an animation: 
 \url{https://www2.yukawa.kyoto-u.ac.jp/~kota.hayashi/Q4B5L-f3D.mp4}
  }
   \label{fig:ang_dist_pflux_Q4B5e16low}
\end{figure*}

Inside the funnel wall, the electromagnetic energy dominates over the rest-mass energy, 
and thus, an approximately force-free magnetosphere is formed. Here, the 
typical ratio of the electromagnetic energy density to the rest-mass energy density is 10--100. 
In such a region, the rotational kinetic energy of the black hole is extracted by 
the Blandford-Znajek mechanism~\cite{blandford1977} and transformed into the 
Poynting flux which propagates outward \addms{(see Appendix~\ref{appendixD} for the result that shows the presence of the outgoing energy flux from the black hole)}.
Figure~\ref{fig:plum} shows the time evolution of $L_{\rm iso}$: 
an isotropic Poynting luminosity, which we define using the Poynting luminosity 
for $\theta < 10^{\circ}$ and $r \approx 1500\,{\rm km}$ as
\begin{eqnarray}
  L_{\rm iso} := \frac{2}{1-\cos(10^{\circ})} L_{\theta < 10^{\circ}, r\approx 1500\,{\rm km}}, 
\end{eqnarray}
where
\begin{eqnarray}
  L_{\theta < 10^{\circ}, r\approx 1500\,{\rm km}} := -\int_{\theta < 10^{\circ},r\approx 1500\,{\rm km}}{T^{\addkh{\mathrm{(EM)}}}}^{~r}_t \sqrt{-g} dS_r.~~\nonumber\\
\end{eqnarray}
\addkh{$T^{\mathrm{(EM)}}_{~\mu\nu}$ denotes the electromagnetic part of the energy-momentum tensor.}
We here choose a particular value ($10^\circ$) for the surface integral because 
the opening angle of the funnel region is initially as narrow as $\sim 10^\circ$ (see Figs.~\ref{fig:snap_xz_mag} and \ref{fig:ang_dist_pflux_Q4B5e16low}). 

Figure~\ref{fig:plum} shows that 
the typical maximum value of $L_{\rm iso}$ is of order $10^{50}\,{\rm erg/s}$ 
and $L_{\rm iso}$ varies with time 
irrespective of the black-hole mass and initial magnetic-field strength. 
This varying isotropic luminosity together with the opening angle of $\theta \sim 10^{\circ}$ 
(cf.~Fig.~\ref{fig:ang_dist_pflux_Q4B5e16low}) is in a fair agreement with those 
for short-hard gamma-ray bursts \addms{in the assumption that the conversion efficiency of the Poynting flux to the gamma-ray radiation is sufficiently high (i.e., close to unity)}~\cite{nakar2007apr,berger2014jun}.\footnote{In the magnetohydrodynamics simulation,
  the flow with low values of $\rho/b^2$ cannot be accurately computed.
  Therefore, it is not possible to reproduce the high Lorentz factor flow in these simulations.} 

The stage with a high value of $L_{\rm iso}\agt 10^{50}$\,erg/s continues 
broadly for 1\,s. Subsequently, the isotropic luminosity starts decreasing. 
This is due to the fact that the opening angle of the funnel region increases 
and the magnetic-flux density is reduced. 
Remember that the funnel region is determined by the gas pressure 
of the thick torus at the funnel wall. In the long-term 
evolution of the accretion torus,  
the rest-mass density and associated gas pressure  
around the funnel wall decrease with time due to the post-merger mass ejection. 
On the other hand, the total magnetic flux penetrating the black hole 
does not significantly decrease in the 
ideal magnetohydrodynamics, and thus, the decrease in the 
magnetic pressure is not as significant as the gas pressure at the funnel wall. 
As the rest-mass density decreases, thus, 
the magnetic pressure exceeds the gas pressure at the original position of the 
funnel wall, and as a result, the funnel wall expands gradually. 

Figure~\ref{fig:snap_xz_mag} displays the snapshot of the toroidal magnetic field together 
with the poloidal magnetic-field lines on the $x$-$z$ plane at selected time slices. 
This indeed shows that the configuration of the magnetic-field lines changes 
from an aligned collimated one near the rotational axis to a more spread one for late time with $t \agt 1$\,s. 
   
Since the collimation of the poloidal magnetic-field lines is loosened, the Poynting flux 
in the vicinity of the rotational axis also decreases gradually. 
Figure~\ref{fig:ang_dist_pflux_Q4B5e16low} shows that the 
opening angle of the strong Poynting-flux ($-{T^{\addkh{\mathrm{(EM)}}}}^{~r}_{t} \sqrt{-g}$) region 
increases from $\alt 10^\circ$ to $\sim 20^\circ$ and the intensity of the Poynting flux 
becomes weak with time. The reason that the peak 
of the Poynting flux is located near the funnel wall is that the 
magnetic-field lines near the funnel wall penetrate the equatorial regions 
of the spinning black hole, and hence, the Blandford-Znajek effect can be more 
efficient. If the Poynting flux indeed determines the luminosity of short-hard gamma-ray bursts, its brightness also should decrease for $t \agt 1$\,s. 
This mechanism could be interpreted as a reason that the timescales of 
short-hard gamma-ray bursts are less than 2\,s with the typical timescale of 
$\sim 1$\,s. Specifically, our numerical results propose that 
the timescale of $\sim 1$\,s is determined by the evolution timescale of 
the accretion disk (torus), which is determined by the neutrino cooling and 
magnetohydrodynamics turbulence (effectively viscous process) that  
control the post-merger mass ejection.

A word of the caution is appropriate here. 
First, the turbulence and dynamo activated by the MRI in the accretion disk 
are stochastic processes.  This implies that the poloidal magnetic-field flux 
penetrating the black hole could not be precisely predicted. For example, 
by the accretion of the magnetic fields with a random polarity, the magnetic 
flux that penetrates the black hole may be smaller than that in the accretion disk. 
Hence, it is 
reasonable that the magnetic-field strength could not be always as strong as the one 
necessary for explaining typical short-hard gamma-ray bursts. 
Indeed, for model Q6B3H, the Poynting luminosity is by one 
order of magnitude lower than those for other models. In this case, 
the magnetic-field strength on the black-hole horizon is about 1/3 of those 
for other models. 
Therefore, broadly speaking, there are two possible cases: 
(1) A magnetosphere with strong poloidal magnetic fields is formed near the rotational 
axis of a spinning black hole. In this case, the maximum isotropic Poynting luminosity of $10^{50}$--$10^{51}~{\rm erg/s}$ consistent with typical short-hard gamma-ray bursts 
can be generated; 
(2) Due to the stochastic process of the MRI-induced turbulent motion,
poloidal magnetic fluxes falling from the disk are not aligned well,
and the poloidal magnetic field formed around the black hole is not strong enough to 
appreciably form a magnetically supported funnel structure (force-free magnetosphere). 
In such a case, the isotropic Poynting luminosity may not be high enough to be consistent with typical short-hard gamma-ray bursts, although a weak Poynting luminosity can be 
generated as in model Q6B3H. For more detailed understanding on this problem, 
a larger number of higher-resolution simulations will be necessary. 
However, this is far beyond the scope of this paper under the current computational 
resource. 

The accretion disks which we find in our simulations do not satisfy the condition for the magnetically-arrested disk~\cite{tchekhovskoy2011}. We calculate the total magnetic flux 
on the upper semi-sphere 
of the horizon $\Phi_{\rm BH}$ and calculate the time evolution of 
$\phi_{\rm BH}:=\Phi_{\rm BH}/\sqrt{\dot M (GM_{\rm BH}/c^2)^2 c}$ 
where $\dot M$ denotes the 
rest-mass accretion rate onto the black-hole horizon calculated by
$-dM_{>{\rm AH}}/dt$ (see the left panel of Fig.~\ref{fig:mdot}), and $c$ and $G$ are recovered to clarify the physical units. In our results, 
$\Phi_{\rm BH}=(1$--$4) \times 10^{27}\,{\rm G\,cm^2}$, and the value of 
the dimensionless quantity, $\phi_{\rm BH}$, increases with the decrease of 
$\dot M$ in time (see the left panel of Fig.~\ref{fig:mdot}). Specifically, irrespective of the initial magnetic-field strength and grid resolution, 
$\phi_{\rm BH}\sim 1$ at $t=100$\,ms and $\sim 5$ at $t=2$\,s for $Q=4$ 
and it is slightly smaller for the runs with $Q=6$. 
Thus in our simulation time, 
$\phi_{\rm BH}$ is much smaller than 50, which is proposed to be necessary to establish the magnetically-arrested disk~\cite{tchekhovskoy2011}. 
In the early stage of its evolution, the accretion disk is the so-called neutrino-dominated 
accretion disk, for which the mass accretion rate is fairly large, the infall magnetic 
fluxes are determined by the equipartition condition in the disk, and thus, it seems to be 
difficult to form a disk which satisfies the condition for the magnetically-arrested disk. 
Our results are quantitatively similar 
to the model BT in Ref.~\cite{christie2019sep}, in which the 
authors considered the evolution of an accretion disk around a spinning black hole 
with the initial condition of a purely toroidal magnetic field. Our results together with 
the results in Ref.~\cite{christie2019sep} suggest that in the absence of an 
extremely strong poloidal magnetic field on the disk from the beginning, 
the magnetically-arrested disk might not be formed as a remnant of neutron-star 
mergers in $t \sim 10$\,s. We emphasize, however, that as we have described in this subsection, an intense Poynting flux can be generated even if the condition for the 
magnetically-arrested disk is not satisfied for the case that 
the rest-mass density along the rotational axis of the 
black hole becomes sufficiently low in a few hundreds ms after the onset of the merger. 

Before closing this section, we note the following points: 
(i) the present simulations are performed imposing the 
equatorial-plane symmetry to save the computational costs. 
In this setting, asymmetric motion in the turbulent state of the accretion disk is neglected. To fully understand the effects of the turbulent motion in the disk 
and resulting formation of the magnetosphere, we need to remove such an unphysical symmetry. To clarify the importance of the asymmetric motion and also to explore the 
case that the orbital angular momentum and black-hole spin are  misaligned, 
we plan to perform a simulation with no plane symmetry in the next step; \addms{(ii) 
the present simulations were started with a poloidal magnetic field confined in the neutron star. The evolution process of the magnetic-field strength is likely to depend on the initial field configuration. In particular, for the case that the field configuration is purely toroidal, the magnetic-field growth rate may be modified significantly in the remnant accretion disk. We also plan to perform simulations with several field configurations in the future work.}

\section{Conclusion} \label{sec:conclusion}

We have reported our new results of general-relativistic neutrino-radiation 
magnetohydrodynamics simulations for the black hole-neutron star merger. 
The mass of the black hole and neutron star are chosen to be \addms{plausible} values 
($M_{\rm BH,0}=5.4$ or $8.1M_\odot$ and $M_{\rm NS}=1.35M_\odot$; cf.~Ref.~\cite{abbott2021jun}), and 
we prepare a rapidly spinning black hole with the dimensionless spin of 0.75 
to consider the case that the neutron star is tidally disrupted in a close orbit. 
The simulations were performed for $\sim 2~{\rm s}$ in the longest case 
to self-consistently explore the dynamical mass ejection, remnant disk evolution, post-merger mass ejection, and collimated Poynting flux generation near the rotational 
axis of the black hole which may be related to short-hard gamma-ray bursts. 

We found that the matter with the mass of $0.04$--$0.05~M_{\odot}$ 
is ejected dynamically right after tidal disruption of the 
neutron star in the timescale of $\alt 10~{\rm ms}$ as found in Ref.~\cite{kyutoku2018jan}. 
Then an accretion disk with the initial rest mass of $0.2$--$0.3M_{\odot}$ is 
formed around the remnant black hole.
In the accretion disk, the magnetohydrodynamics effects such as MRI and winding 
amplify the magnetic field within the timescale of order $10$\,ms, 
and the angular-momentum transport caused by the turbulent motion 
initially induces the mass accretion onto the black hole and disk expansion.
In the turbulent process, the thermal energy is generated and in the first 
$\sim 300$--500\,ms, the thermal energy is dissipated by the neutrino emission.

However, with the expansion of the accretion disk due to the angular-momentum 
transport and magnetic pressure, 
the neutrino luminosity eventually drops below $\sim 10^{51.5}\,{\rm erg/s}$. 
Then, the neutrino cooling does not play a role for carrying away the thermal energy 
from the accretion disk, and the thermal energy generated by the turbulent 
(effectively viscous) process can be fully used for the mass ejection. 
Then, the post-merger mass ejection sets in. In the present study, the 
rest mass of the post-merger ejecta is 
$\sim 0.035M_\odot$ and $\sim 0.020M_{\odot}$ for the models with $Q=4$ and 6, 
respectively. 
This post-merger mass ejection continues from $t \sim 0.3~{\rm s}$ to $\sim 1~{\rm s}$.

Before the post-merger mass ejection sets in,
a low rest-mass density funnel with aligned magnetic-field lines is formed 
near the rotational axis of the spinning black hole. 
This funnel region is magnetically dominant and is approximately in a force-free state. 
In this region, the Blandford-Znajek mechanism extracts the rotation kinetic energy 
of the rapidly spinning black hole, and a collimated Poynting flux is generated 
with the openning angle of $\sim 10^\circ$. 
The estimated maximum isotropic Poynting luminosity is $10^{50}$--$10^{51}\,{\rm erg/s}$.
Together with the opening angle of the Poynting flux with $\sim 10^{\circ}$, 
these numbers are in a fair agreement with the typical short-hard gamma-ray bursts~\cite{nakar2007apr,berger2014jun}. 
The high Poynting luminosity stage continues for $\sim 1~{\rm s}$ and 
the luminosity subsequently 
decreases with time due to the expansion of the funnel wall and resulting 
decrease of the magnetic-flux density. The expansion of the funnel region is 
caused by the decrease of the rest-mass density and gas pressure around the funnel wall 
which takes place due to the post-merger mass ejection. As already mentioned, 
the post-merger mass ejection sets in after the neutrino luminosity drops 
and the duration of the post-merger mass ejection is determined by 
the \addms{angular-momentum transport} timescale of the accretion disk. 
Therefore our present 
results propose that the typical duration of short-hard gamma-ray bursts may be 
determined by the evolution timescale of the accretion disk. Specifically, 
the timescales of the neutrino cooling and viscous evolution in the accretion disk 
(torus) determine the duration of short-hard gamma-ray bursts. 

As we have demonstrated in this paper, seconds-long simulations for the merger of neutron-star binaries 
are inevitable to self-consistently explore the entire merger and post-merger processes. 
This is the case not only for black hole-neutron star binaries but also for 
binary neutron stars. 
We need to focus our effort along this line in the future. 
For the case that a black hole is formed soon after the merger, we expect that 
the \addms{long-term} evolution process \addms{(from the merger to the post-merger mass ejection)} would be qualitatively the same as that found in this paper, 
although the quantitative properties of the post-merger ejecta such as the mass and 
the typical electron fraction are likely to depend sensitively on the mass of the 
remnant black hole and disk. For the case of binary neutron star mergers resulting 
in a massive neutron star, the post-merger evolution process can be 
influenced significantly by the presence of it. If strong global magnetic-field lines 
anchored by the massive neutron star are formed soon after the merger, 
the post-merger mass ejection is likely to be significantly influenced 
by the associated magnetohydrodynamics effects such as the magneto-centrifugal effect~\cite{shibata2021sep}. To explore this possibility, we need to consistently 
follow the evolution of the magnetic-field configuration from 
the merger throughout the post-merger stages. Long-term accurate  
magnetohydrodynamics simulations that can clarify the evolution of the magnetic-field 
structure is in particular desired in the future.

\begin{acknowledgments}
  We thank Kunihito Ioka and Shinya Wanajo for useful discussions. 
  Numerical simulations were performed on
  Sakura, Cobra, and Raven clusters at Max Planck Computing and Data Facility,
  Yukawa-21 at Yukawa Institute for Theoretical Physics of Kyoto University,
  and Cray XC50 at CfCA of National Astronomical Observatory of Japan.
  This work was in part supported by Grant-in-Aid for Scientific Research 
  (grant Nos. 18H01213, 19K14720, and 20H00158) of Japanese MEXT/JSPS.
  \addkh{Kota Hayashi was supported by JST SPRING (grant No. JPMJSP2110).}
\end{acknowledgments}

\appendix

\section{Extension of the equation of state}
\label{app:eos}
Here, we describe our method for extending the tabulated nuclear EOS to lower density and temperature.
The original DD2 EOS~\cite{banik2014sep} covers a range of the rest-mass density of $[1.66\times10^3:1.66\times10^{16}]$\,g/cm$^3$ and temperature of $[0.1:158]$\,MeV, respectively.
Hereafter, we will call DD2 the original table.
We extend the original EOS to low-density and low-temperature sides using the Timmes EOS~\cite{timmes2000}.
One guiding principle of our extension is to make the internal energy continuous. Otherwise, the primitive recovery procedure in the simulation converges to unphysical values or fails.
Here the Timmes EOS contains the effect of not only electrons and positrons with any degeneracy in the non-relativistic to the highly relativistic regime, but also a nuclei component as an ideal gas and the (photon) radiation component.

While the original table returns thermodynamical quantities as functions of density, temperature, and electron fraction assuming the nuclear statistical equilibrium (NSE), the Timmes EOS requires the mean molecular weight of the nuclei as an additional argument.
The internal energy of the original EOS table includes the contribution of the nuclear binding energy coming from the fact that the reference mass of a baryon is assumed to be the atomic mass unit ($\approx 931$\,MeV/$c^2$).
However, the Timmes EOS calculates only the thermal part of the internal energy for nuclei, and hence, we have to add the contribution of the nuclear binding energy.
We first define the contribution of the ``nuclear binding energy" $\varepsilon_\mathrm{nuc}$ to the specific internal energy by
\begin{align}
\varepsilon_\mathrm{nuc}(\rho,T,Y_\mathrm{e}) &:= \varepsilon^\mathrm{DD2}(\rho,T, Y_\mathrm{e}) \notag\\ 
&- \varepsilon^\mathrm{Timmes}(\rho,T, Y_\mathrm{e},\mu^\mathrm{DD2}(\rho,T,Y_\mathrm{e})),
\end{align}
where $\varepsilon^\mathrm{DD2}$ and $\mu^\mathrm{DD2}$ are the specific internal energy and mean molecular weight of the nuclei, respectively, given in the original table, and $\varepsilon^\mathrm{Timmes}$ is the specific internal energy derived from the Timmes EOS.
We note that $\varepsilon^\mathrm{DD2}$ also includes the contribution of the Coulomb energy, and hence, $\varepsilon_\mathrm{nuc}$ also has its contribution.
Using $\varepsilon_\mathrm{nuc}$ defined above, we define the specific internal energy in the extended region of $(\rho,T)$ by
\begin{align}
&\varepsilon^\mathrm{DD2(extended)}(\rho,T,Y_\mathrm{e}) :=\notag\\
&\varepsilon^\mathrm{Timmes}(\rho,T,Y_\mathrm{e},\mu^\mathrm{DD2}(\rho^*,T^*,Y_\mathrm{e})) +\varepsilon_\mathrm{nuc}(\rho^*,T^*,Y_\mathrm{e}),
\end{align}
where $\rho^*$ and $T^*$ are defined by
\begin{align}
\rho^* = \mathrm{max} (\rho,\rho^\mathrm{DD2}_\mathrm{min}), \\
T^* = \mathrm{max} (T,T^\mathrm{DD2}_\mathrm{min})
\end{align}
with the minimum density and temperature of the original EOS table $\rho^\mathrm{DD2}_\mathrm{min}\approx \SI{1.66e3}{g/cm^3}$ and $kT^\mathrm{DD2}_\mathrm{min}=\SI{0.1}{MeV}$.
The mass fractions of free nucleons and heavy nuclei $X_i$ ($i=$ neutrons, protons and heavy nuclei), and the average atomic and mass numbers of heavy nuclei $\langle Z\rangle_\mathrm{heavy}$ and $\langle A\rangle_\mathrm{heavy}$ are defined by
\begin{align}
X_i^\mathrm{DD2(extended)}(\rho,T,Y_\mathrm{e}) &= X_i^\mathrm{DD2}(\rho^*,T^*,Y_\mathrm{e}),\\
\langle Z\rangle_\mathrm{heavy}^\mathrm{DD2(extended)}(\rho,T,Y_\mathrm{e}) &= \langle Z\rangle_\mathrm{heavy}^\mathrm{DD2}(\rho^*,T^*,Y_\mathrm{e}),\\
\langle A\rangle_\mathrm{heavy}^\mathrm{DD2(extended)}(\rho,T,Y_\mathrm{e}) &= \langle A\rangle_\mathrm{heavy}^\mathrm{DD2}(\rho^*,T^*,Y_\mathrm{e}).
\end{align}
Here, the heavy nuclei are referred to the nuclei with mass number larger than 4.

This procedure of the extension of the internal energy assumes that the nuclear composition and the contribution of the nuclear binding energy at the low-density or low-temperature region are the same as those at the closest point of the original table in the $\rho$--$T$ plane with the same value of $Y_\mathrm{e}$.
The method guarantees that $\varepsilon^\mathrm{DD2(extended)}$ and $\varepsilon^\mathrm{DD2}$ are continuously connected at $\rho=\rho^\mathrm{DD2}_\mathrm{min}$ or $T=T^\mathrm{DD2}_\mathrm{min}$.
Other thermodynamical quantities, such as the pressure, are obtained using the Timmes EOS as functions of density, temperature, electron fraction, and mean molecular weight of the nuclei.
It is a reasonable approximation to assume that nuclei are the ideal gas in a low-density region. Hence the extension of the pressure can be done in this region without any artificial reprocessing.
We note that all thermodynamical quantities are derived by using the original table in the parameter region where it is valid.

In this method, the thermodynamical consistency is no longer strictly satisfied in the extended region because we modified the specific internal energy, and nuclear composition in the NSE is not calculated.
However, since the original thermodynamically-consistent EOS is used in the high-density ($\ge \SI{1.66e3}{g/cm^3}$) and high-temperature ($\ge \SI{0.1}{MeV}$) region where the important magnetohydrodynamics processes proceed, this extension does not affect at least the hydrodynamics in the disk and the launch of post-merger ejecta.
In addition, even in the low-density or low-temperature region for which the extended EOS is used, we believe that the effect of the violation of thermodynamical consistency on the hydrodynamics should be minor because we do not artificially modify the pressure in such a region.

With the procedure shown above we extended the original EOS to the ranges of the density of $[0.166:1.66\times10^{16}]$\,g/cm$^3$ and the temperature of $[0.001:158]$\,MeV.
We can set a very-low-density artificial atmosphere with the extended EOS.
This is particularly beneficial to investigate the long-term ejecta dynamics from the mergers of neutron star binaries because its dynamics could be affected if a dense artificial atmosphere were present.
We set the artificial atmosphere density to 1\,g/cm$^3$ in the far region of the computational domain of our simulation.
In this work, the volume of the computational domain is (0.3--5)$\times 10^{29}$\,cm$^3$, and thus the atmosphere density results in (0.16--3)$\times10^{-4}M_\odot$.
On the other hand, the mass of the dynamical ejecta, which can first suffer from the effects of the atmosphere, is $\agt 0.04M_\odot$, which is larger by more than two orders of magnitude than the total mass in the atmosphere.
Hence, the effects of the artificial atmosphere are expected to be minor.

\section{Heating due to the nuclear burning}\label{appendB}

The EOS that we used in this work is derived assuming the NSE for the nuclear composition. Here, we demonstrate how the energy released by the nuclear burning is taken into account in such an EOS.

The total energy density $e_\mathrm{tot}$ in the fluid rest frame including the rest-mass energy is written as
\begin{align}
e_\mathrm{tot} = \sum_i m_ic^2 n_i + e_\mathrm{int}, \label{eq:etot-def}
\end{align}
where $m_i$ and $n_i$ are the mass and number density of $i$-th nuclear species, and $e_\mathrm{int}$ is the ``pure" internal energy density (i.e., without the rest-mass origin).
In our formulation of hydrodynamics and also in constructing the EOS, we define the reference mass of baryons as the atomic unit mass $m_\mathrm{u}$($\approx931$\,MeV/$c^2$), and thus, the rest-mass density is written by
\begin{align}
\rho = m_\mathrm{u} n_\mathrm{b} = m_\mathrm{u} \sum_i A_i n_i,
\end{align}
where $A_i$ is the mass number of $i$-th nuclear species and $n_\mathrm{b}$ is the baryon number density.
Using the definition of $\rho$, Eq.~\eqref{eq:etot-def} can be rewritten as
\begin{align}
e_\mathrm{tot} &= \rho c^2 + \biggl[ \sum_i m_ic^2 n_i - \rho c^2 + e_\mathrm{int} \biggr]\notag\\
&= \rho c^2 + \rho \biggl[\frac{ \langle \Delta m \rangle c^2}{m_\mathrm{u}} +  \varepsilon_\mathrm{int}\biggr], \label{eq:etot_massexcess}
\end{align}
where $\varepsilon_\mathrm{int} = e_\mathrm{int}/\rho$ and 
\begin{align}
\langle \Delta m \rangle = \sum_i (m_i - A_i m_\mathrm{u})\frac{n_i}{n_\mathrm{b}}
\end{align}
is the average mass excess per baryon for a given nuclear composition.
The quantity inside the bracket of the second term in Eq.~\eqref{eq:etot_massexcess}, $e_\mathrm{tot}/\rho-c^2$,  is the ``specific internal energy" $\varepsilon$ in our formulation.

Then we consider the variation only of the nuclear composition.
That is, we consider the change of $\langle \Delta m\rangle$ with keeping the total energy density $e_\mathrm{tot}$ fixed.
From Eq.~\eqref{eq:etot_massexcess}, we have
\begin{align}
0=de_\mathrm{tot}|_\mathrm{comp} = \rho\biggl[\frac{c^2}{m_\mathrm{u}} d\langle \Delta m \rangle  +  d\varepsilon_\mathrm{int}|_\mathrm{comp}\biggr],
\end{align}
where $dX|_\mathrm{comp}$ is the variation of the quantity $X$ associated only with the change in the composition.
We note that $\rho=m_\mathrm{u}n_\mathrm{b}$ does not change by the modification of the nuclear composition.
As a result,
\begin{align}
d\varepsilon_\mathrm{int}|_\mathrm{comp} = -\frac{c^2}{m_\mathrm{u}} d\langle \Delta m \rangle. \label{eq:nuc-heating}
\end{align}
Equation~\eqref{eq:nuc-heating} clearly shows that the nuclear burning, which results in the modification of the nuclear composition, simply changes the internal energy, and thus, this is the net specific heating by the nuclear burning.
For example, $\langle \Delta m \rangle \approx \SI{7.4}{MeV}/c^2$ for the production of irons from the matter composed of half free protons and half free neutrons, and $\langle \Delta m \rangle \approx\SI{1.4}{MeV}/c^2$ from the matter only with alpha particles.
Thus, if protons and neutrons completely recombine into alpha particles in the matter with $Y_\mathrm{e}=0.5$, 6.0\,MeV per baryon is released to increase $\varepsilon_\mathrm{int}$.

It is important to note that the heating by the nuclear burning is automatically incorporated without adding any source term in the equations of hydrodynamics if the change in the average mass excess by the nuclear burning is taken into account.
The net specific heating rate by the nuclear reaction is written by
\begin{align}
\frac{d\varepsilon_\mathrm{int}}{dt}\bigg|_\mathrm{comp}=-\frac{c^2}{m_\mathrm{u}}\frac{d\langle \Delta m \rangle}{dt}.
\end{align}
The time derivative of $\langle \Delta m \rangle$ is in general written by
\begin{align}
\frac{d\langle \Delta m \rangle}{dt} = \sum_i (m_i - A_i m_\mathrm{u})\frac{1}{n_\mathrm{b}}\frac{dn_i}{dt},
\end{align}
where $dn_i/dt$ is the time derivative of the number density of $i$-th species (this method is employed in, e.g., Ref.~\cite{uchida2017oct}).
On the other hand, in the NSE, $\langle \Delta m \rangle$ is a function of the baryon number density (or $\rho$), temperature and electron fraction, and thus, the net heating rate by the nuclear burning can be expressed as
\begin{align}
\frac{d\langle \Delta m \rangle}{dt}= \frac{\partial \langle \Delta m \rangle}{\partial \rho}\frac{d\rho}{dt} + \frac{\partial \langle \Delta m \rangle}{\partial T}\frac{dT}{dt} + \frac{\partial \langle \Delta m \rangle}{\partial Y_\mathrm{e}}\frac{dY_\mathrm{e}}{dt}.
\end{align}


\section{On the resolution of MRI}\label{appendixC}

\begin{figure*}[t]
  \begin{tabular}{cc}
    \begin{minipage}[t]{1.0\hsize}
      \begin{center}
        \includegraphics[scale=0.23]{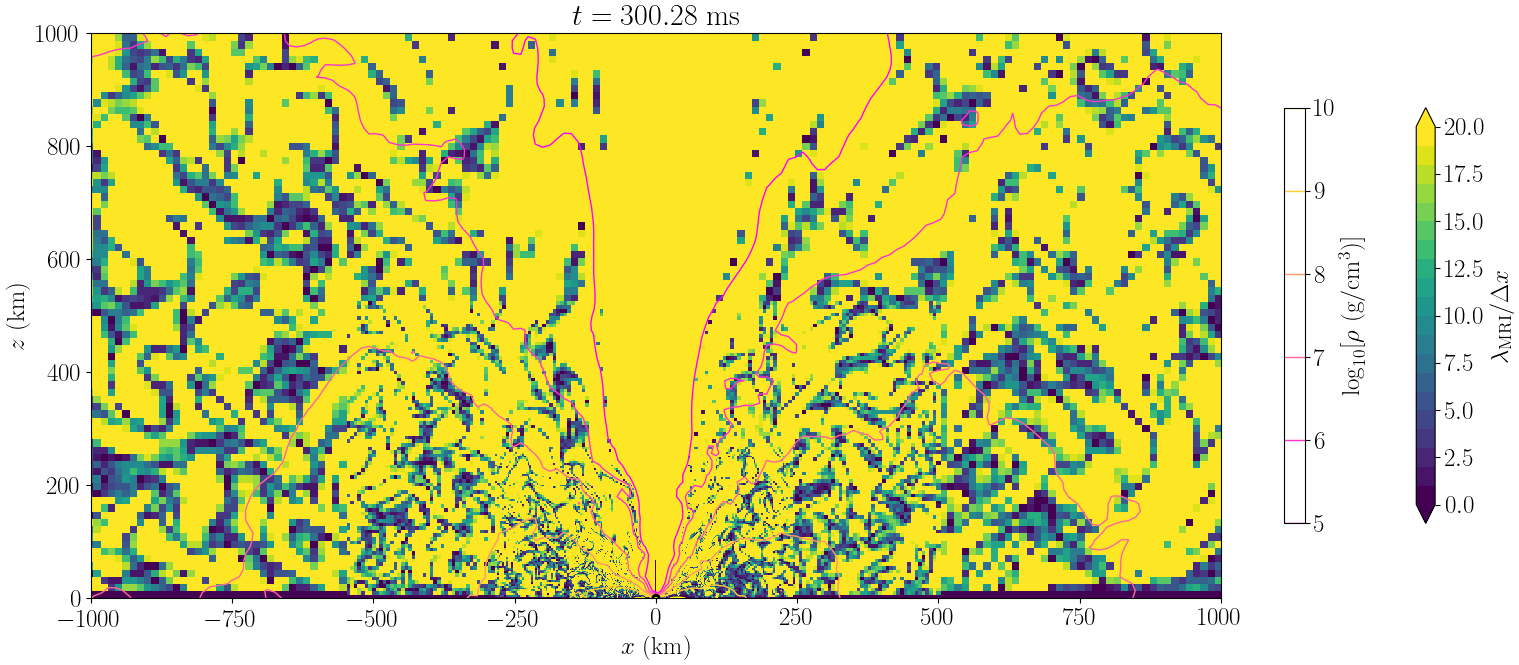}
        \includegraphics[scale=0.23]{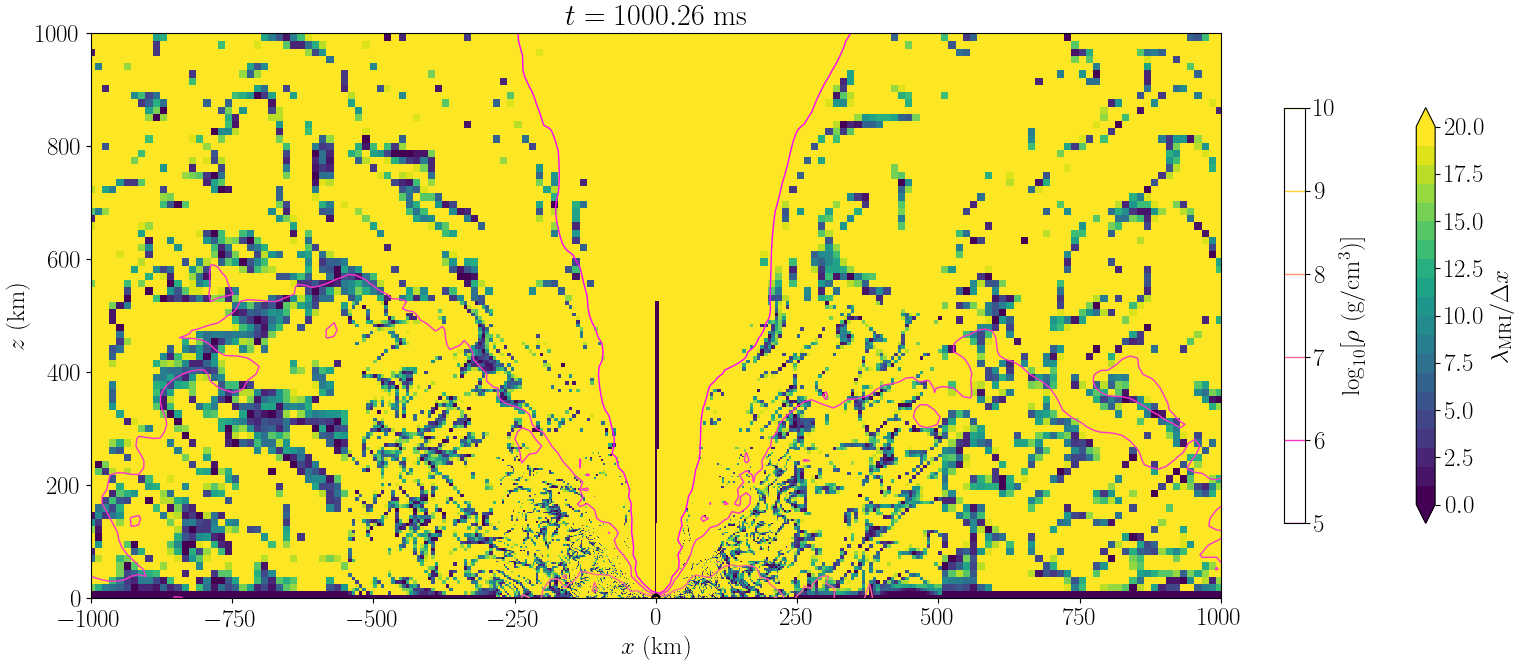}
      \end{center}
    \end{minipage}
  \end{tabular}
  \caption{The snapshots of the MRI quality factor (color profile) together with the rest-mass density (contour) on the $x$-$z$ plane at $t \approx 300$ and 1000\,ms for model Q4B5L.}
  \label{fig:mri_z_qual}
\end{figure*}

\begin{figure*}[t]
  \begin{tabular}{cc}
    \begin{minipage}[t]{1.0\hsize}
      \begin{center}
        \includegraphics[scale=0.22]{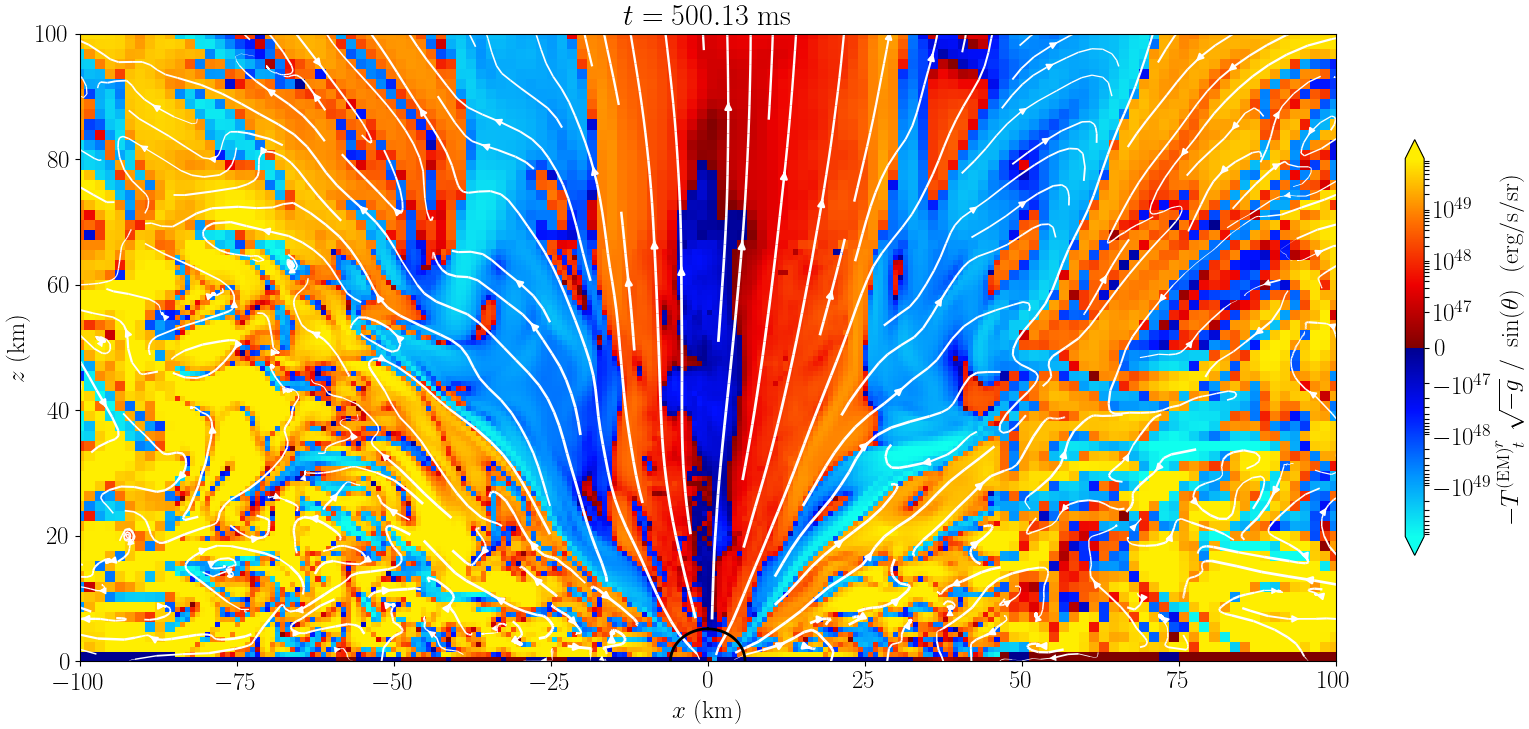}
        \includegraphics[scale=0.22]{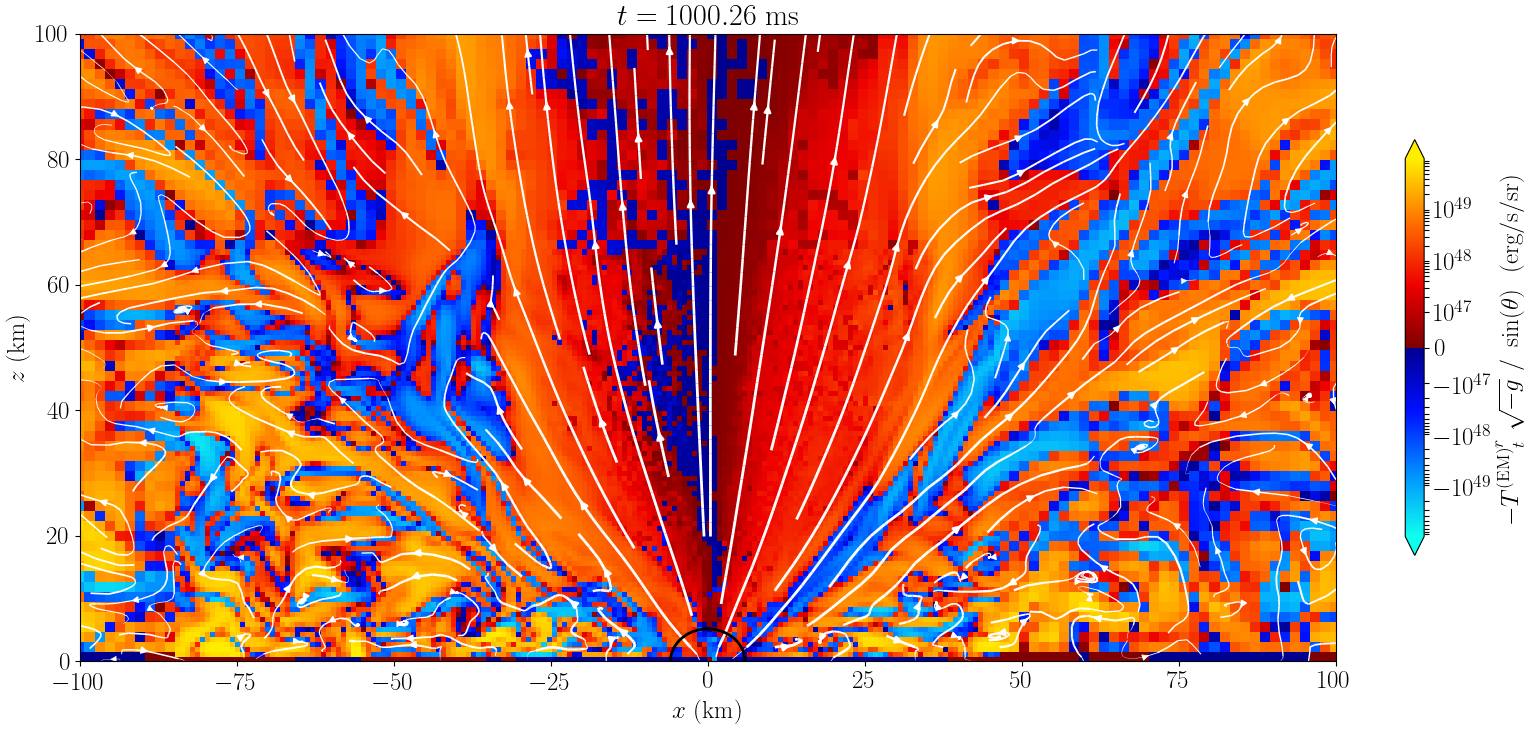}
      \end{center}
    \end{minipage}
  \end{tabular}
  \caption{The snapshots of the outgoing Poynting flux per steradian (color profile) together with poloidal magnetic-field lines (white curves) near the apparent horizon on the $x$--$z$ plane at $t\approx 500$ and 1000\,ms for model Q4B5L.
  The apparent horizon is shown with the black circle.}
  \label{fig:pflux_ah}
\end{figure*}

Figure~\ref{fig:mri_z_qual} shows the snapshots of the MRI quality factor defined by ${\lambda_z}_{\rm MRI}/{\Delta}x$ on the $x$-$z$ plane for model Q4B5L.
Here, ${\lambda_z}_{\rm MRI}$ is the wavelength of the fastest growing mode of the axisymmetric MRI, which is defined by 
\begin{eqnarray}
{\lambda_z}_{\rm MRI} &=& \frac{b_{z}}{\sqrt{4{\pi}{\rho}h+b^{\mu}b_{\mu}}}\frac{2{\pi}}{\Omega},
\end{eqnarray}
where $\Omega$ denotes the local angular velocity and $z$-direction is the direction of the rotation axis. We note that for other time slices, the similar feature is also found. 
Figure~\ref{fig:mri_z_qual} shows that the fastest growing mode is covered by more than 20 grid points in the large portion of the disk even for the low-resolution run (${\lambda_z}_{\rm MRI} \sim 10\,{\rm km}$ in the inner region of the disk), and thus, we consider that the fastest growing mode of the MRI is resolved with a reasonable accuracy in the present work.

\section{Evidence for the Blandford-Znajek mechanism}\label{appendixD}
Figure~\ref{fig:pflux_ah} shows the snapshots for an outgoing Poynting flux per steradian near the apparent horizon defined by $-{T^{\mathrm{(EM)}}}^{~r}_{t}\sqrt{-g}/\sin{\theta}$.
Along the poloidal magnetic-field line, the outgoing Poynting flux is distributed from the apparent horizon to the mangnetosphere around the rotational axis of the black hole.
This indicates that energy is extracted from the black hole through the magnetic field, and thus, we can interpret that the Blandford-Znajek mechanism is in operation. In addition, we find that the total Poynting luminosity on the apparent horizon is $\sim10^{49}\,{\rm erg/s}$. This value is consistent with the luminosity expected from the formula for the Blandford-Znajek mechanism~\cite{blandford1977} for the resultant values of the magnetic-field strength, black-hole mass, and spin.

~\\

\bibliography{paper}

\begin{thebibliography}{103}%
\makeatletter
\providecommand \@ifxundefined [1]{%
 \@ifx{#1\undefined}
}%
\providecommand \@ifnum [1]{%
 \ifnum #1\expandafter \@firstoftwo
 \else \expandafter \@secondoftwo
 \fi
}%
\providecommand \@ifx [1]{%
 \ifx #1\expandafter \@firstoftwo
 \else \expandafter \@secondoftwo
 \fi
}%
\providecommand \natexlab [1]{#1}%
\providecommand \enquote  [1]{``#1''}%
\providecommand \bibnamefont  [1]{#1}%
\providecommand \bibfnamefont [1]{#1}%
\providecommand \citenamefont [1]{#1}%
\providecommand \href@noop [0]{\@secondoftwo}%
\providecommand \href [0]{\begingroup \@sanitize@url \@href}%
\providecommand \@href[1]{\@@startlink{#1}\@@href}%
\providecommand \@@href[1]{\endgroup#1\@@endlink}%
\providecommand \@sanitize@url [0]{\catcode `\\12\catcode `\$12\catcode
  `\&12\catcode `\#12\catcode `\^12\catcode `\_12\catcode `\%12\relax}%
\providecommand \@@startlink[1]{}%
\providecommand \@@endlink[0]{}%
\providecommand \url  [0]{\begingroup\@sanitize@url \@url }%
\providecommand \@url [1]{\endgroup\@href {#1}{\urlprefix }}%
\providecommand \urlprefix  [0]{URL }%
\providecommand \Eprint [0]{\href }%
\providecommand \doibase [0]{http://dx.doi.org/}%
\providecommand \selectlanguage [0]{\@gobble}%
\providecommand \bibinfo  [0]{\@secondoftwo}%
\providecommand \bibfield  [0]{\@secondoftwo}%
\providecommand \translation [1]{[#1]}%
\providecommand \BibitemOpen [0]{}%
\providecommand \bibitemStop [0]{}%
\providecommand \bibitemNoStop [0]{.\EOS\space}%
\providecommand \EOS [0]{\spacefactor3000\relax}%
\providecommand \BibitemShut  [1]{\csname bibitem#1\endcsname}%
\let\auto@bib@innerbib\@empty
\bibitem [{\citenamefont {Abbott}\ \emph {et~al.}(2016)\citenamefont {Abbott}
  \emph {et~al.}}]{abbott2016feb}%
  \BibitemOpen
  \bibfield  {author} {\bibinfo {author} {\bibfnamefont {B.~P.}\ \bibnamefont
  {Abbott}} \emph {et~al.} (\bibinfo {collaboration} {LIGO Scientific
  Collaboration and Virgo Collaboration}),\ }\href {\doibase
  10.1103/PhysRevLett.116.061102} {\bibfield  {journal} {\bibinfo  {journal}
  {Phys. Rev. Lett.}\ }\textbf {\bibinfo {volume} {116}},\ \bibinfo {pages}
  {061102} (\bibinfo {year} {2016})}\BibitemShut {NoStop}%
\bibitem [{\citenamefont {Abbott}\ \emph
  {et~al.}(2021{\natexlab{a}})\citenamefont {Abbott} \emph
  {et~al.}}]{abbott2021jun}%
  \BibitemOpen
  \bibfield  {author} {\bibinfo {author} {\bibfnamefont {R.}~\bibnamefont
  {Abbott}} \emph {et~al.} (\bibinfo {collaboration} {LIGO Scientific
  Collaboration and Virgo Collaboration}),\ }\href {\doibase
  10.1103/PhysRevX.11.021053} {\bibfield  {journal} {\bibinfo  {journal} {Phys.
  Rev. X}\ }\textbf {\bibinfo {volume} {11}},\ \bibinfo {pages} {021053}
  (\bibinfo {year} {2021}{\natexlab{a}})}\BibitemShut {NoStop}%
\bibitem [{\citenamefont {Abbott}\ \emph
  {et~al.}(2021{\natexlab{b}})\citenamefont {Abbott} \emph {et~al.}}]{ligoo3b}%
  \BibitemOpen
  \bibfield  {author} {\bibinfo {author} {\bibfnamefont {R.}~\bibnamefont
  {Abbott}} \emph {et~al.},\ }\href@noop {} {\  (\bibinfo {year}
  {2021}{\natexlab{b}})},\ \Eprint {http://arxiv.org/abs/2111.03606}
  {arXiv:2111.03606 [gr-qc]} \BibitemShut {NoStop}%
\bibitem [{\citenamefont {Abbott}\ \emph
  {et~al.}(2017{\natexlab{a}})\citenamefont {Abbott} \emph
  {et~al.}}]{abbott2017oct1}%
  \BibitemOpen
  \bibfield  {author} {\bibinfo {author} {\bibfnamefont {B.~P.}\ \bibnamefont
  {Abbott}} \emph {et~al.} (\bibinfo {collaboration} {LIGO Scientific
  Collaboration and Virgo Collaboration}),\ }\href {\doibase
  10.1103/PhysRevLett.119.161101} {\bibfield  {journal} {\bibinfo  {journal}
  {Phys. Rev. Lett.}\ }\textbf {\bibinfo {volume} {119}},\ \bibinfo {pages}
  {161101} (\bibinfo {year} {2017}{\natexlab{a}})}\BibitemShut {NoStop}%
\bibitem [{\citenamefont {Abbott}\ \emph
  {et~al.}(2017{\natexlab{b}})\citenamefont {Abbott} \emph
  {et~al.}}]{abbott2017oct2}%
  \BibitemOpen
  \bibfield  {author} {\bibinfo {author} {\bibfnamefont {B.~P.}\ \bibnamefont
  {Abbott}} \emph {et~al.},\ }\href {\doibase 10.3847/2041-8213/aa91c9}
  {\bibfield  {journal} {\bibinfo  {journal} {\apj}\ }\textbf {\bibinfo
  {volume} {848}},\ \bibinfo {pages} {L12} (\bibinfo {year}
  {2017}{\natexlab{b}})}\BibitemShut {NoStop}%
\bibitem [{\citenamefont {Abbott}\ \emph
  {et~al.}(2017{\natexlab{c}})\citenamefont {Abbott} \emph
  {et~al.}}]{abbott2017oct3}%
  \BibitemOpen
  \bibfield  {author} {\bibinfo {author} {\bibfnamefont {B.~P.}\ \bibnamefont
  {Abbott}} \emph {et~al.},\ }\href {\doibase 10.3847/2041-8213/aa920c}
  {\bibfield  {journal} {\bibinfo  {journal} {\apj}\ }\textbf {\bibinfo
  {volume} {848}},\ \bibinfo {pages} {L13} (\bibinfo {year}
  {2017}{\natexlab{c}})}\BibitemShut {NoStop}%
\bibitem [{\citenamefont {Abbott}\ \emph
  {et~al.}(2021{\natexlab{c}})\citenamefont {Abbott} \emph
  {et~al.}}]{abbott2021jun2}%
  \BibitemOpen
  \bibfield  {author} {\bibinfo {author} {\bibfnamefont {R.}~\bibnamefont
  {Abbott}} \emph {et~al.},\ }\href {\doibase 10.3847/2041-8213/ac082e}
  {\bibfield  {journal} {\bibinfo  {journal} {\apj}\ }\textbf {\bibinfo
  {volume} {915}},\ \bibinfo {pages} {L5} (\bibinfo {year}
  {2021}{\natexlab{c}})}\BibitemShut {NoStop}%
\bibitem [{\citenamefont {Shibata}\ and\ \citenamefont
  {Taniguchi}(2011)}]{shibata2011aug}%
  \BibitemOpen
  \bibfield  {author} {\bibinfo {author} {\bibfnamefont {M.}~\bibnamefont
  {Shibata}}\ and\ \bibinfo {author} {\bibfnamefont {K.}~\bibnamefont
  {Taniguchi}},\ }\href {https://doi.org/10.12942/lrr-2011-6} {\bibfield
  {journal} {\bibinfo  {journal} {Living Rev. Relativity}\ }\textbf {\bibinfo
  {volume} {14}},\ \bibinfo {pages} {6} (\bibinfo {year} {2011})}\BibitemShut
  {NoStop}%
\bibitem [{\citenamefont {{Kyutoku}}\ \emph {et~al.}(2021)\citenamefont
  {{Kyutoku}}, \citenamefont {{Shibata}},\ and\ \citenamefont
  {{Taniguchi}}}]{KST2021}%
  \BibitemOpen
  \bibfield  {author} {\bibinfo {author} {\bibfnamefont {K.}~\bibnamefont
  {{Kyutoku}}}, \bibinfo {author} {\bibfnamefont {M.}~\bibnamefont
  {{Shibata}}}, \ and\ \bibinfo {author} {\bibfnamefont {K.}~\bibnamefont
  {{Taniguchi}}},\ }\href {\doibase 10.1007/s41114-021-00033-4} {\bibfield
  {journal} {\bibinfo  {journal} {Living Rev. Relativity}\ }\textbf {\bibinfo
  {volume} {24}},\ \bibinfo {eid} {5} (\bibinfo {year} {2021})}\BibitemShut
  {NoStop}%
\bibitem [{\citenamefont {{Eichler}}\ \emph {et~al.}(1989)\citenamefont
  {{Eichler}}, \citenamefont {{Livio}}, \citenamefont {{Piran}},\ and\
  \citenamefont {{Schramm}}}]{eichler1989}%
  \BibitemOpen
  \bibfield  {author} {\bibinfo {author} {\bibfnamefont {D.}~\bibnamefont
  {{Eichler}}}, \bibinfo {author} {\bibfnamefont {M.}~\bibnamefont {{Livio}}},
  \bibinfo {author} {\bibfnamefont {T.}~\bibnamefont {{Piran}}}, \ and\
  \bibinfo {author} {\bibfnamefont {D.~N.}\ \bibnamefont {{Schramm}}},\ }\href
  {\doibase 10.1038/340126a0} {\bibfield  {journal} {\bibinfo  {journal}
  {\nat}\ }\textbf {\bibinfo {volume} {340}},\ \bibinfo {pages} {126} (\bibinfo
  {year} {1989})}\BibitemShut {NoStop}%
\bibitem [{\citenamefont {Nakar}(2007)}]{nakar2007apr}%
  \BibitemOpen
  \bibfield  {author} {\bibinfo {author} {\bibfnamefont {E.}~\bibnamefont
  {Nakar}},\ }\href {\doibase https://doi.org/10.1016/j.physrep.2007.02.005}
  {\bibfield  {journal} {\bibinfo  {journal} {Phys. Rep.}\ }\textbf {\bibinfo
  {volume} {442}},\ \bibinfo {pages} {166 } (\bibinfo {year} {2007})},\
  \bibinfo {note} {the Hans Bethe Centennial Volume 1906-2006}\BibitemShut
  {NoStop}%
\bibitem [{\citenamefont {Berger}(2014)}]{berger2014jun}%
  \BibitemOpen
  \bibfield  {author} {\bibinfo {author} {\bibfnamefont {E.}~\bibnamefont
  {Berger}},\ }\href {\doibase 10.1146/annurev-astro-081913-035926} {\bibfield
  {journal} {\bibinfo  {journal} {Annu. Rev. Astron. Astrophys.}\ }\textbf
  {\bibinfo {volume} {52}},\ \bibinfo {pages} {43} (\bibinfo {year}
  {2014})}\BibitemShut {NoStop}%
\bibitem [{\citenamefont {{Lattimer}}\ and\ \citenamefont
  {{Schramm}}(1974)}]{lattimer1974}%
  \BibitemOpen
  \bibfield  {author} {\bibinfo {author} {\bibfnamefont {J.~M.}\ \bibnamefont
  {{Lattimer}}}\ and\ \bibinfo {author} {\bibfnamefont {D.~N.}\ \bibnamefont
  {{Schramm}}},\ }\href {\doibase 10.1086/181612} {\bibfield  {journal}
  {\bibinfo  {journal} {Astrophys. J. Lett.}\ }\textbf {\bibinfo {volume}
  {192}},\ \bibinfo {pages} {L145} (\bibinfo {year} {1974})}\BibitemShut
  {NoStop}%
\bibitem [{\citenamefont {Li}\ and\ \citenamefont
  {Paczy{\'{n}}ski}(1998)}]{li1998nov}%
  \BibitemOpen
  \bibfield  {author} {\bibinfo {author} {\bibfnamefont {L.-X.}\ \bibnamefont
  {Li}}\ and\ \bibinfo {author} {\bibfnamefont {B.}~\bibnamefont
  {Paczy{\'{n}}ski}},\ }\href {\doibase 10.1086/311680} {\bibfield  {journal}
  {\bibinfo  {journal} {Astrophys. J.}\ }\textbf {\bibinfo {volume} {507}},\
  \bibinfo {pages} {L59} (\bibinfo {year} {1998})}\BibitemShut {NoStop}%
\bibitem [{\citenamefont {Metzger}\ \emph {et~al.}(2010)\citenamefont
  {Metzger}, \citenamefont {Mart{\'{i}}nez-Pinedo}, \citenamefont {Darbha},
  \citenamefont {Quataert}, \citenamefont {Arcones}, \citenamefont {Kasen},
  \citenamefont {Thomas}, \citenamefont {Nugent}, \citenamefont {Panov},\ and\
  \citenamefont {Zinner}}]{metzger2010jun}%
  \BibitemOpen
  \bibfield  {author} {\bibinfo {author} {\bibfnamefont {B.~D.}\ \bibnamefont
  {Metzger}}, \bibinfo {author} {\bibfnamefont {G.}~\bibnamefont
  {Mart{\'{i}}nez-Pinedo}}, \bibinfo {author} {\bibfnamefont {S.}~\bibnamefont
  {Darbha}}, \bibinfo {author} {\bibfnamefont {E.}~\bibnamefont {Quataert}},
  \bibinfo {author} {\bibfnamefont {A.}~\bibnamefont {Arcones}}, \bibinfo
  {author} {\bibfnamefont {D.}~\bibnamefont {Kasen}}, \bibinfo {author}
  {\bibfnamefont {R.}~\bibnamefont {Thomas}}, \bibinfo {author} {\bibfnamefont
  {P.}~\bibnamefont {Nugent}}, \bibinfo {author} {\bibfnamefont {I.~V.}\
  \bibnamefont {Panov}}, \ and\ \bibinfo {author} {\bibfnamefont {N.~T.}\
  \bibnamefont {Zinner}},\ }\href {\doibase 10.1111/j.1365-2966.2010.16864.x}
  {\bibfield  {journal} {\bibinfo  {journal} {Mon. Not. R. Astron. Soc.}\
  }\textbf {\bibinfo {volume} {406}},\ \bibinfo {pages} {2650} (\bibinfo {year}
  {2010})}\BibitemShut {NoStop}%
\bibitem [{\citenamefont {Shibata}\ and\ \citenamefont
  {Ury\ifmmode~\bar{u}\else \={u}\fi{}}(2006)}]{shibata2006dec}%
  \BibitemOpen
  \bibfield  {author} {\bibinfo {author} {\bibfnamefont {M.}~\bibnamefont
  {Shibata}}\ and\ \bibinfo {author} {\bibfnamefont {K.}~\bibnamefont
  {Ury\ifmmode~\bar{u}\else \={u}\fi{}}},\ }\href {\doibase
  10.1103/PhysRevD.74.121503} {\bibfield  {journal} {\bibinfo  {journal} {Phys.
  Rev. D}\ }\textbf {\bibinfo {volume} {74}},\ \bibinfo {pages} {121503}
  (\bibinfo {year} {2006})}\BibitemShut {NoStop}%
\bibitem [{\citenamefont {Shibata}\ and\ \citenamefont
  {Ury{\={u}}}(2007)}]{shibata2007may}%
  \BibitemOpen
  \bibfield  {author} {\bibinfo {author} {\bibfnamefont {M.}~\bibnamefont
  {Shibata}}\ and\ \bibinfo {author} {\bibfnamefont {K.}~\bibnamefont
  {Ury{\={u}}}},\ }\href {\doibase 10.1088/0264-9381/24/12/s09} {\bibfield
  {journal} {\bibinfo  {journal} {Classical Quantum Gravity}\ }\textbf
  {\bibinfo {volume} {24}},\ \bibinfo {pages} {S125} (\bibinfo {year}
  {2007})}\BibitemShut {NoStop}%
\bibitem [{\citenamefont {Shibata}\ and\ \citenamefont
  {Taniguchi}(2008)}]{shibata2008apr}%
  \BibitemOpen
  \bibfield  {author} {\bibinfo {author} {\bibfnamefont {M.}~\bibnamefont
  {Shibata}}\ and\ \bibinfo {author} {\bibfnamefont {K.}~\bibnamefont
  {Taniguchi}},\ }\href {\doibase 10.1103/PhysRevD.77.084015} {\bibfield
  {journal} {\bibinfo  {journal} {Phys. Rev. D}\ }\textbf {\bibinfo {volume}
  {77}},\ \bibinfo {pages} {084015} (\bibinfo {year} {2008})}\BibitemShut
  {NoStop}%
\bibitem [{\citenamefont {Etienne}\ \emph {et~al.}(2008)\citenamefont
  {Etienne}, \citenamefont {Faber}, \citenamefont {Liu}, \citenamefont
  {Shapiro}, \citenamefont {Taniguchi},\ and\ \citenamefont
  {Baumgarte}}]{etienne2008apr}%
  \BibitemOpen
  \bibfield  {author} {\bibinfo {author} {\bibfnamefont {Z.~B.}\ \bibnamefont
  {Etienne}}, \bibinfo {author} {\bibfnamefont {J.~A.}\ \bibnamefont {Faber}},
  \bibinfo {author} {\bibfnamefont {Y.~T.}\ \bibnamefont {Liu}}, \bibinfo
  {author} {\bibfnamefont {S.~L.}\ \bibnamefont {Shapiro}}, \bibinfo {author}
  {\bibfnamefont {K.}~\bibnamefont {Taniguchi}}, \ and\ \bibinfo {author}
  {\bibfnamefont {T.~W.}\ \bibnamefont {Baumgarte}},\ }\href {\doibase
  10.1103/PhysRevD.77.084002} {\bibfield  {journal} {\bibinfo  {journal} {Phys.
  Rev. D}\ }\textbf {\bibinfo {volume} {77}},\ \bibinfo {pages} {084002}
  (\bibinfo {year} {2008})}\BibitemShut {NoStop}%
\bibitem [{\citenamefont {Duez}\ \emph {et~al.}(2008)\citenamefont {Duez},
  \citenamefont {Foucart}, \citenamefont {Kidder}, \citenamefont {Pfeiffer},
  \citenamefont {Scheel},\ and\ \citenamefont {Teukolsky}}]{duez2008nov}%
  \BibitemOpen
  \bibfield  {author} {\bibinfo {author} {\bibfnamefont {M.~D.}\ \bibnamefont
  {Duez}}, \bibinfo {author} {\bibfnamefont {F.}~\bibnamefont {Foucart}},
  \bibinfo {author} {\bibfnamefont {L.~E.}\ \bibnamefont {Kidder}}, \bibinfo
  {author} {\bibfnamefont {H.~P.}\ \bibnamefont {Pfeiffer}}, \bibinfo {author}
  {\bibfnamefont {M.~A.}\ \bibnamefont {Scheel}}, \ and\ \bibinfo {author}
  {\bibfnamefont {S.~A.}\ \bibnamefont {Teukolsky}},\ }\href {\doibase
  10.1103/PhysRevD.78.104015} {\bibfield  {journal} {\bibinfo  {journal} {Phys.
  Rev. D}\ }\textbf {\bibinfo {volume} {78}},\ \bibinfo {pages} {104015}
  (\bibinfo {year} {2008})}\BibitemShut {NoStop}%
\bibitem [{\citenamefont {Shibata}\ \emph {et~al.}(2009)\citenamefont
  {Shibata}, \citenamefont {Kyutoku}, \citenamefont {Yamamoto},\ and\
  \citenamefont {Taniguchi}}]{shibata2009feb}%
  \BibitemOpen
  \bibfield  {author} {\bibinfo {author} {\bibfnamefont {M.}~\bibnamefont
  {Shibata}}, \bibinfo {author} {\bibfnamefont {K.}~\bibnamefont {Kyutoku}},
  \bibinfo {author} {\bibfnamefont {T.}~\bibnamefont {Yamamoto}}, \ and\
  \bibinfo {author} {\bibfnamefont {K.}~\bibnamefont {Taniguchi}},\ }\href
  {\doibase 10.1103/PhysRevD.79.044030} {\bibfield  {journal} {\bibinfo
  {journal} {Phys. Rev. D}\ }\textbf {\bibinfo {volume} {79}},\ \bibinfo
  {pages} {044030} (\bibinfo {year} {2009})}\BibitemShut {NoStop}%
\bibitem [{\citenamefont {Etienne}\ \emph {et~al.}(2009)\citenamefont
  {Etienne}, \citenamefont {Liu}, \citenamefont {Shapiro},\ and\ \citenamefont
  {Baumgarte}}]{etienne2009feb}%
  \BibitemOpen
  \bibfield  {author} {\bibinfo {author} {\bibfnamefont {Z.~B.}\ \bibnamefont
  {Etienne}}, \bibinfo {author} {\bibfnamefont {Y.~T.}\ \bibnamefont {Liu}},
  \bibinfo {author} {\bibfnamefont {S.~L.}\ \bibnamefont {Shapiro}}, \ and\
  \bibinfo {author} {\bibfnamefont {T.~W.}\ \bibnamefont {Baumgarte}},\ }\href
  {\doibase 10.1103/PhysRevD.79.044024} {\bibfield  {journal} {\bibinfo
  {journal} {Phys. Rev. D}\ }\textbf {\bibinfo {volume} {79}},\ \bibinfo
  {pages} {044024} (\bibinfo {year} {2009})}\BibitemShut {NoStop}%
\bibitem [{\citenamefont {Chawla}\ \emph {et~al.}(2010)\citenamefont {Chawla},
  \citenamefont {Anderson}, \citenamefont {Besselman}, \citenamefont {Lehner},
  \citenamefont {Liebling}, \citenamefont {Motl},\ and\ \citenamefont
  {Neilsen}}]{chawla2010sep}%
  \BibitemOpen
  \bibfield  {author} {\bibinfo {author} {\bibfnamefont {S.}~\bibnamefont
  {Chawla}}, \bibinfo {author} {\bibfnamefont {M.}~\bibnamefont {Anderson}},
  \bibinfo {author} {\bibfnamefont {M.}~\bibnamefont {Besselman}}, \bibinfo
  {author} {\bibfnamefont {L.}~\bibnamefont {Lehner}}, \bibinfo {author}
  {\bibfnamefont {S.~L.}\ \bibnamefont {Liebling}}, \bibinfo {author}
  {\bibfnamefont {P.~M.}\ \bibnamefont {Motl}}, \ and\ \bibinfo {author}
  {\bibfnamefont {D.}~\bibnamefont {Neilsen}},\ }\href {\doibase
  10.1103/PhysRevLett.105.111101} {\bibfield  {journal} {\bibinfo  {journal}
  {Phys. Rev. Lett.}\ }\textbf {\bibinfo {volume} {105}},\ \bibinfo {pages}
  {111101} (\bibinfo {year} {2010})}\BibitemShut {NoStop}%
\bibitem [{\citenamefont {Duez}\ \emph {et~al.}(2010)\citenamefont {Duez},
  \citenamefont {Foucart}, \citenamefont {Kidder}, \citenamefont {Ott},\ and\
  \citenamefont {Teukolsky}}]{duez2010may}%
  \BibitemOpen
  \bibfield  {author} {\bibinfo {author} {\bibfnamefont {M.~D.}\ \bibnamefont
  {Duez}}, \bibinfo {author} {\bibfnamefont {F.}~\bibnamefont {Foucart}},
  \bibinfo {author} {\bibfnamefont {L.~E.}\ \bibnamefont {Kidder}}, \bibinfo
  {author} {\bibfnamefont {C.~D.}\ \bibnamefont {Ott}}, \ and\ \bibinfo
  {author} {\bibfnamefont {S.~A.}\ \bibnamefont {Teukolsky}},\ }\href {\doibase
  10.1088/0264-9381/27/11/114106} {\bibfield  {journal} {\bibinfo  {journal}
  {Classical Quantum Gravity}\ }\textbf {\bibinfo {volume} {27}},\ \bibinfo
  {pages} {114106} (\bibinfo {year} {2010})}\BibitemShut {NoStop}%
\bibitem [{\citenamefont {Kyutoku}\ \emph {et~al.}(2010)\citenamefont
  {Kyutoku}, \citenamefont {Shibata},\ and\ \citenamefont
  {Taniguchi}}]{kyutoku2010aug}%
  \BibitemOpen
  \bibfield  {author} {\bibinfo {author} {\bibfnamefont {K.}~\bibnamefont
  {Kyutoku}}, \bibinfo {author} {\bibfnamefont {M.}~\bibnamefont {Shibata}}, \
  and\ \bibinfo {author} {\bibfnamefont {K.}~\bibnamefont {Taniguchi}},\ }\href
  {\doibase 10.1103/PhysRevD.82.044049} {\bibfield  {journal} {\bibinfo
  {journal} {Phys. Rev. D}\ }\textbf {\bibinfo {volume} {82}},\ \bibinfo
  {pages} {044049} (\bibinfo {year} {2010})}\BibitemShut {NoStop}%
\bibitem [{\citenamefont {Kyutoku}\ \emph {et~al.}(2011)\citenamefont
  {Kyutoku}, \citenamefont {Okawa}, \citenamefont {Shibata},\ and\
  \citenamefont {Taniguchi}}]{kyutoku2011sep}%
  \BibitemOpen
  \bibfield  {author} {\bibinfo {author} {\bibfnamefont {K.}~\bibnamefont
  {Kyutoku}}, \bibinfo {author} {\bibfnamefont {H.}~\bibnamefont {Okawa}},
  \bibinfo {author} {\bibfnamefont {M.}~\bibnamefont {Shibata}}, \ and\
  \bibinfo {author} {\bibfnamefont {K.}~\bibnamefont {Taniguchi}},\ }\href
  {\doibase 10.1103/PhysRevD.84.064018} {\bibfield  {journal} {\bibinfo
  {journal} {Phys. Rev. D}\ }\textbf {\bibinfo {volume} {84}},\ \bibinfo
  {pages} {064018} (\bibinfo {year} {2011})}\BibitemShut {NoStop}%
\bibitem [{\citenamefont {Foucart}\ \emph {et~al.}(2011)\citenamefont
  {Foucart}, \citenamefont {Duez}, \citenamefont {Kidder},\ and\ \citenamefont
  {Teukolsky}}]{foucart2011jan}%
  \BibitemOpen
  \bibfield  {author} {\bibinfo {author} {\bibfnamefont {F.}~\bibnamefont
  {Foucart}}, \bibinfo {author} {\bibfnamefont {M.~D.}\ \bibnamefont {Duez}},
  \bibinfo {author} {\bibfnamefont {L.~E.}\ \bibnamefont {Kidder}}, \ and\
  \bibinfo {author} {\bibfnamefont {S.~A.}\ \bibnamefont {Teukolsky}},\ }\href
  {\doibase 10.1103/PhysRevD.83.024005} {\bibfield  {journal} {\bibinfo
  {journal} {Phys. Rev. D}\ }\textbf {\bibinfo {volume} {83}},\ \bibinfo
  {pages} {024005} (\bibinfo {year} {2011})}\BibitemShut {NoStop}%
\bibitem [{\citenamefont {Foucart}\ \emph {et~al.}(2012)\citenamefont
  {Foucart}, \citenamefont {Duez}, \citenamefont {Kidder}, \citenamefont
  {Scheel}, \citenamefont {Szilagyi},\ and\ \citenamefont
  {Teukolsky}}]{foucart2012feb}%
  \BibitemOpen
  \bibfield  {author} {\bibinfo {author} {\bibfnamefont {F.}~\bibnamefont
  {Foucart}}, \bibinfo {author} {\bibfnamefont {M.~D.}\ \bibnamefont {Duez}},
  \bibinfo {author} {\bibfnamefont {L.~E.}\ \bibnamefont {Kidder}}, \bibinfo
  {author} {\bibfnamefont {M.~A.}\ \bibnamefont {Scheel}}, \bibinfo {author}
  {\bibfnamefont {B.}~\bibnamefont {Szilagyi}}, \ and\ \bibinfo {author}
  {\bibfnamefont {S.~A.}\ \bibnamefont {Teukolsky}},\ }\href {\doibase
  10.1103/PhysRevD.85.044015} {\bibfield  {journal} {\bibinfo  {journal} {Phys.
  Rev. D}\ }\textbf {\bibinfo {volume} {85}},\ \bibinfo {pages} {044015}
  (\bibinfo {year} {2012})}\BibitemShut {NoStop}%
\bibitem [{\citenamefont {Etienne}\ \emph
  {et~al.}(2012{\natexlab{a}})\citenamefont {Etienne}, \citenamefont {Liu},
  \citenamefont {Paschalidis},\ and\ \citenamefont {Shapiro}}]{etienne2012mar}%
  \BibitemOpen
  \bibfield  {author} {\bibinfo {author} {\bibfnamefont {Z.~B.}\ \bibnamefont
  {Etienne}}, \bibinfo {author} {\bibfnamefont {Y.~T.}\ \bibnamefont {Liu}},
  \bibinfo {author} {\bibfnamefont {V.}~\bibnamefont {Paschalidis}}, \ and\
  \bibinfo {author} {\bibfnamefont {S.~L.}\ \bibnamefont {Shapiro}},\ }\href
  {\doibase 10.1103/PhysRevD.85.064029} {\bibfield  {journal} {\bibinfo
  {journal} {Phys. Rev. D}\ }\textbf {\bibinfo {volume} {85}},\ \bibinfo
  {pages} {064029} (\bibinfo {year} {2012}{\natexlab{a}})}\BibitemShut
  {NoStop}%
\bibitem [{\citenamefont {Etienne}\ \emph
  {et~al.}(2012{\natexlab{b}})\citenamefont {Etienne}, \citenamefont
  {Paschalidis},\ and\ \citenamefont {Shapiro}}]{etienne2012oct}%
  \BibitemOpen
  \bibfield  {author} {\bibinfo {author} {\bibfnamefont {Z.~B.}\ \bibnamefont
  {Etienne}}, \bibinfo {author} {\bibfnamefont {V.}~\bibnamefont
  {Paschalidis}}, \ and\ \bibinfo {author} {\bibfnamefont {S.~L.}\ \bibnamefont
  {Shapiro}},\ }\href {\doibase 10.1103/PhysRevD.86.084026} {\bibfield
  {journal} {\bibinfo  {journal} {Phys. Rev. D}\ }\textbf {\bibinfo {volume}
  {86}},\ \bibinfo {pages} {084026} (\bibinfo {year}
  {2012}{\natexlab{b}})}\BibitemShut {NoStop}%
\bibitem [{\citenamefont {Kyutoku}\ \emph {et~al.}(2013)\citenamefont
  {Kyutoku}, \citenamefont {Ioka},\ and\ \citenamefont
  {Shibata}}]{kyutoku2013aug}%
  \BibitemOpen
  \bibfield  {author} {\bibinfo {author} {\bibfnamefont {K.}~\bibnamefont
  {Kyutoku}}, \bibinfo {author} {\bibfnamefont {K.}~\bibnamefont {Ioka}}, \
  and\ \bibinfo {author} {\bibfnamefont {M.}~\bibnamefont {Shibata}},\ }\href
  {\doibase 10.1103/PhysRevD.88.041503} {\bibfield  {journal} {\bibinfo
  {journal} {Phys. Rev. D}\ }\textbf {\bibinfo {volume} {88}},\ \bibinfo
  {pages} {041503} (\bibinfo {year} {2013})}\BibitemShut {NoStop}%
\bibitem [{\citenamefont {Kyutoku}\ \emph {et~al.}(2015)\citenamefont
  {Kyutoku}, \citenamefont {Ioka}, \citenamefont {Okawa}, \citenamefont
  {Shibata},\ and\ \citenamefont {Taniguchi}}]{kyutoku2015aug}%
  \BibitemOpen
  \bibfield  {author} {\bibinfo {author} {\bibfnamefont {K.}~\bibnamefont
  {Kyutoku}}, \bibinfo {author} {\bibfnamefont {K.}~\bibnamefont {Ioka}},
  \bibinfo {author} {\bibfnamefont {H.}~\bibnamefont {Okawa}}, \bibinfo
  {author} {\bibfnamefont {M.}~\bibnamefont {Shibata}}, \ and\ \bibinfo
  {author} {\bibfnamefont {K.}~\bibnamefont {Taniguchi}},\ }\href {\doibase
  10.1103/PhysRevD.92.044028} {\bibfield  {journal} {\bibinfo  {journal} {Phys.
  Rev. D}\ }\textbf {\bibinfo {volume} {92}},\ \bibinfo {pages} {044028}
  (\bibinfo {year} {2015})}\BibitemShut {NoStop}%
\bibitem [{\citenamefont {Foucart}\ \emph {et~al.}(2013)\citenamefont
  {Foucart}, \citenamefont {Deaton}, \citenamefont {Duez}, \citenamefont
  {Kidder}, \citenamefont {MacDonald}, \citenamefont {Ott}, \citenamefont
  {Pfeiffer}, \citenamefont {Scheel}, \citenamefont {Szilagyi},\ and\
  \citenamefont {Teukolsky}}]{foucart2013apr}%
  \BibitemOpen
  \bibfield  {author} {\bibinfo {author} {\bibfnamefont {F.}~\bibnamefont
  {Foucart}}, \bibinfo {author} {\bibfnamefont {M.~B.}\ \bibnamefont {Deaton}},
  \bibinfo {author} {\bibfnamefont {M.~D.}\ \bibnamefont {Duez}}, \bibinfo
  {author} {\bibfnamefont {L.~E.}\ \bibnamefont {Kidder}}, \bibinfo {author}
  {\bibfnamefont {I.}~\bibnamefont {MacDonald}}, \bibinfo {author}
  {\bibfnamefont {C.~D.}\ \bibnamefont {Ott}}, \bibinfo {author} {\bibfnamefont
  {H.~P.}\ \bibnamefont {Pfeiffer}}, \bibinfo {author} {\bibfnamefont {M.~A.}\
  \bibnamefont {Scheel}}, \bibinfo {author} {\bibfnamefont {B.}~\bibnamefont
  {Szilagyi}}, \ and\ \bibinfo {author} {\bibfnamefont {S.~A.}\ \bibnamefont
  {Teukolsky}},\ }\href {\doibase 10.1103/PhysRevD.87.084006} {\bibfield
  {journal} {\bibinfo  {journal} {Phys. Rev. D}\ }\textbf {\bibinfo {volume}
  {87}},\ \bibinfo {pages} {084006} (\bibinfo {year} {2013})}\BibitemShut
  {NoStop}%
\bibitem [{\citenamefont {Lovelace}\ \emph {et~al.}(2013)\citenamefont
  {Lovelace}, \citenamefont {Duez}, \citenamefont {Foucart}, \citenamefont
  {Kidder}, \citenamefont {Pfeiffer}, \citenamefont {Scheel},\ and\
  \citenamefont {Szil{\'{a}}gyi}}]{lovelace2013jun}%
  \BibitemOpen
  \bibfield  {author} {\bibinfo {author} {\bibfnamefont {G.}~\bibnamefont
  {Lovelace}}, \bibinfo {author} {\bibfnamefont {M.~D.}\ \bibnamefont {Duez}},
  \bibinfo {author} {\bibfnamefont {F.}~\bibnamefont {Foucart}}, \bibinfo
  {author} {\bibfnamefont {L.~E.}\ \bibnamefont {Kidder}}, \bibinfo {author}
  {\bibfnamefont {H.~P.}\ \bibnamefont {Pfeiffer}}, \bibinfo {author}
  {\bibfnamefont {M.~A.}\ \bibnamefont {Scheel}}, \ and\ \bibinfo {author}
  {\bibfnamefont {B.}~\bibnamefont {Szil{\'{a}}gyi}},\ }\href {\doibase
  10.1088/0264-9381/30/13/135004} {\bibfield  {journal} {\bibinfo  {journal}
  {Classical Quantum Gravity}\ }\textbf {\bibinfo {volume} {30}},\ \bibinfo
  {pages} {135004} (\bibinfo {year} {2013})}\BibitemShut {NoStop}%
\bibitem [{\citenamefont {Deaton}\ \emph {et~al.}(2013)\citenamefont {Deaton},
  \citenamefont {Duez}, \citenamefont {Foucart}, \citenamefont {O'Connor},
  \citenamefont {Ott}, \citenamefont {Kidder}, \citenamefont {Muhlberger},
  \citenamefont {Scheel},\ and\ \citenamefont {Szilagyi}}]{deaton2013sep}%
  \BibitemOpen
  \bibfield  {author} {\bibinfo {author} {\bibfnamefont {M.~B.}\ \bibnamefont
  {Deaton}}, \bibinfo {author} {\bibfnamefont {M.~D.}\ \bibnamefont {Duez}},
  \bibinfo {author} {\bibfnamefont {F.}~\bibnamefont {Foucart}}, \bibinfo
  {author} {\bibfnamefont {E.}~\bibnamefont {O'Connor}}, \bibinfo {author}
  {\bibfnamefont {C.~D.}\ \bibnamefont {Ott}}, \bibinfo {author} {\bibfnamefont
  {L.~E.}\ \bibnamefont {Kidder}}, \bibinfo {author} {\bibfnamefont {C.~D.}\
  \bibnamefont {Muhlberger}}, \bibinfo {author} {\bibfnamefont {M.~A.}\
  \bibnamefont {Scheel}}, \ and\ \bibinfo {author} {\bibfnamefont
  {B.}~\bibnamefont {Szilagyi}},\ }\href {\doibase 10.1088/0004-637x/776/1/47}
  {\bibfield  {journal} {\bibinfo  {journal} {Astrophys. J.}\ }\textbf
  {\bibinfo {volume} {776}},\ \bibinfo {pages} {47} (\bibinfo {year}
  {2013})}\BibitemShut {NoStop}%
\bibitem [{\citenamefont {Foucart}\ \emph {et~al.}(2014)\citenamefont
  {Foucart}, \citenamefont {Deaton}, \citenamefont {Duez}, \citenamefont
  {O'Connor}, \citenamefont {Ott}, \citenamefont {Haas}, \citenamefont
  {Kidder}, \citenamefont {Pfeiffer}, \citenamefont {Scheel},\ and\
  \citenamefont {Szilagyi}}]{foucart2014jul}%
  \BibitemOpen
  \bibfield  {author} {\bibinfo {author} {\bibfnamefont {F.}~\bibnamefont
  {Foucart}}, \bibinfo {author} {\bibfnamefont {M.~B.}\ \bibnamefont {Deaton}},
  \bibinfo {author} {\bibfnamefont {M.~D.}\ \bibnamefont {Duez}}, \bibinfo
  {author} {\bibfnamefont {E.}~\bibnamefont {O'Connor}}, \bibinfo {author}
  {\bibfnamefont {C.~D.}\ \bibnamefont {Ott}}, \bibinfo {author} {\bibfnamefont
  {R.}~\bibnamefont {Haas}}, \bibinfo {author} {\bibfnamefont {L.~E.}\
  \bibnamefont {Kidder}}, \bibinfo {author} {\bibfnamefont {H.~P.}\
  \bibnamefont {Pfeiffer}}, \bibinfo {author} {\bibfnamefont {M.~A.}\
  \bibnamefont {Scheel}}, \ and\ \bibinfo {author} {\bibfnamefont
  {B.}~\bibnamefont {Szilagyi}},\ }\href {\doibase 10.1103/PhysRevD.90.024026}
  {\bibfield  {journal} {\bibinfo  {journal} {Phys. Rev. D}\ }\textbf {\bibinfo
  {volume} {90}},\ \bibinfo {pages} {024026} (\bibinfo {year}
  {2014})}\BibitemShut {NoStop}%
\bibitem [{\citenamefont {Paschalidis}\ \emph {et~al.}(2015)\citenamefont
  {Paschalidis}, \citenamefont {Ruiz},\ and\ \citenamefont
  {Shapiro}}]{paschalidis2015jun}%
  \BibitemOpen
  \bibfield  {author} {\bibinfo {author} {\bibfnamefont {V.}~\bibnamefont
  {Paschalidis}}, \bibinfo {author} {\bibfnamefont {M.}~\bibnamefont {Ruiz}}, \
  and\ \bibinfo {author} {\bibfnamefont {S.~L.}\ \bibnamefont {Shapiro}},\
  }\href {\doibase 10.1088/2041-8205/806/1/l14} {\bibfield  {journal} {\bibinfo
   {journal} {Astrophys. J.}\ }\textbf {\bibinfo {volume} {806}},\ \bibinfo
  {pages} {L14} (\bibinfo {year} {2015})}\BibitemShut {NoStop}%
\bibitem [{\citenamefont {Kawaguchi}\ \emph {et~al.}(2015)\citenamefont
  {Kawaguchi}, \citenamefont {Kyutoku}, \citenamefont {Nakano}, \citenamefont
  {Okawa}, \citenamefont {Shibata},\ and\ \citenamefont
  {Taniguchi}}]{kawaguchi2015jul}%
  \BibitemOpen
  \bibfield  {author} {\bibinfo {author} {\bibfnamefont {K.}~\bibnamefont
  {Kawaguchi}}, \bibinfo {author} {\bibfnamefont {K.}~\bibnamefont {Kyutoku}},
  \bibinfo {author} {\bibfnamefont {H.}~\bibnamefont {Nakano}}, \bibinfo
  {author} {\bibfnamefont {H.}~\bibnamefont {Okawa}}, \bibinfo {author}
  {\bibfnamefont {M.}~\bibnamefont {Shibata}}, \ and\ \bibinfo {author}
  {\bibfnamefont {K.}~\bibnamefont {Taniguchi}},\ }\href {\doibase
  10.1103/PhysRevD.92.024014} {\bibfield  {journal} {\bibinfo  {journal} {Phys.
  Rev. D}\ }\textbf {\bibinfo {volume} {92}},\ \bibinfo {pages} {024014}
  (\bibinfo {year} {2015})}\BibitemShut {NoStop}%
\bibitem [{\citenamefont {Kiuchi}\ \emph {et~al.}(2015)\citenamefont {Kiuchi},
  \citenamefont {Sekiguchi}, \citenamefont {Kyutoku}, \citenamefont {Shibata},
  \citenamefont {Taniguchi},\ and\ \citenamefont {Wada}}]{kiuchi2015sep}%
  \BibitemOpen
  \bibfield  {author} {\bibinfo {author} {\bibfnamefont {K.}~\bibnamefont
  {Kiuchi}}, \bibinfo {author} {\bibfnamefont {Y.}~\bibnamefont {Sekiguchi}},
  \bibinfo {author} {\bibfnamefont {K.}~\bibnamefont {Kyutoku}}, \bibinfo
  {author} {\bibfnamefont {M.}~\bibnamefont {Shibata}}, \bibinfo {author}
  {\bibfnamefont {K.}~\bibnamefont {Taniguchi}}, \ and\ \bibinfo {author}
  {\bibfnamefont {T.}~\bibnamefont {Wada}},\ }\href {\doibase
  10.1103/PhysRevD.92.064034} {\bibfield  {journal} {\bibinfo  {journal} {Phys.
  Rev. D}\ }\textbf {\bibinfo {volume} {92}},\ \bibinfo {pages} {064034}
  (\bibinfo {year} {2015})}\BibitemShut {NoStop}%
\bibitem [{\citenamefont {Foucart}\ \emph {et~al.}(2017)\citenamefont
  {Foucart}, \citenamefont {Desai}, \citenamefont {Brege}, \citenamefont
  {Duez}, \citenamefont {Kasen}, \citenamefont {Hemberger}, \citenamefont
  {Kidder}, \citenamefont {Pfeiffer},\ and\ \citenamefont
  {Scheel}}]{foucart2017jan}%
  \BibitemOpen
  \bibfield  {author} {\bibinfo {author} {\bibfnamefont {F.}~\bibnamefont
  {Foucart}}, \bibinfo {author} {\bibfnamefont {D.}~\bibnamefont {Desai}},
  \bibinfo {author} {\bibfnamefont {W.}~\bibnamefont {Brege}}, \bibinfo
  {author} {\bibfnamefont {M.~D.}\ \bibnamefont {Duez}}, \bibinfo {author}
  {\bibfnamefont {D.}~\bibnamefont {Kasen}}, \bibinfo {author} {\bibfnamefont
  {D.~A.}\ \bibnamefont {Hemberger}}, \bibinfo {author} {\bibfnamefont {L.~E.}\
  \bibnamefont {Kidder}}, \bibinfo {author} {\bibfnamefont {H.~P.}\
  \bibnamefont {Pfeiffer}}, \ and\ \bibinfo {author} {\bibfnamefont {M.~A.}\
  \bibnamefont {Scheel}},\ }\href {\doibase 10.1088/1361-6382/aa573b}
  {\bibfield  {journal} {\bibinfo  {journal} {Classical Quantum Gravity}\
  }\textbf {\bibinfo {volume} {34}},\ \bibinfo {pages} {044002} (\bibinfo
  {year} {2017})}\BibitemShut {NoStop}%
\bibitem [{\citenamefont {Kyutoku}\ \emph {et~al.}(2018)\citenamefont
  {Kyutoku}, \citenamefont {Kiuchi}, \citenamefont {Sekiguchi}, \citenamefont
  {Shibata},\ and\ \citenamefont {Taniguchi}}]{kyutoku2018jan}%
  \BibitemOpen
  \bibfield  {author} {\bibinfo {author} {\bibfnamefont {K.}~\bibnamefont
  {Kyutoku}}, \bibinfo {author} {\bibfnamefont {K.}~\bibnamefont {Kiuchi}},
  \bibinfo {author} {\bibfnamefont {Y.}~\bibnamefont {Sekiguchi}}, \bibinfo
  {author} {\bibfnamefont {M.}~\bibnamefont {Shibata}}, \ and\ \bibinfo
  {author} {\bibfnamefont {K.}~\bibnamefont {Taniguchi}},\ }\href {\doibase
  10.1103/PhysRevD.97.023009} {\bibfield  {journal} {\bibinfo  {journal} {Phys.
  Rev. D}\ }\textbf {\bibinfo {volume} {97}},\ \bibinfo {pages} {023009}
  (\bibinfo {year} {2018})}\BibitemShut {NoStop}%
\bibitem [{\citenamefont {Brege}\ \emph {et~al.}(2018)\citenamefont {Brege},
  \citenamefont {Duez}, \citenamefont {Foucart}, \citenamefont {Deaton},
  \citenamefont {Caro}, \citenamefont {Hemberger}, \citenamefont {Kidder},
  \citenamefont {O'Connor}, \citenamefont {Pfeiffer},\ and\ \citenamefont
  {Scheel}}]{brege2018sep}%
  \BibitemOpen
  \bibfield  {author} {\bibinfo {author} {\bibfnamefont {W.}~\bibnamefont
  {Brege}}, \bibinfo {author} {\bibfnamefont {M.~D.}\ \bibnamefont {Duez}},
  \bibinfo {author} {\bibfnamefont {F.}~\bibnamefont {Foucart}}, \bibinfo
  {author} {\bibfnamefont {M.~B.}\ \bibnamefont {Deaton}}, \bibinfo {author}
  {\bibfnamefont {J.}~\bibnamefont {Caro}}, \bibinfo {author} {\bibfnamefont
  {D.~A.}\ \bibnamefont {Hemberger}}, \bibinfo {author} {\bibfnamefont {L.~E.}\
  \bibnamefont {Kidder}}, \bibinfo {author} {\bibfnamefont {E.}~\bibnamefont
  {O'Connor}}, \bibinfo {author} {\bibfnamefont {H.~P.}\ \bibnamefont
  {Pfeiffer}}, \ and\ \bibinfo {author} {\bibfnamefont {M.~A.}\ \bibnamefont
  {Scheel}},\ }\href {\doibase 10.1103/PhysRevD.98.063009} {\bibfield
  {journal} {\bibinfo  {journal} {Phys. Rev. D}\ }\textbf {\bibinfo {volume}
  {98}},\ \bibinfo {pages} {063009} (\bibinfo {year} {2018})}\BibitemShut
  {NoStop}%
\bibitem [{\citenamefont {Ruiz}\ \emph {et~al.}(2018)\citenamefont {Ruiz},
  \citenamefont {Shapiro},\ and\ \citenamefont {Tsokaros}}]{ruiz2018dec}%
  \BibitemOpen
  \bibfield  {author} {\bibinfo {author} {\bibfnamefont {M.}~\bibnamefont
  {Ruiz}}, \bibinfo {author} {\bibfnamefont {S.~L.}\ \bibnamefont {Shapiro}}, \
  and\ \bibinfo {author} {\bibfnamefont {A.}~\bibnamefont {Tsokaros}},\ }\href
  {\doibase 10.1103/PhysRevD.98.123017} {\bibfield  {journal} {\bibinfo
  {journal} {Phys. Rev. D}\ }\textbf {\bibinfo {volume} {98}},\ \bibinfo
  {pages} {123017} (\bibinfo {year} {2018})}\BibitemShut {NoStop}%
\bibitem [{\citenamefont {Foucart}\ \emph
  {et~al.}(2019{\natexlab{a}})\citenamefont {Foucart}, \citenamefont {Duez},
  \citenamefont {Hinderer}, \citenamefont {Caro}, \citenamefont {Williamson},
  \citenamefont {Boyle}, \citenamefont {Buonanno}, \citenamefont {Haas},
  \citenamefont {Hemberger}, \citenamefont {Kidder}, \citenamefont {Pfeiffer},\
  and\ \citenamefont {Scheel}}]{foucart2019feb}%
  \BibitemOpen
  \bibfield  {author} {\bibinfo {author} {\bibfnamefont {F.}~\bibnamefont
  {Foucart}}, \bibinfo {author} {\bibfnamefont {M.~D.}\ \bibnamefont {Duez}},
  \bibinfo {author} {\bibfnamefont {T.}~\bibnamefont {Hinderer}}, \bibinfo
  {author} {\bibfnamefont {J.}~\bibnamefont {Caro}}, \bibinfo {author}
  {\bibfnamefont {A.~R.}\ \bibnamefont {Williamson}}, \bibinfo {author}
  {\bibfnamefont {M.}~\bibnamefont {Boyle}}, \bibinfo {author} {\bibfnamefont
  {A.}~\bibnamefont {Buonanno}}, \bibinfo {author} {\bibfnamefont
  {R.}~\bibnamefont {Haas}}, \bibinfo {author} {\bibfnamefont {D.~A.}\
  \bibnamefont {Hemberger}}, \bibinfo {author} {\bibfnamefont {L.~E.}\
  \bibnamefont {Kidder}}, \bibinfo {author} {\bibfnamefont {H.~P.}\
  \bibnamefont {Pfeiffer}}, \ and\ \bibinfo {author} {\bibfnamefont {M.~A.}\
  \bibnamefont {Scheel}},\ }\href {\doibase 10.1103/PhysRevD.99.044008}
  {\bibfield  {journal} {\bibinfo  {journal} {Phys. Rev. D}\ }\textbf {\bibinfo
  {volume} {99}},\ \bibinfo {pages} {044008} (\bibinfo {year}
  {2019}{\natexlab{a}})}\BibitemShut {NoStop}%
\bibitem [{\citenamefont {Foucart}\ \emph
  {et~al.}(2019{\natexlab{b}})\citenamefont {Foucart}, \citenamefont {Duez},
  \citenamefont {Kidder}, \citenamefont {Nissanke}, \citenamefont {Pfeiffer},\
  and\ \citenamefont {Scheel}}]{foucart2019may}%
  \BibitemOpen
  \bibfield  {author} {\bibinfo {author} {\bibfnamefont {F.}~\bibnamefont
  {Foucart}}, \bibinfo {author} {\bibfnamefont {M.~D.}\ \bibnamefont {Duez}},
  \bibinfo {author} {\bibfnamefont {L.~E.}\ \bibnamefont {Kidder}}, \bibinfo
  {author} {\bibfnamefont {S.~M.}\ \bibnamefont {Nissanke}}, \bibinfo {author}
  {\bibfnamefont {H.~P.}\ \bibnamefont {Pfeiffer}}, \ and\ \bibinfo {author}
  {\bibfnamefont {M.~A.}\ \bibnamefont {Scheel}},\ }\href {\doibase
  10.1103/PhysRevD.99.103025} {\bibfield  {journal} {\bibinfo  {journal} {Phys.
  Rev. D}\ }\textbf {\bibinfo {volume} {99}},\ \bibinfo {pages} {103025}
  (\bibinfo {year} {2019}{\natexlab{b}})}\BibitemShut {NoStop}%
\bibitem [{\citenamefont {Hinderer}\ \emph {et~al.}(2019)\citenamefont
  {Hinderer}, \citenamefont {Nissanke}, \citenamefont {Foucart}, \citenamefont
  {Hotokezaka}, \citenamefont {Vincent}, \citenamefont {Kasliwal},
  \citenamefont {Schmidt}, \citenamefont {Williamson}, \citenamefont {Nichols},
  \citenamefont {Duez}, \citenamefont {Kidder}, \citenamefont {Pfeiffer},\ and\
  \citenamefont {Scheel}}]{hinderer2019sep}%
  \BibitemOpen
  \bibfield  {author} {\bibinfo {author} {\bibfnamefont {T.}~\bibnamefont
  {Hinderer}}, \bibinfo {author} {\bibfnamefont {S.}~\bibnamefont {Nissanke}},
  \bibinfo {author} {\bibfnamefont {F.}~\bibnamefont {Foucart}}, \bibinfo
  {author} {\bibfnamefont {K.}~\bibnamefont {Hotokezaka}}, \bibinfo {author}
  {\bibfnamefont {T.}~\bibnamefont {Vincent}}, \bibinfo {author} {\bibfnamefont
  {M.}~\bibnamefont {Kasliwal}}, \bibinfo {author} {\bibfnamefont
  {P.}~\bibnamefont {Schmidt}}, \bibinfo {author} {\bibfnamefont {A.~R.}\
  \bibnamefont {Williamson}}, \bibinfo {author} {\bibfnamefont {D.~A.}\
  \bibnamefont {Nichols}}, \bibinfo {author} {\bibfnamefont {M.~D.}\
  \bibnamefont {Duez}}, \bibinfo {author} {\bibfnamefont {L.~E.}\ \bibnamefont
  {Kidder}}, \bibinfo {author} {\bibfnamefont {H.~P.}\ \bibnamefont
  {Pfeiffer}}, \ and\ \bibinfo {author} {\bibfnamefont {M.~A.}\ \bibnamefont
  {Scheel}},\ }\href {\doibase 10.1103/PhysRevD.100.063021} {\bibfield
  {journal} {\bibinfo  {journal} {Phys. Rev. D}\ }\textbf {\bibinfo {volume}
  {100}},\ \bibinfo {pages} {063021} (\bibinfo {year} {2019})}\BibitemShut
  {NoStop}%
\bibitem [{\citenamefont {Hayashi}\ \emph {et~al.}(2021)\citenamefont
  {Hayashi}, \citenamefont {Kawaguchi}, \citenamefont {Kiuchi}, \citenamefont
  {Kyutoku},\ and\ \citenamefont {Shibata}}]{hayashi2021feb}%
  \BibitemOpen
  \bibfield  {author} {\bibinfo {author} {\bibfnamefont {K.}~\bibnamefont
  {Hayashi}}, \bibinfo {author} {\bibfnamefont {K.}~\bibnamefont {Kawaguchi}},
  \bibinfo {author} {\bibfnamefont {K.}~\bibnamefont {Kiuchi}}, \bibinfo
  {author} {\bibfnamefont {K.}~\bibnamefont {Kyutoku}}, \ and\ \bibinfo
  {author} {\bibfnamefont {M.}~\bibnamefont {Shibata}},\ }\href {\doibase
  10.1103/PhysRevD.103.043007} {\bibfield  {journal} {\bibinfo  {journal}
  {Phys. Rev. D}\ }\textbf {\bibinfo {volume} {103}},\ \bibinfo {pages}
  {043007} (\bibinfo {year} {2021})}\BibitemShut {NoStop}%
\bibitem [{\citenamefont {Foucart}\ \emph {et~al.}(2021)\citenamefont
  {Foucart}, \citenamefont {Chernoglazov}, \citenamefont {Boyle}, \citenamefont
  {Hinderer}, \citenamefont {Miller}, \citenamefont {Moxon}, \citenamefont
  {Scheel}, \citenamefont {Deppe}, \citenamefont {Duez}, \citenamefont
  {H\'ebert}, \citenamefont {Kidder}, \citenamefont {Throwe},\ and\
  \citenamefont {Pfeiffer}}]{foucart2021mar}%
  \BibitemOpen
  \bibfield  {author} {\bibinfo {author} {\bibfnamefont {F.}~\bibnamefont
  {Foucart}}, \bibinfo {author} {\bibfnamefont {A.}~\bibnamefont
  {Chernoglazov}}, \bibinfo {author} {\bibfnamefont {M.}~\bibnamefont {Boyle}},
  \bibinfo {author} {\bibfnamefont {T.}~\bibnamefont {Hinderer}}, \bibinfo
  {author} {\bibfnamefont {M.}~\bibnamefont {Miller}}, \bibinfo {author}
  {\bibfnamefont {J.}~\bibnamefont {Moxon}}, \bibinfo {author} {\bibfnamefont
  {M.~A.}\ \bibnamefont {Scheel}}, \bibinfo {author} {\bibfnamefont
  {N.}~\bibnamefont {Deppe}}, \bibinfo {author} {\bibfnamefont {M.~D.}\
  \bibnamefont {Duez}}, \bibinfo {author} {\bibfnamefont {F.}~\bibnamefont
  {H\'ebert}}, \bibinfo {author} {\bibfnamefont {L.~E.}\ \bibnamefont
  {Kidder}}, \bibinfo {author} {\bibfnamefont {W.}~\bibnamefont {Throwe}}, \
  and\ \bibinfo {author} {\bibfnamefont {H.~P.}\ \bibnamefont {Pfeiffer}},\
  }\href {\doibase 10.1103/PhysRevD.103.064007} {\bibfield  {journal} {\bibinfo
   {journal} {Phys. Rev. D}\ }\textbf {\bibinfo {volume} {103}},\ \bibinfo
  {pages} {064007} (\bibinfo {year} {2021})}\BibitemShut {NoStop}%
\bibitem [{\citenamefont {Chaurasia}\ \emph {et~al.}(2021)\citenamefont
  {Chaurasia}, \citenamefont {Dietrich},\ and\ \citenamefont
  {Rosswog}}]{chaurasia2021oct}%
  \BibitemOpen
  \bibfield  {author} {\bibinfo {author} {\bibfnamefont {S.~V.}\ \bibnamefont
  {Chaurasia}}, \bibinfo {author} {\bibfnamefont {T.}~\bibnamefont {Dietrich}},
  \ and\ \bibinfo {author} {\bibfnamefont {S.}~\bibnamefont {Rosswog}},\ }\href
  {\doibase 10.1103/PhysRevD.104.084010} {\bibfield  {journal} {\bibinfo
  {journal} {Phys. Rev. D}\ }\textbf {\bibinfo {volume} {104}},\ \bibinfo
  {pages} {084010} (\bibinfo {year} {2021})}\BibitemShut {NoStop}%
\bibitem [{\citenamefont {Most}\ \emph {et~al.}(2021)\citenamefont {Most},
  \citenamefont {Papenfort}, \citenamefont {Tootle},\ and\ \citenamefont
  {Rezzolla}}]{most2021jul}%
  \BibitemOpen
  \bibfield  {author} {\bibinfo {author} {\bibfnamefont {E.~R.}\ \bibnamefont
  {Most}}, \bibinfo {author} {\bibfnamefont {L.~J.}\ \bibnamefont {Papenfort}},
  \bibinfo {author} {\bibfnamefont {S.~D.}\ \bibnamefont {Tootle}}, \ and\
  \bibinfo {author} {\bibfnamefont {L.}~\bibnamefont {Rezzolla}},\ }\href
  {\doibase 10.1093/mnras/stab1824} {\bibfield  {journal} {\bibinfo  {journal}
  {Mon. Not. R. Astron. Soc.}\ }\textbf {\bibinfo {volume} {506}},\ \bibinfo
  {pages} {3511} (\bibinfo {year} {2021})}\BibitemShut {NoStop}%
\bibitem [{\citenamefont {Fern{\'{a}}ndez}\ and\ \citenamefont
  {Metzger}(2013)}]{fernamdez2013aug}%
  \BibitemOpen
  \bibfield  {author} {\bibinfo {author} {\bibfnamefont {R.}~\bibnamefont
  {Fern{\'{a}}ndez}}\ and\ \bibinfo {author} {\bibfnamefont {B.~D.}\
  \bibnamefont {Metzger}},\ }\href {\doibase 10.1093/mnras/stt1312} {\bibfield
  {journal} {\bibinfo  {journal} {Mon. Not. R. Astron. Soc.}\ }\textbf
  {\bibinfo {volume} {435}},\ \bibinfo {pages} {502} (\bibinfo {year}
  {2013})}\BibitemShut {NoStop}%
\bibitem [{\citenamefont {Metzger}\ and\ \citenamefont
  {Fern{\'{a}}ndez}(2014)}]{metzger2014may}%
  \BibitemOpen
  \bibfield  {author} {\bibinfo {author} {\bibfnamefont {B.~D.}\ \bibnamefont
  {Metzger}}\ and\ \bibinfo {author} {\bibfnamefont {R.}~\bibnamefont
  {Fern{\'{a}}ndez}},\ }\href {\doibase 10.1093/mnras/stu802} {\bibfield
  {journal} {\bibinfo  {journal} {Mon. Not. R. Astron. Soc.}\ }\textbf
  {\bibinfo {volume} {441}},\ \bibinfo {pages} {3444} (\bibinfo {year}
  {2014})}\BibitemShut {NoStop}%
\bibitem [{\citenamefont {Just}\ \emph {et~al.}(2015)\citenamefont {Just},
  \citenamefont {Bauswein}, \citenamefont {Pulpillo}, \citenamefont {Goriely},\
  and\ \citenamefont {Janka}}]{just2015feb}%
  \BibitemOpen
  \bibfield  {author} {\bibinfo {author} {\bibfnamefont {O.}~\bibnamefont
  {Just}}, \bibinfo {author} {\bibfnamefont {A.}~\bibnamefont {Bauswein}},
  \bibinfo {author} {\bibfnamefont {R.~A.}\ \bibnamefont {Pulpillo}}, \bibinfo
  {author} {\bibfnamefont {S.}~\bibnamefont {Goriely}}, \ and\ \bibinfo
  {author} {\bibfnamefont {H.-T.}\ \bibnamefont {Janka}},\ }\href {\doibase
  10.1093/mnras/stv009} {\bibfield  {journal} {\bibinfo  {journal} {Mon. Not.
  R. Astron. Soc.}\ }\textbf {\bibinfo {volume} {448}},\ \bibinfo {pages} {541}
  (\bibinfo {year} {2015})}\BibitemShut {NoStop}%
\bibitem [{\citenamefont {Fern{\'{a}}ndez}\ \emph {et~al.}(2015)\citenamefont
  {Fern{\'{a}}ndez}, \citenamefont {Quataert}, \citenamefont {Schwab},
  \citenamefont {Kasen},\ and\ \citenamefont {Rosswog}}]{fernandez2015mar}%
  \BibitemOpen
  \bibfield  {author} {\bibinfo {author} {\bibfnamefont {R.}~\bibnamefont
  {Fern{\'{a}}ndez}}, \bibinfo {author} {\bibfnamefont {E.}~\bibnamefont
  {Quataert}}, \bibinfo {author} {\bibfnamefont {J.}~\bibnamefont {Schwab}},
  \bibinfo {author} {\bibfnamefont {D.}~\bibnamefont {Kasen}}, \ and\ \bibinfo
  {author} {\bibfnamefont {S.}~\bibnamefont {Rosswog}},\ }\href {\doibase
  10.1093/mnras/stv238} {\bibfield  {journal} {\bibinfo  {journal} {Mon. Not.
  R. Astron. Soc.}\ }\textbf {\bibinfo {volume} {449}},\ \bibinfo {pages} {390}
  (\bibinfo {year} {2015})}\BibitemShut {NoStop}%
\bibitem [{\citenamefont {Fern{\'{a}}ndez}\ \emph {et~al.}(2017)\citenamefont
  {Fern{\'{a}}ndez}, \citenamefont {Foucart}, \citenamefont {Kasen},
  \citenamefont {Lippuner}, \citenamefont {Desai},\ and\ \citenamefont
  {Roberts}}]{fernandez2017jul}%
  \BibitemOpen
  \bibfield  {author} {\bibinfo {author} {\bibfnamefont {R.}~\bibnamefont
  {Fern{\'{a}}ndez}}, \bibinfo {author} {\bibfnamefont {F.}~\bibnamefont
  {Foucart}}, \bibinfo {author} {\bibfnamefont {D.}~\bibnamefont {Kasen}},
  \bibinfo {author} {\bibfnamefont {J.}~\bibnamefont {Lippuner}}, \bibinfo
  {author} {\bibfnamefont {D.}~\bibnamefont {Desai}}, \ and\ \bibinfo {author}
  {\bibfnamefont {L.~F.}\ \bibnamefont {Roberts}},\ }\href {\doibase
  10.1088/1361-6382/aa7a77} {\bibfield  {journal} {\bibinfo  {journal}
  {Classical Quantum Gravity}\ }\textbf {\bibinfo {volume} {34}},\ \bibinfo
  {pages} {154001} (\bibinfo {year} {2017})}\BibitemShut {NoStop}%
\bibitem [{\citenamefont {Siegel}\ and\ \citenamefont
  {Metzger}(2018)}]{daniel2018may}%
  \BibitemOpen
  \bibfield  {author} {\bibinfo {author} {\bibfnamefont {D.~M.}\ \bibnamefont
  {Siegel}}\ and\ \bibinfo {author} {\bibfnamefont {B.~D.}\ \bibnamefont
  {Metzger}},\ }\href {\doibase 10.3847/1538-4357/aabaec} {\bibfield  {journal}
  {\bibinfo  {journal} {\apj}\ }\textbf {\bibinfo {volume} {858}},\ \bibinfo
  {pages} {52} (\bibinfo {year} {2018})}\BibitemShut {NoStop}%
\bibitem [{\citenamefont {Fern{\'{a}}ndez}\ \emph {et~al.}(2019)\citenamefont
  {Fern{\'{a}}ndez}, \citenamefont {Tchekhovskoy}, \citenamefont {Quataert},
  \citenamefont {Foucart},\ and\ \citenamefont {Kasen}}]{fernandez2018oct}%
  \BibitemOpen
  \bibfield  {author} {\bibinfo {author} {\bibfnamefont {R.}~\bibnamefont
  {Fern{\'{a}}ndez}}, \bibinfo {author} {\bibfnamefont {A.}~\bibnamefont
  {Tchekhovskoy}}, \bibinfo {author} {\bibfnamefont {E.}~\bibnamefont
  {Quataert}}, \bibinfo {author} {\bibfnamefont {F.}~\bibnamefont {Foucart}}, \
  and\ \bibinfo {author} {\bibfnamefont {D.}~\bibnamefont {Kasen}},\ }\href
  {\doibase 10.1093/mnras/sty2932} {\bibfield  {journal} {\bibinfo  {journal}
  {Mon. Not. R. Astron. Soc.}\ }\textbf {\bibinfo {volume} {482}},\ \bibinfo
  {pages} {3373} (\bibinfo {year} {2019})}\BibitemShut {NoStop}%
\bibitem [{\citenamefont {Janiuk}(2019)}]{agnieszka2019sep}%
  \BibitemOpen
  \bibfield  {author} {\bibinfo {author} {\bibfnamefont {A.}~\bibnamefont
  {Janiuk}},\ }\href {\doibase 10.3847/1538-4357/ab3349} {\bibfield  {journal}
  {\bibinfo  {journal} {\apj}\ }\textbf {\bibinfo {volume} {882}},\ \bibinfo
  {pages} {163} (\bibinfo {year} {2019})}\BibitemShut {NoStop}%
\bibitem [{\citenamefont {Christie}\ \emph {et~al.}(2019)\citenamefont
  {Christie}, \citenamefont {Lalakos}, \citenamefont {Tchekhovskoy},
  \citenamefont {Fern{\'{a}}ndez}, \citenamefont {Foucart}, \citenamefont
  {Quataert},\ and\ \citenamefont {Kasen}}]{christie2019sep}%
  \BibitemOpen
  \bibfield  {author} {\bibinfo {author} {\bibfnamefont {I.~M.}\ \bibnamefont
  {Christie}}, \bibinfo {author} {\bibfnamefont {A.}~\bibnamefont {Lalakos}},
  \bibinfo {author} {\bibfnamefont {A.}~\bibnamefont {Tchekhovskoy}}, \bibinfo
  {author} {\bibfnamefont {R.}~\bibnamefont {Fern{\'{a}}ndez}}, \bibinfo
  {author} {\bibfnamefont {F.}~\bibnamefont {Foucart}}, \bibinfo {author}
  {\bibfnamefont {E.}~\bibnamefont {Quataert}}, \ and\ \bibinfo {author}
  {\bibfnamefont {D.}~\bibnamefont {Kasen}},\ }\href {\doibase
  10.1093/mnras/stz2552} {\bibfield  {journal} {\bibinfo  {journal} {Mon. Not.
  R. Astron. Soc.}\ }\textbf {\bibinfo {volume} {490}},\ \bibinfo {pages}
  {4811} (\bibinfo {year} {2019})}\BibitemShut {NoStop}%
\bibitem [{\citenamefont {Miller}\ \emph {et~al.}(2019)\citenamefont {Miller},
  \citenamefont {Ryan}, \citenamefont {Dolence}, \citenamefont {Burrows},
  \citenamefont {Fontes}, \citenamefont {Fryer}, \citenamefont {Korobkin},
  \citenamefont {Lippuner}, \citenamefont {Mumpower},\ and\ \citenamefont
  {Wollaeger}}]{miller2019jul}%
  \BibitemOpen
  \bibfield  {author} {\bibinfo {author} {\bibfnamefont {J.~M.}\ \bibnamefont
  {Miller}}, \bibinfo {author} {\bibfnamefont {B.~R.}\ \bibnamefont {Ryan}},
  \bibinfo {author} {\bibfnamefont {J.~C.}\ \bibnamefont {Dolence}}, \bibinfo
  {author} {\bibfnamefont {A.}~\bibnamefont {Burrows}}, \bibinfo {author}
  {\bibfnamefont {C.~J.}\ \bibnamefont {Fontes}}, \bibinfo {author}
  {\bibfnamefont {C.~L.}\ \bibnamefont {Fryer}}, \bibinfo {author}
  {\bibfnamefont {O.}~\bibnamefont {Korobkin}}, \bibinfo {author}
  {\bibfnamefont {J.}~\bibnamefont {Lippuner}}, \bibinfo {author}
  {\bibfnamefont {M.~R.}\ \bibnamefont {Mumpower}}, \ and\ \bibinfo {author}
  {\bibfnamefont {R.~T.}\ \bibnamefont {Wollaeger}},\ }\href {\doibase
  10.1103/PhysRevD.100.023008} {\bibfield  {journal} {\bibinfo  {journal}
  {Phys. Rev. D}\ }\textbf {\bibinfo {volume} {100}},\ \bibinfo {pages}
  {023008} (\bibinfo {year} {2019})}\BibitemShut {NoStop}%
\bibitem [{\citenamefont {Fujibayashi}\ \emph
  {et~al.}(2020{\natexlab{a}})\citenamefont {Fujibayashi}, \citenamefont
  {Shibata}, \citenamefont {Wanajo}, \citenamefont {Kiuchi}, \citenamefont
  {Kyutoku},\ and\ \citenamefont {Sekiguchi}}]{fujibayashi2020apr}%
  \BibitemOpen
  \bibfield  {author} {\bibinfo {author} {\bibfnamefont {S.}~\bibnamefont
  {Fujibayashi}}, \bibinfo {author} {\bibfnamefont {M.}~\bibnamefont
  {Shibata}}, \bibinfo {author} {\bibfnamefont {S.}~\bibnamefont {Wanajo}},
  \bibinfo {author} {\bibfnamefont {K.}~\bibnamefont {Kiuchi}}, \bibinfo
  {author} {\bibfnamefont {K.}~\bibnamefont {Kyutoku}}, \ and\ \bibinfo
  {author} {\bibfnamefont {Y.}~\bibnamefont {Sekiguchi}},\ }\href {\doibase
  10.1103/PhysRevD.101.083029} {\bibfield  {journal} {\bibinfo  {journal}
  {Phys. Rev. D}\ }\textbf {\bibinfo {volume} {101}},\ \bibinfo {pages}
  {083029} (\bibinfo {year} {2020}{\natexlab{a}})}\BibitemShut {NoStop}%
\bibitem [{\citenamefont {Fujibayashi}\ \emph
  {et~al.}(2020{\natexlab{b}})\citenamefont {Fujibayashi}, \citenamefont
  {Shibata}, \citenamefont {Wanajo}, \citenamefont {Kiuchi}, \citenamefont
  {Kyutoku},\ and\ \citenamefont {Sekiguchi}}]{fujibayashi2020dec}%
  \BibitemOpen
  \bibfield  {author} {\bibinfo {author} {\bibfnamefont {S.}~\bibnamefont
  {Fujibayashi}}, \bibinfo {author} {\bibfnamefont {M.}~\bibnamefont
  {Shibata}}, \bibinfo {author} {\bibfnamefont {S.}~\bibnamefont {Wanajo}},
  \bibinfo {author} {\bibfnamefont {K.}~\bibnamefont {Kiuchi}}, \bibinfo
  {author} {\bibfnamefont {K.}~\bibnamefont {Kyutoku}}, \ and\ \bibinfo
  {author} {\bibfnamefont {Y.}~\bibnamefont {Sekiguchi}},\ }\href {\doibase
  10.1103/PhysRevD.102.123014} {\bibfield  {journal} {\bibinfo  {journal}
  {Phys. Rev. D}\ }\textbf {\bibinfo {volume} {102}},\ \bibinfo {pages}
  {123014} (\bibinfo {year} {2020}{\natexlab{b}})}\BibitemShut {NoStop}%
\bibitem [{\citenamefont {Li}\ and\ \citenamefont {Siegel}(2021)}]{Li2021jun}%
  \BibitemOpen
  \bibfield  {author} {\bibinfo {author} {\bibfnamefont {X.}~\bibnamefont
  {Li}}\ and\ \bibinfo {author} {\bibfnamefont {D.~M.}\ \bibnamefont
  {Siegel}},\ }\href {\doibase 10.1103/PhysRevLett.126.251101} {\bibfield
  {journal} {\bibinfo  {journal} {Phys. Rev. Lett.}\ }\textbf {\bibinfo
  {volume} {126}},\ \bibinfo {pages} {251101} (\bibinfo {year}
  {2021})}\BibitemShut {NoStop}%
\bibitem [{\citenamefont {Fern{\'{a}}ndez}\ \emph {et~al.}(2020)\citenamefont
  {Fern{\'{a}}ndez}, \citenamefont {Foucart},\ and\ \citenamefont
  {Lippuner}}]{fernandez2020jul}%
  \BibitemOpen
  \bibfield  {author} {\bibinfo {author} {\bibfnamefont {R.}~\bibnamefont
  {Fern{\'{a}}ndez}}, \bibinfo {author} {\bibfnamefont {F.}~\bibnamefont
  {Foucart}}, \ and\ \bibinfo {author} {\bibfnamefont {J.}~\bibnamefont
  {Lippuner}},\ }\href {\doibase 10.1093/mnras/staa2209} {\bibfield  {journal}
  {\bibinfo  {journal} {Mon. Not. R. Astron. Soc.}\ }\textbf {\bibinfo {volume}
  {497}},\ \bibinfo {pages} {3221} (\bibinfo {year} {2020})}\BibitemShut
  {NoStop}%
\bibitem [{\citenamefont {{Just}}\ \emph {et~al.}(2022)\citenamefont {{Just}},
  \citenamefont {{Goriely}}, \citenamefont {{Janka}}, \citenamefont
  {{Nagataki}},\ and\ \citenamefont {{Bauswein}}}]{just2022jan}%
  \BibitemOpen
  \bibfield  {author} {\bibinfo {author} {\bibfnamefont {O.}~\bibnamefont
  {{Just}}}, \bibinfo {author} {\bibfnamefont {S.}~\bibnamefont {{Goriely}}},
  \bibinfo {author} {\bibfnamefont {H.~T.}\ \bibnamefont {{Janka}}}, \bibinfo
  {author} {\bibfnamefont {S.}~\bibnamefont {{Nagataki}}}, \ and\ \bibinfo
  {author} {\bibfnamefont {A.}~\bibnamefont {{Bauswein}}},\ }\href {\doibase
  10.1093/mnras/stab2861} {\bibfield  {journal} {\bibinfo  {journal} {Mon. Not.
  R. Astron. Soc.}\ }\textbf {\bibinfo {volume} {509}},\ \bibinfo {pages}
  {1377} (\bibinfo {year} {2022})}\BibitemShut {NoStop}%
\bibitem [{\citenamefont {Shibata}\ \emph {et~al.}(2021)\citenamefont
  {Shibata}, \citenamefont {Fujibayashi},\ and\ \citenamefont
  {Sekiguchi}}]{shibata2021sep}%
  \BibitemOpen
  \bibfield  {author} {\bibinfo {author} {\bibfnamefont {M.}~\bibnamefont
  {Shibata}}, \bibinfo {author} {\bibfnamefont {S.}~\bibnamefont
  {Fujibayashi}}, \ and\ \bibinfo {author} {\bibfnamefont {Y.}~\bibnamefont
  {Sekiguchi}},\ }\href {\doibase 10.1103/PhysRevD.104.063026} {\bibfield
  {journal} {\bibinfo  {journal} {Phys. Rev. D}\ }\textbf {\bibinfo {volume}
  {104}},\ \bibinfo {pages} {063026} (\bibinfo {year} {2021})}\BibitemShut
  {NoStop}%
\bibitem [{\citenamefont {Shibata}\ and\ \citenamefont
  {Nakamura}(1995)}]{shibata1995nov}%
  \BibitemOpen
  \bibfield  {author} {\bibinfo {author} {\bibfnamefont {M.}~\bibnamefont
  {Shibata}}\ and\ \bibinfo {author} {\bibfnamefont {T.}~\bibnamefont
  {Nakamura}},\ }\href {\doibase 10.1103/PhysRevD.52.5428} {\bibfield
  {journal} {\bibinfo  {journal} {Phys. Rev. D}\ }\textbf {\bibinfo {volume}
  {52}},\ \bibinfo {pages} {5428} (\bibinfo {year} {1995})}\BibitemShut
  {NoStop}%
\bibitem [{\citenamefont {Baumgarte}\ and\ \citenamefont
  {Shapiro}(1998)}]{baumgarte1998dec}%
  \BibitemOpen
  \bibfield  {author} {\bibinfo {author} {\bibfnamefont {T.~W.}\ \bibnamefont
  {Baumgarte}}\ and\ \bibinfo {author} {\bibfnamefont {S.~L.}\ \bibnamefont
  {Shapiro}},\ }\href {\doibase 10.1103/PhysRevD.59.024007} {\bibfield
  {journal} {\bibinfo  {journal} {Phys. Rev. D}\ }\textbf {\bibinfo {volume}
  {59}},\ \bibinfo {pages} {024007} (\bibinfo {year} {1998})}\BibitemShut
  {NoStop}%
\bibitem [{\citenamefont {Campanelli}\ \emph {et~al.}(2006)\citenamefont
  {Campanelli}, \citenamefont {Lousto}, \citenamefont {Marronetti},\ and\
  \citenamefont {Zlochower}}]{campanelli2006mar}%
  \BibitemOpen
  \bibfield  {author} {\bibinfo {author} {\bibfnamefont {M.}~\bibnamefont
  {Campanelli}}, \bibinfo {author} {\bibfnamefont {C.~O.}\ \bibnamefont
  {Lousto}}, \bibinfo {author} {\bibfnamefont {P.}~\bibnamefont {Marronetti}},
  \ and\ \bibinfo {author} {\bibfnamefont {Y.}~\bibnamefont {Zlochower}},\
  }\href {\doibase 10.1103/PhysRevLett.96.111101} {\bibfield  {journal}
  {\bibinfo  {journal} {Phys. Rev. Lett.}\ }\textbf {\bibinfo {volume} {96}},\
  \bibinfo {pages} {111101} (\bibinfo {year} {2006})}\BibitemShut {NoStop}%
\bibitem [{\citenamefont {Baker}\ \emph {et~al.}(2006)\citenamefont {Baker},
  \citenamefont {Centrella}, \citenamefont {Choi}, \citenamefont {Koppitz},\
  and\ \citenamefont {van Meter}}]{baker2006mar}%
  \BibitemOpen
  \bibfield  {author} {\bibinfo {author} {\bibfnamefont {J.~G.}\ \bibnamefont
  {Baker}}, \bibinfo {author} {\bibfnamefont {J.}~\bibnamefont {Centrella}},
  \bibinfo {author} {\bibfnamefont {D.-I.}\ \bibnamefont {Choi}}, \bibinfo
  {author} {\bibfnamefont {M.}~\bibnamefont {Koppitz}}, \ and\ \bibinfo
  {author} {\bibfnamefont {J.}~\bibnamefont {van Meter}},\ }\href {\doibase
  10.1103/PhysRevLett.96.111102} {\bibfield  {journal} {\bibinfo  {journal}
  {Phys. Rev. Lett.}\ }\textbf {\bibinfo {volume} {96}},\ \bibinfo {pages}
  {111102} (\bibinfo {year} {2006})}\BibitemShut {NoStop}%
\bibitem [{\citenamefont {Marronetti}\ \emph {et~al.}(2008)\citenamefont
  {Marronetti}, \citenamefont {Tichy}, \citenamefont {Br\"ugmann},
  \citenamefont {Gonz\'alez},\ and\ \citenamefont
  {Sperhake}}]{marronetti2008mar}%
  \BibitemOpen
  \bibfield  {author} {\bibinfo {author} {\bibfnamefont {P.}~\bibnamefont
  {Marronetti}}, \bibinfo {author} {\bibfnamefont {W.}~\bibnamefont {Tichy}},
  \bibinfo {author} {\bibfnamefont {B.}~\bibnamefont {Br\"ugmann}}, \bibinfo
  {author} {\bibfnamefont {J.}~\bibnamefont {Gonz\'alez}}, \ and\ \bibinfo
  {author} {\bibfnamefont {U.}~\bibnamefont {Sperhake}},\ }\href {\doibase
  10.1103/PhysRevD.77.064010} {\bibfield  {journal} {\bibinfo  {journal} {Phys.
  Rev. D}\ }\textbf {\bibinfo {volume} {77}},\ \bibinfo {pages} {064010}
  (\bibinfo {year} {2008})}\BibitemShut {NoStop}%
\bibitem [{\citenamefont {Hilditch}\ \emph {et~al.}(2013)\citenamefont
  {Hilditch}, \citenamefont {Bernuzzi}, \citenamefont {Thierfelder},
  \citenamefont {Cao}, \citenamefont {Tichy},\ and\ \citenamefont
  {Br\"ugmann}}]{hilditch2013oct}%
  \BibitemOpen
  \bibfield  {author} {\bibinfo {author} {\bibfnamefont {D.}~\bibnamefont
  {Hilditch}}, \bibinfo {author} {\bibfnamefont {S.}~\bibnamefont {Bernuzzi}},
  \bibinfo {author} {\bibfnamefont {M.}~\bibnamefont {Thierfelder}}, \bibinfo
  {author} {\bibfnamefont {Z.}~\bibnamefont {Cao}}, \bibinfo {author}
  {\bibfnamefont {W.}~\bibnamefont {Tichy}}, \ and\ \bibinfo {author}
  {\bibfnamefont {B.}~\bibnamefont {Br\"ugmann}},\ }\href {\doibase
  10.1103/PhysRevD.88.084057} {\bibfield  {journal} {\bibinfo  {journal} {Phys.
  Rev. D}\ }\textbf {\bibinfo {volume} {88}},\ \bibinfo {pages} {084057}
  (\bibinfo {year} {2013})}\BibitemShut {NoStop}%
\bibitem [{\citenamefont {Shibata}\ and\ \citenamefont
  {Sekiguchi}(2005)}]{shibata2005aug}%
  \BibitemOpen
  \bibfield  {author} {\bibinfo {author} {\bibfnamefont {M.}~\bibnamefont
  {Shibata}}\ and\ \bibinfo {author} {\bibfnamefont {Y.}~\bibnamefont
  {Sekiguchi}},\ }\href {\doibase 10.1103/PhysRevD.72.044014} {\bibfield
  {journal} {\bibinfo  {journal} {Phys. Rev. D}\ }\textbf {\bibinfo {volume}
  {72}},\ \bibinfo {pages} {044014} (\bibinfo {year} {2005})}\BibitemShut
  {NoStop}%
\bibitem [{\citenamefont {Shibata}\ \emph {et~al.}(2007)\citenamefont
  {Shibata}, \citenamefont {Sekiguchi},\ and\ \citenamefont
  {Takahashi}}]{shibata2007aug}%
  \BibitemOpen
  \bibfield  {author} {\bibinfo {author} {\bibfnamefont {M.}~\bibnamefont
  {Shibata}}, \bibinfo {author} {\bibfnamefont {Y.}~\bibnamefont {Sekiguchi}},
  \ and\ \bibinfo {author} {\bibfnamefont {R.}~\bibnamefont {Takahashi}},\
  }\href {\doibase 10.1143/PTP.118.257} {\bibfield  {journal} {\bibinfo
  {journal} {Prog. Theor. Phys.}\ }\textbf {\bibinfo {volume} {118}},\ \bibinfo
  {pages} {257} (\bibinfo {year} {2007})}\BibitemShut {NoStop}%
\bibitem [{\citenamefont {Kiuchi}\ \emph {et~al.}(2012)\citenamefont {Kiuchi},
  \citenamefont {Kyutoku},\ and\ \citenamefont {Shibata}}]{kiuchi2012sep}%
  \BibitemOpen
  \bibfield  {author} {\bibinfo {author} {\bibfnamefont {K.}~\bibnamefont
  {Kiuchi}}, \bibinfo {author} {\bibfnamefont {K.}~\bibnamefont {Kyutoku}}, \
  and\ \bibinfo {author} {\bibfnamefont {M.}~\bibnamefont {Shibata}},\ }\href
  {\doibase 10.1103/PhysRevD.86.064008} {\bibfield  {journal} {\bibinfo
  {journal} {Phys. Rev. D}\ }\textbf {\bibinfo {volume} {86}},\ \bibinfo
  {pages} {064008} (\bibinfo {year} {2012})}\BibitemShut {NoStop}%
\bibitem [{\citenamefont {{Evans}}\ and\ \citenamefont
  {{Hawley}}(1988)}]{evans1988sep}%
  \BibitemOpen
  \bibfield  {author} {\bibinfo {author} {\bibfnamefont {C.~R.}\ \bibnamefont
  {{Evans}}}\ and\ \bibinfo {author} {\bibfnamefont {J.~F.}\ \bibnamefont
  {{Hawley}}},\ }\href {\doibase 10.1086/166684} {\bibfield  {journal}
  {\bibinfo  {journal} {\apj}\ }\textbf {\bibinfo {volume} {332}},\ \bibinfo
  {pages} {659} (\bibinfo {year} {1988})}\BibitemShut {NoStop}%
\bibitem [{\citenamefont {Balsara}(2009)}]{balsara2009}%
  \BibitemOpen
  \bibfield  {author} {\bibinfo {author} {\bibfnamefont {D.~S.}\ \bibnamefont
  {Balsara}},\ }\href {\doibase https://doi.org/10.1016/j.jcp.2009.03.038}
  {\bibfield  {journal} {\bibinfo  {journal} {J. Comput. Phys.}\ }\textbf
  {\bibinfo {volume} {228}},\ \bibinfo {pages} {5040} (\bibinfo {year}
  {2009})}\BibitemShut {NoStop}%
\bibitem [{\citenamefont {Sekiguchi}\ \emph {et~al.}(2012)\citenamefont
  {Sekiguchi}, \citenamefont {Kiuchi}, \citenamefont {Kyutoku},\ and\
  \citenamefont {Shibata}}]{sekiguchi2012oct}%
  \BibitemOpen
  \bibfield  {author} {\bibinfo {author} {\bibfnamefont {Y.}~\bibnamefont
  {Sekiguchi}}, \bibinfo {author} {\bibfnamefont {K.}~\bibnamefont {Kiuchi}},
  \bibinfo {author} {\bibfnamefont {K.}~\bibnamefont {Kyutoku}}, \ and\
  \bibinfo {author} {\bibfnamefont {M.}~\bibnamefont {Shibata}},\ }\href@noop
  {} {\bibfield  {journal} {\bibinfo  {journal} {Prog. Theor. Exp. Phys.}\
  }\textbf {\bibinfo {volume} {2012}},\ \bibinfo {pages} {01A304} (\bibinfo
  {year} {2012})}\BibitemShut {NoStop}%
\bibitem [{\citenamefont {Thorne}(1981)}]{thorne1981feb}%
  \BibitemOpen
  \bibfield  {author} {\bibinfo {author} {\bibfnamefont {K.~S.}\ \bibnamefont
  {Thorne}},\ }\href {\doibase 10.1093/mnras/194.2.439} {\bibfield  {journal}
  {\bibinfo  {journal} {Mon. Not. R. Astron. Soc.}\ }\textbf {\bibinfo {volume}
  {194}},\ \bibinfo {pages} {439} (\bibinfo {year} {1981})}\BibitemShut
  {NoStop}%
\bibitem [{\citenamefont {Shibata}\ \emph {et~al.}(2011)\citenamefont
  {Shibata}, \citenamefont {Kiuchi}, \citenamefont {Sekiguchi},\ and\
  \citenamefont {Suwa}}]{shibata2011jun}%
  \BibitemOpen
  \bibfield  {author} {\bibinfo {author} {\bibfnamefont {M.}~\bibnamefont
  {Shibata}}, \bibinfo {author} {\bibfnamefont {K.}~\bibnamefont {Kiuchi}},
  \bibinfo {author} {\bibfnamefont {Y.}~\bibnamefont {Sekiguchi}}, \ and\
  \bibinfo {author} {\bibfnamefont {Y.}~\bibnamefont {Suwa}},\ }\href {\doibase
  10.1143/PTP.125.1255} {\bibfield  {journal} {\bibinfo  {journal} {Prog.
  Theor. Phys.}\ }\textbf {\bibinfo {volume} {125}},\ \bibinfo {pages} {1255}
  (\bibinfo {year} {2011})}\BibitemShut {NoStop}%
\bibitem [{\citenamefont {Fujibayashi}\ \emph
  {et~al.}(2020{\natexlab{c}})\citenamefont {Fujibayashi}, \citenamefont
  {Wanajo}, \citenamefont {Kiuchi}, \citenamefont {Kyutoku}, \citenamefont
  {Sekiguchi},\ and\ \citenamefont {Shibata}}]{fujibayashi2020sep}%
  \BibitemOpen
  \bibfield  {author} {\bibinfo {author} {\bibfnamefont {S.}~\bibnamefont
  {Fujibayashi}}, \bibinfo {author} {\bibfnamefont {S.}~\bibnamefont {Wanajo}},
  \bibinfo {author} {\bibfnamefont {K.}~\bibnamefont {Kiuchi}}, \bibinfo
  {author} {\bibfnamefont {K.}~\bibnamefont {Kyutoku}}, \bibinfo {author}
  {\bibfnamefont {Y.}~\bibnamefont {Sekiguchi}}, \ and\ \bibinfo {author}
  {\bibfnamefont {M.}~\bibnamefont {Shibata}},\ }\href {\doibase
  10.3847/1538-4357/abafc2} {\bibfield  {journal} {\bibinfo  {journal} {\apj}\
  }\textbf {\bibinfo {volume} {901}},\ \bibinfo {pages} {122} (\bibinfo {year}
  {2020}{\natexlab{c}})}\BibitemShut {NoStop}%
\bibitem [{\citenamefont {Banik}\ \emph {et~al.}(2014)\citenamefont {Banik},
  \citenamefont {Hempel},\ and\ \citenamefont {Bandyopadhyay}}]{banik2014sep}%
  \BibitemOpen
  \bibfield  {author} {\bibinfo {author} {\bibfnamefont {S.}~\bibnamefont
  {Banik}}, \bibinfo {author} {\bibfnamefont {M.}~\bibnamefont {Hempel}}, \
  and\ \bibinfo {author} {\bibfnamefont {D.}~\bibnamefont {Bandyopadhyay}},\
  }\href {\doibase 10.1088/0067-0049/214/2/22} {\bibfield  {journal} {\bibinfo
  {journal} {\apj}\ }\textbf {\bibinfo {volume} {214}},\ \bibinfo {pages} {22}
  (\bibinfo {year} {2014})}\BibitemShut {NoStop}%
\bibitem [{\citenamefont {Timmes}\ and\ \citenamefont
  {Swesty}(2000)}]{timmes2000}%
  \BibitemOpen
  \bibfield  {author} {\bibinfo {author} {\bibfnamefont {F.~X.}\ \bibnamefont
  {Timmes}}\ and\ \bibinfo {author} {\bibfnamefont {F.~D.}\ \bibnamefont
  {Swesty}},\ }\href {\doibase 10.1086/313304} {\bibfield  {journal} {\bibinfo
  {journal} {\apj}\ }\textbf {\bibinfo {volume} {126}},\ \bibinfo {pages} {501}
  (\bibinfo {year} {2000})}\BibitemShut {NoStop}%
\bibitem [{\citenamefont {{Miller}}\ \emph {et~al.}(2019)\citenamefont
  {{Miller}}, \citenamefont {{Lamb}}, \citenamefont {{Dittmann}}, \citenamefont
  {{Bogdanov}}, \citenamefont {{Arzoumanian}}, \citenamefont {{Gendreau}},
  \citenamefont {{Guillot}}, \citenamefont {{Harding}}, \citenamefont {{Ho}},
  \citenamefont {{Lattimer}}, \citenamefont {{Ludlam}}, \citenamefont
  {{Mahmoodifar}}, \citenamefont {{Morsink}}, \citenamefont {{Ray}},
  \citenamefont {{Strohmayer}}, \citenamefont {{Wood}}, \citenamefont
  {{Enoto}}, \citenamefont {{Foster}}, \citenamefont {{Okajima}}, \citenamefont
  {{Prigozhin}},\ and\ \citenamefont {{Soong}}}]{miller2019dec}%
  \BibitemOpen
  \bibfield  {author} {\bibinfo {author} {\bibfnamefont {M.~C.}\ \bibnamefont
  {{Miller}}}, \bibinfo {author} {\bibfnamefont {F.~K.}\ \bibnamefont
  {{Lamb}}}, \bibinfo {author} {\bibfnamefont {A.~J.}\ \bibnamefont
  {{Dittmann}}}, \bibinfo {author} {\bibfnamefont {S.}~\bibnamefont
  {{Bogdanov}}}, \bibinfo {author} {\bibfnamefont {Z.}~\bibnamefont
  {{Arzoumanian}}}, \bibinfo {author} {\bibfnamefont {K.~C.}\ \bibnamefont
  {{Gendreau}}}, \bibinfo {author} {\bibfnamefont {S.}~\bibnamefont
  {{Guillot}}}, \bibinfo {author} {\bibfnamefont {A.~K.}\ \bibnamefont
  {{Harding}}}, \bibinfo {author} {\bibfnamefont {W.~C.~G.}\ \bibnamefont
  {{Ho}}}, \bibinfo {author} {\bibfnamefont {J.~M.}\ \bibnamefont
  {{Lattimer}}}, \bibinfo {author} {\bibfnamefont {R.~M.}\ \bibnamefont
  {{Ludlam}}}, \bibinfo {author} {\bibfnamefont {S.}~\bibnamefont
  {{Mahmoodifar}}}, \bibinfo {author} {\bibfnamefont {S.~M.}\ \bibnamefont
  {{Morsink}}}, \bibinfo {author} {\bibfnamefont {P.~S.}\ \bibnamefont
  {{Ray}}}, \bibinfo {author} {\bibfnamefont {T.~E.}\ \bibnamefont
  {{Strohmayer}}}, \bibinfo {author} {\bibfnamefont {K.~S.}\ \bibnamefont
  {{Wood}}}, \bibinfo {author} {\bibfnamefont {T.}~\bibnamefont {{Enoto}}},
  \bibinfo {author} {\bibfnamefont {R.}~\bibnamefont {{Foster}}}, \bibinfo
  {author} {\bibfnamefont {T.}~\bibnamefont {{Okajima}}}, \bibinfo {author}
  {\bibfnamefont {G.}~\bibnamefont {{Prigozhin}}}, \ and\ \bibinfo {author}
  {\bibfnamefont {Y.}~\bibnamefont {{Soong}}},\ }\href {\doibase
  10.3847/2041-8213/ab50c5} {\bibfield  {journal} {\bibinfo  {journal}
  {Astrophys. J. Lett.}\ }\textbf {\bibinfo {volume} {887}},\ \bibinfo {eid}
  {L24} (\bibinfo {year} {2019})}\BibitemShut {NoStop}%
\bibitem [{\citenamefont {{Balbus}}\ and\ \citenamefont
  {{Hawley}}(1991)}]{balbus1991}%
  \BibitemOpen
  \bibfield  {author} {\bibinfo {author} {\bibfnamefont {S.~A.}\ \bibnamefont
  {{Balbus}}}\ and\ \bibinfo {author} {\bibfnamefont {J.~F.}\ \bibnamefont
  {{Hawley}}},\ }\href {\doibase 10.1086/170270} {\bibfield  {journal}
  {\bibinfo  {journal} {\apj}\ }\textbf {\bibinfo {volume} {376}},\ \bibinfo
  {pages} {214} (\bibinfo {year} {1991})}\BibitemShut {NoStop}%
\bibitem [{\citenamefont {Balbus}\ and\ \citenamefont
  {Hawley}(1998)}]{balbus1998}%
  \BibitemOpen
  \bibfield  {author} {\bibinfo {author} {\bibfnamefont {S.~A.}\ \bibnamefont
  {Balbus}}\ and\ \bibinfo {author} {\bibfnamefont {J.~F.}\ \bibnamefont
  {Hawley}},\ }\href {\doibase 10.1103/RevModPhys.70.1} {\bibfield  {journal}
  {\bibinfo  {journal} {Rev. Mod. Phys.}\ }\textbf {\bibinfo {volume} {70}},\
  \bibinfo {pages} {1} (\bibinfo {year} {1998})}\BibitemShut {NoStop}%
\bibitem [{\citenamefont {Masada}\ and\ \citenamefont
  {Sano}(2008)}]{masada2008}%
  \BibitemOpen
  \bibfield  {author} {\bibinfo {author} {\bibfnamefont {Y.}~\bibnamefont
  {Masada}}\ and\ \bibinfo {author} {\bibfnamefont {T.}~\bibnamefont {Sano}},\
  }\href {\doibase 10.1086/592601} {\bibfield  {journal} {\bibinfo  {journal}
  {\apj}\ }\textbf {\bibinfo {volume} {689}},\ \bibinfo {pages} {1234}
  (\bibinfo {year} {2008})}\BibitemShut {NoStop}%
\bibitem [{\citenamefont {Guilet}\ \emph {et~al.}(2017)\citenamefont {Guilet},
  \citenamefont {Bauswein}, \citenamefont {Just},\ and\ \citenamefont
  {Janka}}]{guilet2016}%
  \BibitemOpen
  \bibfield  {author} {\bibinfo {author} {\bibfnamefont {J.}~\bibnamefont
  {Guilet}}, \bibinfo {author} {\bibfnamefont {A.}~\bibnamefont {Bauswein}},
  \bibinfo {author} {\bibfnamefont {O.}~\bibnamefont {Just}}, \ and\ \bibinfo
  {author} {\bibfnamefont {H.-T.}\ \bibnamefont {Janka}},\ }\href {\doibase
  10.1093/mnras/stx1739} {\bibfield  {journal} {\bibinfo  {journal} {Mon. Not.
  R. Astron. Soc.}\ }\textbf {\bibinfo {volume} {471}},\ \bibinfo {pages}
  {1879} (\bibinfo {year} {2017})}\BibitemShut {NoStop}%
\bibitem [{\citenamefont {Brandenburg}\ and\ \citenamefont
  {Subramanian}(2005)}]{Brandenburg2005}%
  \BibitemOpen
  \bibfield  {author} {\bibinfo {author} {\bibfnamefont {A.}~\bibnamefont
  {Brandenburg}}\ and\ \bibinfo {author} {\bibfnamefont {K.}~\bibnamefont
  {Subramanian}},\ }\href {\doibase
  https://doi.org/10.1016/j.physrep.2005.06.005} {\bibfield  {journal}
  {\bibinfo  {journal} {Phys. Rep.}\ }\textbf {\bibinfo {volume} {417}},\
  \bibinfo {pages} {1} (\bibinfo {year} {2005})}\BibitemShut {NoStop}%
\bibitem [{\citenamefont {Blandford}\ and\ \citenamefont
  {Payne}(1982)}]{blandford1982}%
  \BibitemOpen
  \bibfield  {author} {\bibinfo {author} {\bibfnamefont {R.~D.}\ \bibnamefont
  {Blandford}}\ and\ \bibinfo {author} {\bibfnamefont {D.~G.}\ \bibnamefont
  {Payne}},\ }\href {\doibase 10.1093/mnras/199.4.883} {\bibfield  {journal}
  {\bibinfo  {journal} {Mon. Not. R. Astron. Soc.}\ }\textbf {\bibinfo {volume}
  {199}},\ \bibinfo {pages} {883} (\bibinfo {year} {1982})}\BibitemShut
  {NoStop}%
\bibitem [{\citenamefont {{Fuller}}\ \emph {et~al.}(1985)\citenamefont
  {{Fuller}}, \citenamefont {{Fowler}},\ and\ \citenamefont
  {{Newman}}}]{fuller1985jun}%
  \BibitemOpen
  \bibfield  {author} {\bibinfo {author} {\bibfnamefont {G.~M.}\ \bibnamefont
  {{Fuller}}}, \bibinfo {author} {\bibfnamefont {W.~A.}\ \bibnamefont
  {{Fowler}}}, \ and\ \bibinfo {author} {\bibfnamefont {M.~J.}\ \bibnamefont
  {{Newman}}},\ }\href {\doibase 10.1086/163208} {\bibfield  {journal}
  {\bibinfo  {journal} {\apj}\ }\textbf {\bibinfo {volume} {293}},\ \bibinfo
  {pages} {1} (\bibinfo {year} {1985})}\BibitemShut {NoStop}%
\bibitem [{\citenamefont {{Metzger}}\ \emph {et~al.}(2008)\citenamefont
  {{Metzger}}, \citenamefont {{Piro}},\ and\ \citenamefont
  {{Quataert}}}]{Metzger2008}%
  \BibitemOpen
  \bibfield  {author} {\bibinfo {author} {\bibfnamefont {B.~D.}\ \bibnamefont
  {{Metzger}}}, \bibinfo {author} {\bibfnamefont {A.~L.}\ \bibnamefont
  {{Piro}}}, \ and\ \bibinfo {author} {\bibfnamefont {E.}~\bibnamefont
  {{Quataert}}},\ }\href {\doibase 10.1111/j.1365-2966.2008.13789.x} {\bibfield
   {journal} {\bibinfo  {journal} {Mon. Not. R. Astron. Soc.}\ }\textbf
  {\bibinfo {volume} {390}},\ \bibinfo {pages} {781} (\bibinfo {year}
  {2008})}\BibitemShut {NoStop}%
\bibitem [{\citenamefont {Liska}\ \emph {et~al.}(2020)\citenamefont {Liska},
  \citenamefont {Tchekhovskoy},\ and\ \citenamefont {Quataert}}]{liska2020}%
  \BibitemOpen
  \bibfield  {author} {\bibinfo {author} {\bibfnamefont {M.}~\bibnamefont
  {Liska}}, \bibinfo {author} {\bibfnamefont {A.}~\bibnamefont {Tchekhovskoy}},
  \ and\ \bibinfo {author} {\bibfnamefont {E.}~\bibnamefont {Quataert}},\
  }\href {\doibase 10.1093/mnras/staa955} {\bibfield  {journal} {\bibinfo
  {journal} {Mon. Not. R. Astron. Soc.}\ }\textbf {\bibinfo {volume} {494}},\
  \bibinfo {pages} {3656} (\bibinfo {year} {2020})}\BibitemShut {NoStop}%
\bibitem [{\citenamefont {Blandford}\ and\ \citenamefont
  {Znajek}(1977)}]{blandford1977}%
  \BibitemOpen
  \bibfield  {author} {\bibinfo {author} {\bibfnamefont {R.~D.}\ \bibnamefont
  {Blandford}}\ and\ \bibinfo {author} {\bibfnamefont {R.~L.}\ \bibnamefont
  {Znajek}},\ }\href {\doibase 10.1093/mnras/179.3.433} {\bibfield  {journal}
  {\bibinfo  {journal} {Mon. Not. R. Astron. Soc.}\ }\textbf {\bibinfo {volume}
  {179}},\ \bibinfo {pages} {433} (\bibinfo {year} {1977})}\BibitemShut
  {NoStop}%
\bibitem [{\citenamefont {Shibata}\ \emph {et~al.}(2006)\citenamefont
  {Shibata}, \citenamefont {Liu}, \citenamefont {Shapiro},\ and\ \citenamefont
  {Stephens}}]{shibata2006}%
  \BibitemOpen
  \bibfield  {author} {\bibinfo {author} {\bibfnamefont {M.}~\bibnamefont
  {Shibata}}, \bibinfo {author} {\bibfnamefont {Y.~T.}\ \bibnamefont {Liu}},
  \bibinfo {author} {\bibfnamefont {S.~L.}\ \bibnamefont {Shapiro}}, \ and\
  \bibinfo {author} {\bibfnamefont {B.~C.}\ \bibnamefont {Stephens}},\ }\href
  {\doibase 10.1103/PhysRevD.74.104026} {\bibfield  {journal} {\bibinfo
  {journal} {Phys. Rev. D}\ }\textbf {\bibinfo {volume} {74}},\ \bibinfo
  {pages} {104026} (\bibinfo {year} {2006})}\BibitemShut {NoStop}%
\bibitem [{\citenamefont {{Bardeen}}\ \emph {et~al.}(1972)\citenamefont
  {{Bardeen}}, \citenamefont {{Press}},\ and\ \citenamefont
  {{Teukolsky}}}]{bardeen1972dec}%
  \BibitemOpen
  \bibfield  {author} {\bibinfo {author} {\bibfnamefont {J.~M.}\ \bibnamefont
  {{Bardeen}}}, \bibinfo {author} {\bibfnamefont {W.~H.}\ \bibnamefont
  {{Press}}}, \ and\ \bibinfo {author} {\bibfnamefont {S.~A.}\ \bibnamefont
  {{Teukolsky}}},\ }\href {\doibase 10.1086/151796} {\bibfield  {journal}
  {\bibinfo  {journal} {Astrophys. J.}\ }\textbf {\bibinfo {volume} {178}},\
  \bibinfo {pages} {347} (\bibinfo {year} {1972})}\BibitemShut {NoStop}%
\bibitem [{\citenamefont {{De}}\ and\ \citenamefont
  {{Siegel}}(2021)}]{de2021nov}%
  \BibitemOpen
  \bibfield  {author} {\bibinfo {author} {\bibfnamefont {S.}~\bibnamefont
  {{De}}}\ and\ \bibinfo {author} {\bibfnamefont {D.~M.}\ \bibnamefont
  {{Siegel}}},\ }\href {\doibase 10.3847/1538-4357/ac110b} {\bibfield
  {journal} {\bibinfo  {journal} {\apj}\ }\textbf {\bibinfo {volume} {921}},\
  \bibinfo {eid} {94} (\bibinfo {year} {2021})}\BibitemShut {NoStop}%
\bibitem [{\citenamefont {{Beloborodov}}(2003)}]{Beloborodov2003may}%
  \BibitemOpen
  \bibfield  {author} {\bibinfo {author} {\bibfnamefont {A.~M.}\ \bibnamefont
  {{Beloborodov}}},\ }\href {\doibase 10.1086/374217} {\bibfield  {journal}
  {\bibinfo  {journal} {\apj}\ }\textbf {\bibinfo {volume} {588}},\ \bibinfo
  {pages} {931} (\bibinfo {year} {2003})}\BibitemShut {NoStop}%
\bibitem [{\citenamefont {Hotokezaka}\ \emph {et~al.}(2013)\citenamefont
  {Hotokezaka}, \citenamefont {Kiuchi}, \citenamefont {Kyutoku}, \citenamefont
  {Okawa}, \citenamefont {Sekiguchi}, \citenamefont {Shibata},\ and\
  \citenamefont {Taniguchi}}]{hotokezaka2013}%
  \BibitemOpen
  \bibfield  {author} {\bibinfo {author} {\bibfnamefont {K.}~\bibnamefont
  {Hotokezaka}}, \bibinfo {author} {\bibfnamefont {K.}~\bibnamefont {Kiuchi}},
  \bibinfo {author} {\bibfnamefont {K.}~\bibnamefont {Kyutoku}}, \bibinfo
  {author} {\bibfnamefont {H.}~\bibnamefont {Okawa}}, \bibinfo {author}
  {\bibfnamefont {Y.}~\bibnamefont {Sekiguchi}}, \bibinfo {author}
  {\bibfnamefont {M.}~\bibnamefont {Shibata}}, \ and\ \bibinfo {author}
  {\bibfnamefont {K.}~\bibnamefont {Taniguchi}},\ }\href {\doibase
  10.1103/PhysRevD.87.024001} {\bibfield  {journal} {\bibinfo  {journal} {Phys.
  Rev. D}\ }\textbf {\bibinfo {volume} {87}},\ \bibinfo {pages} {024001}
  (\bibinfo {year} {2013})}\BibitemShut {NoStop}%
\bibitem [{\citenamefont {Wanajo}\ \emph {et~al.}(2014)\citenamefont {Wanajo},
  \citenamefont {Sekiguchi}, \citenamefont {Nishimura}, \citenamefont {Kiuchi},
  \citenamefont {Kyutoku},\ and\ \citenamefont {Shibata}}]{wanajo2014}%
  \BibitemOpen
  \bibfield  {author} {\bibinfo {author} {\bibfnamefont {S.}~\bibnamefont
  {Wanajo}}, \bibinfo {author} {\bibfnamefont {Y.}~\bibnamefont {Sekiguchi}},
  \bibinfo {author} {\bibfnamefont {N.}~\bibnamefont {Nishimura}}, \bibinfo
  {author} {\bibfnamefont {K.}~\bibnamefont {Kiuchi}}, \bibinfo {author}
  {\bibfnamefont {K.}~\bibnamefont {Kyutoku}}, \ and\ \bibinfo {author}
  {\bibfnamefont {M.}~\bibnamefont {Shibata}},\ }\href {\doibase
  10.1088/2041-8205/789/2/l39} {\bibfield  {journal} {\bibinfo  {journal}
  {\apj}\ }\textbf {\bibinfo {volume} {789}},\ \bibinfo {pages} {L39} (\bibinfo
  {year} {2014})}\BibitemShut {NoStop}%
\bibitem [{\citenamefont {Kawaguchi}\ \emph {et~al.}(2020)\citenamefont
  {Kawaguchi}, \citenamefont {Shibata},\ and\ \citenamefont
  {Tanaka}}]{kawaguchi2020}%
  \BibitemOpen
  \bibfield  {author} {\bibinfo {author} {\bibfnamefont {K.}~\bibnamefont
  {Kawaguchi}}, \bibinfo {author} {\bibfnamefont {M.}~\bibnamefont {Shibata}},
  \ and\ \bibinfo {author} {\bibfnamefont {M.}~\bibnamefont {Tanaka}},\ }\href
  {\doibase 10.3847/1538-4357/ab61f6} {\bibfield  {journal} {\bibinfo
  {journal} {\apj}\ }\textbf {\bibinfo {volume} {889}},\ \bibinfo {pages} {171}
  (\bibinfo {year} {2020})}\BibitemShut {NoStop}%
\bibitem [{\citenamefont {Tchekhovskoy}\ \emph {et~al.}(2011)\citenamefont
  {Tchekhovskoy}, \citenamefont {Narayan},\ and\ \citenamefont
  {McKinney}}]{tchekhovskoy2011}%
  \BibitemOpen
  \bibfield  {author} {\bibinfo {author} {\bibfnamefont {A.}~\bibnamefont
  {Tchekhovskoy}}, \bibinfo {author} {\bibfnamefont {R.}~\bibnamefont
  {Narayan}}, \ and\ \bibinfo {author} {\bibfnamefont {J.~C.}\ \bibnamefont
  {McKinney}},\ }\href {\doibase 10.1111/j.1745-3933.2011.01147.x} {\bibfield
  {journal} {\bibinfo  {journal} {Mon. Not. R. Astron. Soc.}\ }\textbf
  {\bibinfo {volume} {418}},\ \bibinfo {pages} {L79} (\bibinfo {year}
  {2011})}\BibitemShut {NoStop}%
\bibitem [{\citenamefont {{Uchida}}\ \emph {et~al.}(2017)\citenamefont
  {{Uchida}}, \citenamefont {{Shibata}}, \citenamefont {{Yoshida}},
  \citenamefont {{Sekiguchi}},\ and\ \citenamefont {{Umeda}}}]{uchida2017oct}%
  \BibitemOpen
  \bibfield  {author} {\bibinfo {author} {\bibfnamefont {H.}~\bibnamefont
  {{Uchida}}}, \bibinfo {author} {\bibfnamefont {M.}~\bibnamefont {{Shibata}}},
  \bibinfo {author} {\bibfnamefont {T.}~\bibnamefont {{Yoshida}}}, \bibinfo
  {author} {\bibfnamefont {Y.}~\bibnamefont {{Sekiguchi}}}, \ and\ \bibinfo
  {author} {\bibfnamefont {H.}~\bibnamefont {{Umeda}}},\ }\href {\doibase
  10.1103/PhysRevD.96.083016} {\bibfield  {journal} {\bibinfo  {journal}
  {\prd}\ }\textbf {\bibinfo {volume} {96}},\ \bibinfo {eid} {083016} (\bibinfo
  {year} {2017})}\BibitemShut {NoStop}%
\end{thebibliography}%

\end{document}